\RequirePackage{ifpdf}
\documentclass[hyper,12pt,A4paper]{JHEP3}

\usepackage{graphicx}
\usepackage{latexsym,amsmath,amsfonts,amssymb}
\usepackage{mathrsfs}
\usepackage[makeroom]{cancel}
\usepackage{bbm}
\usepackage{bm}
\usepackage{subfigure}
\usepackage{paralist}
\usepackage{url}

\numberwithin{equation}{section}

\newcommand{\nn}{\nonumber}
\newcommand{\mat}[1]{\begin{pmatrix} #1 \end{pmatrix}}

\newcommand{\be}{\begin{equation}} 
\newcommand{\ee}{\end{equation}}
\newcommand{\bea}{\begin{equation} \begin{aligned}} \newcommand{\eea}{\end{aligned} \end{equation}}

\newcommand{\bit}{\begin{itemize}} 
\newcommand{\eit}{\end{itemize}}

\newcommand{\cF}{\mathcal{F}}

\newcommand{\cL}{\mathcal{L}}

\newcommand{\bW}{\mathbb{W}}

\newcommand{\Z}{\mathbb{Z}}
\newcommand{\C}{\mathbb{C}}
\newcommand{\R}{\mathbb{R}}

\renewcommand{\t}{\widetilde }
\renewcommand{\d}{\partial }

\renewcommand{\b}{\bar }

\newcommand{\alphadot}{{\dot\alpha}}
\newcommand{\betadot}{{\dot\beta}}

\newcommand{\half}{{1\over 2}}

\newcommand{\bz}{{\b z}}
\newcommand{\bw}{{\b w}}

\newcommand{\CA}{\mathcal{A}}

\newcommand{\CC}{\mathcal{C}}

\newcommand{\CE}{\mathcal{E}}
\newcommand{\CF}{\mathcal{F}}

\newcommand{\CH}{\mathcal{H}}
\newcommand{\CI}{\mathcal{I}}
\newcommand{\CJ}{\mathcal{J}}

\newcommand{\CL}{\mathcal{L}}
\newcommand{\CM}{\mathcal{M}}
\newcommand{\CN}{\mathcal{N}}
\newcommand{\CO}{\mathcal{O}}
\newcommand{\CP}{\mathcal{P}}
\newcommand{\CQ}{\mathcal{Q}}
\newcommand{\CR}{\mathcal{R}}
\newcommand{\CS}{\mathcal{S}}
\newcommand{\CT}{\mathcal{T}}

\newcommand{\CV}{\mathcal{V}}
\newcommand{\CW}{\mathcal{W}}

\newcommand{\CY}{\mathcal{Y}}

\newcommand{\FR}{\mathfrak{R}}
\newcommand{\Fg}{\mathfrak{g}}
\newcommand{\Fh}{\mathfrak{h}}

\newcommand{\GG}{\mathbf{G}}

\newcommand{\rk}{{{\rm rk}(\GG)}}

\newcommand{\m}{\mathfrak{m}}
\newcommand{\n}{\mathfrak{n}}

\newcommand{\h}{\hat}

\newcommand{\Mgp}{\CM_{g,p}}

\DeclareMathOperator{\Tr}{Tr}
\DeclareMathOperator{\tr}{tr}

\newcommand{\eps}{\epsilon}
\newcommand{\ep}{\varepsilon}

\newcommand{\SL}{{\mathscr L}}

\newcommand{\ov}{\over}

\newcommand{\fM}{\mathfrak{M}}

\newcommand{\dilog}{{\text{Li}_2}}

\newcommand{\bA}{{\bf A}}
\newcommand{\bV}{{\bf V}}
\newcommand{\bH}{{\bf H}}

\newcommand{\by}{{\bf y}}
\title{ 't Hooft anomalies and the holomorphy of supersymmetric partition functions
}

\author{Cyril Closset,$^{\flat}$ Lorenzo Di Pietro,$^\sharp$  Heeyeon Kim$^{\flat}$  \\

{}$^{\flat}$ Mathematical Institute, University of Oxford\\ Woodstock Road, Oxford, OX2 6GG, United Kingdom\\
{}$^{\sharp}$ Perimeter Institute for Theoretical Physics\\
31 Caroline Street North, Waterloo, N2L 2Y5, Ontario, Canada
}

\preprint{}

\abstract{We study the dependence of supersymmetric partition functions on continuous parameters for the flavor symmetry group, $G_F$, for 2d $\mathcal{N} = (0,2)$ and 4d $\mathcal{N}=1$ supersymmetric quantum field theories. In any diffeomorphism-invariant scheme and in the presence of $G_F$ 't Hooft anomalies, the supersymmetric Ward identities imply that the partition function has a non-holomorphic dependence on the flavor parameters. We show this explicitly for the 2d torus partition function, $Z_{T^2}$, and for a large class of 4d partition functions on half-BPS four-manifolds, $Z_{\mathcal{M}_4}$---in particular, for $\mathcal{M}_4=S^3 \times S^1$ and $\CM_4=\Sigma_g \times T^2$.  We propose a new expression for $Z_{\mathcal{M}_{d-1} \times S^1}$, which differs from earlier holomorphic results by the introduction of a non-holomorphic ``Casimir'' pre-factor. The latter is fixed by studying the ``high temperature'' limit of  the partition function. Our proposal agrees with the supersymmetric Ward identities, and with explicit calculations of the absolute value of the partition function using a gauge-invariant zeta-function regularization.}

\begin{document}

\section{Introduction}
 In this paper, we explore intriguing properties of supersymmetric partition functions in even space-time dimensions.%
 \footnote{For a review of exact results for supersymmetric partition functions, see {\it e.g.}  \protect\cite{Pestun:2016zxk}.} In two dimensions, we consider the $T^2$ partition function of 2d $\CN=(0,2)$ supersymmetric theories, or elliptic genus~\cite{Witten:1986bf}. In four dimensions, we consider the $S^3\times S^1$ supersymmetric partition function, which computes the $\CN=1$ supersymmetric index \cite{Romelsberger:2005eg, Kinney:2005ej}.%
 \footnote{We will focus on the ${\bf q}={\bf p}$ limit of the three-sphere index. We will also consider a more general class of generalized indices, replacing the $S^3$ with $\CM_{g,p}$, a degree-$p$ circle bundle over a Riemann surface of genus $g$ \protect\cite{Closset:2017bse}.} In both cases, the partition functions depend non-trivially on some continuous parameters:
\be\label{Z holo intro}
Z_{T^2}(\nu, \tau)~, \qquad \qquad Z_{S^3\times S^1}(\nu, \tau)~.
\ee 
Here, there is one ``flavor parameter,'' $\nu$, for each $U(1)$ current, with $\prod_\alpha U(1)_\alpha$ the maximal torus of the flavor (non-$R$) symmetry group, $G_F$, of the  field theory. The parameter $\tau$, on the other hand, denotes a ``geometric parameter,'' corresponding to the complex structure modulus of the torus in $2d$, or to a complex structure modulus on $S^3 \times S^1$ seen as a Hopf surface \cite{Closset:2013vra}. This setup preserves two supercharges of opposite $R$-charges, $\CQ$ and $\t \CQ$.

Through a number of explicit computations---see in particular \cite{Benini:2013nda, Benini:2013xpa, Closset:2013sxa, Assel:2014paa, Nishioka:2014zpa}---, it was found that these supersymmetric partition functions are locally {\it holomorphic} in the complex parameters $\nu$ and $\tau$. Moreover, there is a simple  {\it a-priori} argument for the holomorphy of \eqref{Z holo intro}, which is that the anti-holomorphic parameters, $\b\nu$ and $\b\tau$, couple to $\CQ$-exact operators in conserved-current supermultiplets \cite{Closset:2013vra}. 

The holomorphy of \eqref{Z holo intro} in the flavor parameters $\nu$, however, clashes with gauge invariance whenever the flavor symmetry, $G_F$, has non-vanishing {\it 't Hooft anomalies.} To see why, consider the effective action for the quantum field theory coupled to a background gauge field for the flavor symmetry:
\begin{align}\label{W def intro}
W[a] & \equiv - \log Z[a]~,\\
Z[a] & = \int [D\varphi] \, e^{- \int d^dx \left( \SL_0[\varphi] + a_\mu j^\mu + \cdots \right)}~,\nonumber
\end{align}
where the ellipsis denotes additional couplings designed to make the full classical action gauge-invariant. A 't Hooft anomaly is a violation of the gauge invariance of the effective action \eqref{W def intro} which cannot be removed by a local counterterm.%
\footnote{See, for instance, section 2 of \protect\cite{Cordova:2018cvg} for a nice and detailed discussion of 't Hooft anomalies.} 
Under a gauge transformation $\delta_\alpha a_\mu = D_\mu \alpha$, with $\alpha=\alpha(x)$ a gauge parameter, it takes the form:
\be\label{delta W intro 1}
\delta_\alpha W[a]=i \CA_F  \int_{\CM_{d} }\alpha\; \omega_{d}(a)~,
\ee
on any $d$-manifold $\CM_d$ (with $d$ even), where the real $d$-form $\omega_{d}(a)$ is local in the background gauge field. The form $\omega_d$ is universal and well-known, in any dimension and for any compact Lie group $G_F$. The coefficient $\CA_F \in \Z$ in \eqref{delta W intro 1} depends on the particular theory; in UV-free theories, it is  determined by the flavor charges of the chiral fermions. 

The crucial aspect of \eqref{delta W intro 1}, for our purpose, is that it is {\it purely imaginary,} given a real background gauge field $a_\mu$. Therefore, the absolute value of the partition function, $|Z|$, is gauge invariant, as in the well-known case of free fermions. In fact, we expect $|Z|$ to be invariant under both small and large gauge transformations, provided we work in a scheme that respects the consistent anomaly equation \eqref{delta W intro 1}. (This expectation will also be borne out by explicit computations.)

On the other hand, the absolute value of a supersymmetric partition function that would be truly {\it holomorphic} in the flavor parameters cannot be fully gauge invariant and consistent with \eqref{delta W intro 1}.
Consider, for instance, the supersymmetric partition function on $S^3 \times S^1$. It depends on the background $U(1)$ gauge field through two real parameters, which we denote by $a_t$ and $a_\psi$. They correspond to holonomies of the gauge field through the $S^1$ factor and through an $S^1$ Hopf fiber $\gamma$ inside $S^3$, respectively:
\be
a_t ={1\ov 2 \pi} \int_{S^1} a~, \qquad\qquad
a_\psi={1\ov 2 \pi} \int_{\gamma} a~.
\ee
The complex flavor parameter $\nu$ is defined in terms of these real parameters as:
\be\label{nu to apsi intro}
\nu = \tau a_\psi - a_t~.
\ee
The parameter $a_t$ is only defined modulo the identification $a_t\sim a_t-1$ (equivalently, $\nu \sim \nu+1$), which corresponds to a {\it large gauge transformation} along the $S^1$.
Under it, the $S^3 \times S^1$ partition function computed in \cite{Assel:2014paa} transforms as:%
 \footnote{There are additional contributions to these anomalous transformations from mixed anomalies between the flavor group and the $R$-symmetry, which we will discuss further in the main text.}
\be\label{IS3 lgt intro}
\CI_{S^3\times S^1}(\nu +1, \tau) = e^{\pi i \left({\CA_{qqq}\ov \tau^2} \left(\nu^2 + \nu + {1\ov 3}\right)  - {\CA_q\ov 6}\right)}\, \CI_{S^3\times S^1}(\nu, \tau)~.
\ee
where $\CA_{qqq}$ and $\CA_q$ are $U(1)$ 't Hooft anomalies (cubic and mixed $U(1)$-gravitational, respectively). Since the whole object $\CI_{S^3 \times S^1}(\nu, \tau)$ is holomorphic in $\nu$, its absolute value is clearly not---and could not be---gauge invariant under the large gauge transformation $\nu \sim \nu+1$. 

A completely analogous puzzle arises in the simpler case of the torus partition function, $Z_{T^2}(\nu, \tau)$, of a 2d $\CN=(0,2)$ gauge theory. The explicit localization result of \cite{Benini:2013nda, Benini:2013xpa} transforms non-trivially under the large gauge transformations $\nu \sim \nu+1$ and $\nu\sim \nu+ \tau$, and its absolute value is not gauge invariant. It is also interesting to study modular transformations of the torus; we expect the absolute value of the partition function to be modular invariant, which is not the case of the explicitly holomorphic answer for the elliptic genus.

The holomorphy of supersymmetric partition functions in the flavor parameters follow from a supersymmetric Ward identity \cite{Closset:2013vra, Closset:2014uda}, as already mentioned. We have:
\be\label{naive SUSY Ward id}
{\d \ov \d\b \nu} W(\nu, \tau) =\langle  \CJ \rangle = \langle \{ \CQ, \t j\} \rangle =0~,
\ee
schematically. Here, the operator $\CJ$, which happens to be the supersymmetry variation of a fermionic operator $\t j$, is the component of the conserved current $j^\mu$ that couples to the anti-holomorphic parameter $\b\nu$. In that last equality, we then use the fact that the expectation value of any $\CQ$-exact operator vanishes.
Thus, we seem to have a tension between:
\begin{itemize}
\item[(i)] Gauge invariance of the absolute value of the supersymmetric partition function $Z_{\CM_d}$, whenever the flavor symmetry has 't Hooft anomalies; then, $Z_{\CM_d}$ should be compatible with the anomalous $G_F$ gauge variation \eqref{delta W intro 1}.
\item[(ii)] Holomorphy of the supersymmetric partition function in the flavor parameters, which follows from supersymmetry according to \eqref{naive SUSY Ward id}.
\end{itemize}
In this paper, we resolve this apparent contradiction. The resolution is based on the fact that, whenever we insist on working in a diffeormorphism-invariant scheme,%
\footnote{Such a scheme always exists in four-dimensions. In 2d, we might have a gravitational anomaly. Our discussion can be generalized to that case as well.}
  the $G_F$ 't Hooft anomalies induce  specific quantum corrections to the supersymmetric Ward identities, which are determined by the Wess-Zumino consistency conditions. As a consequence, the relation \eqref{naive SUSY Ward id} is modified, and the partition function acquires a non-holomorphic dependence on the flavor parameters, that is precisely as required to make its absolute value gauge-invariant.

These quantum corrections to supersymmetric Ward identities, also referred to as ``supersymmetry anomalies,''  were studied long ago in \cite{Itoyama:1985qi, Itoyama:1985ni}, and more recently in  \cite{Papadimitriou:2017kzw, An:2017ihs, Katsianis:2019hhg, Papadimitriou:2019gel, An:2019zok, An:2018roi, Papadimitriou:2019yug}. They follow from an anomalous transformation of the effective action:
\be\label{susy ward id intro notananomaly}
\delta_\zeta W \neq 0~,
\ee
where $\delta_\zeta$ denotes a supersymmetry variation in the Wess-Zumino (WZ) gauge  \cite{Wess:1974tw} for the background vector multiplet.
This requires some explaining, however, to avoid possible confusions. First, the anomalous supersymmetry variation \eqref{susy ward id intro notananomaly} does not appear with any  new anomaly coefficient; instead, the non-vanishing value of $\delta_\zeta W$ is entirely determined by the bosonic 't~Hooft anomalies for the flavor currents.  Secondly, one can understand \eqref{susy ward id intro notananomaly} as a consequence of working in the Wess-Zumino gauge, as explained {\it e.g.} in~\cite{Zumino:1985vr}. Recall that, in the WZ gauge, the supersymmetry transformations themselves depend on the (background) gauge field, because they are defined as the ``bare'' supersymmetry transformation, $\delta_\zeta^{(0)}$, plus a compensating gauge transformation, $\delta_{\Omega(\zeta)}$, namely:
\be
\delta_\zeta =\delta_\zeta^{(0)}+\delta_{\Omega(\zeta)}~.
\ee
Here, $\delta_{\Omega(\zeta)}$ is a specific supersymmetric gauge transformation whose chiral-multiplet-valued gauge parameter, $\Omega(\zeta)$, depends explicitly on $a_\mu$. In this language, the supersymmetry variation \eqref{susy ward id intro notananomaly} follows from:
\be\label{del Omega zeta intro}
\delta_\zeta^{(0)}W=0~, \qquad \qquad \delta_{\Omega(\zeta)}W   \propto \CA_F \neq 0~,
\ee
as we will explain in detail---in this language, \eqref{susy ward id intro notananomaly} is a straightforward consequence of the ordinary 't Hooft anomaly in the presence of supersymmetry. Of course, our physical conclusions below must be the same in any gauge; it is merely convenient to fix the WZ gauge from the start, as usually done in the discussion of supersymmetric partition functions. 

In general, the supersymmetric Ward identity \eqref{susy ward id intro notananomaly} receives contributions from both the ``flavor sector'' and the ``geometric sector'' (in particular, from the $U(1)_R$ symmetry 't Hooft anomalies). The background gauge fields for the former sector sit in vector multiplets, while the background fields for the latter (including the $U(1)_R$ gauge field) are part of a supergravity multiplet.%
\footnote{The geometric-sector contributions to $\delta_\zeta W$ can be understood as arising from fixing the WZ gauge in supergravity, similarly to the flavor sector contributions.} 
 In this work, for simplicity, we focus on the flavor sector and, thus, on the consequences of the flavor 't Hooft anomalies in the presence of supersymmetry.%

\subsection*{'t Hooft anomalies and supersymmetry}
Let us then study supersymmetric partition functions with backgrounds vector multiplets in the Wess-Zumino (WZ) gauge \cite{Wess:1974tw}.
Any background vector multiplet, $\CV_F$, has components:
\be
\CV_F= (a_\mu~,\, \lambda~,\, \t\lambda~,\, D)~,
\ee
in WZ gauge (in 2d or 4d), with $\lambda, \t\lambda$ the gauginos.
Then, the supersymmetry variation \eqref{susy ward id intro notananomaly} of the effective action takes the schematic form:
\be\label{susy anomaly intro}
\delta_\zeta W= \CA_F \int d^dx \sqrt{g} \, \zeta \Psi(\lambda, \t\lambda, a)~,\qquad\quad
\ee
where $\zeta$ is a supersymmetry parameter of $R$-charge $\pm 1$ and dimension $-\half$, and the  coefficient, $\CA_F$, is the same 't Hooft anomaly coefficient that appears in \eqref{delta W intro 1}. The local function $\Psi(\lambda, \t\lambda, a)$ in \eqref{susy anomaly intro} must be fermionic and of engineering dimension $d+\half$; in particular, it is at least linear in the gauginos (it is quadratic in the vector multiplet in 2d, and cubic in 4d).
This term was first computed  long ago \cite{Itoyama:1985qi, Itoyama:1985ni}, while a systematic analysis in 4d $\CN=1$ theories was recently carried out in \cite{Papadimitriou:2019yug}. 

In the WZ gauge  for $\CV_F$, the existence of the anomalous supersymmetry variation \eqref{susy anomaly intro} is implied by the Wess-Zumino consistency conditions for the anomalies \cite{Itoyama:1985qi}:
\be
[\delta_\alpha, \delta_\zeta] W=0~.
\ee
Since the supersymmetry variation of the gauge anomaly \eqref{delta W intro 1} is non-trivial:
\be
\delta_\zeta (\delta_\alpha W) \neq 0~,
\ee
it follows that $\delta_\alpha(\delta_\zeta W) \neq 0$, therefore $\delta_\zeta W$  itself must be non-zero.
Alternatively, one can simply compute \eqref{susy anomaly intro} by fixing the WZ gauge, as in \eqref{del Omega zeta intro}.

We then study $\delta_\zeta W$ on a {\it fixed} geometric background---more precisely, a fixed supergravity background {\it \`a la} Festuccia-Seiberg \cite{Festuccia:2011ws}. We will consider the torus in 2d, and half-BPS four-manifolds $\CM_4$ in 4d~\cite{Dumitrescu:2012ha, Klare:2012gn}. For instance, for 4d $\CN=1$ theories, the cubic 't Hooft anomaly:
\be\label{U1 anomaly 4d intro}
\delta_\alpha W = {i \CA_{qqq} \ov 24\pi^2} \int_{\CM_4} \alpha \, f \wedge f~,
\ee
for a $U(1)$ flavor symmetry, with $f=da$ the field strength of the background gauge field $a_\mu$ and $ \CA_{qqq}\in \Z$ the cubic anomaly coefficient, implies:
\be\label{susy anomaly Aqqq intro}
\delta_\zeta W = -{ \CA_{qqq} \ov 24\pi^2} \int_{\CM_4} d^4x \sqrt{g} \, \left(\epsilon^{\mu\nu\rho\sigma} \zeta \sigma_\mu \t\lambda \, a_\nu f_{\rho\sigma} -  3i \zeta \lambdaÂ \, \t\lambda\t\lambda \right)~,
\ee
which happens to take the same form as in flat space \cite{Itoyama:1985qi}.

\subsection*{The ``gauge-invariant,'' non-holomorphic partition function}
In this work, we distinguish between two distinct expressions for the supersymmetric partition function on $\CM_d$:
\begin{itemize}
\item[(i)] The ``gauge-invariant partition function,'' which we will generally simply call ``the partition function,'' and denote by $Z(\nu, \tau)$. Here, by ``gauge-invariant,'' we mean that $|Z(\nu, \tau)|^2$ is completely gauge invariant. This partition function is computed in a diff-invariant scheme compatible with the 't Hooft anomaly~\eqref{delta W intro 1}.
\item[(ii)] The ``holomorphic partition function,'' which is the one studied so far in the literature. We denote it by $\CI(\nu, \tau)$, to distinguish it from $Z(\nu, \tau)$. It is implicitly computed in a scheme that violates gauge and diffeomorphism invariance. 
\end{itemize}
In all the examples we will consider, the supersymmetric background is of the form $\CM_d \cong \CM_{d-1} \times S^1$, and the supersymmetric partition function can then be computed as an index, by canonical quantization on $\CM_{d-1}$. In 4d, the holomorphic partition function $\CI_{\CM_3 \times S^1}$ can be understood as a generalized 4d $\CN=1$ index, as discussed {\it e.g.} in \cite{Benini:2011nc, Nishioka:2014zpa, Benini:2015noa, Closset:2017bse} for a large class of geometries. For the $S^3 \times S^1$ partition function, we have the holomorphic result:
\be\label{CI S3 intro}
\CI_{S^3\times S^1}(\nu, \tau)= e^{2\pi i \tau \CE(\nu, \tau)} \, {\bf I}_{S^3}(\nu, \tau)~.
\ee
Here,  ${\bf I}_{S^3}(\nu, \tau)$ is the ordinary $\CN=1$ supersymmetric index \cite{Romelsberger:2005eg, Kinney:2005ej}, and the function $\CE(\nu, \tau)$ in \eqref{CI S3 intro} is the supersymmetric Casimir energy  \cite{Assel:2014paa, Assel:2015nca, Bobev:2015kza}:
\be
 \CE(\nu, \tau)= {\CA_{qqq}\nu^3 \ov 6 \tau^3}- {\CA_q \nu \ov 12 \tau}~.
 \ee
 The non-trivial large-gauge transformation in \eqref{IS3 lgt intro} comes entirely from the latter.%
 \footnote{This is specific to this simple case. On spaces of slightly more general topology, such as $\CM_{g,p} \times S^1$, there are other large gauge transformations under which the ``pure index,'' ${\bf I}_{\CM_3}$, also transforms non-trivially.}

We should compare this result to the expected properties of the ``gauge invariant'' partition function. Working in a diff-invariant scheme, we have an anomalous supersymmetry variation \eqref{susy anomaly intro}, which leads to an anomaly in the decoupling of some $\CQ$-exact operators  \cite{Papadimitriou:2017kzw}. Indeed, consider the naive Ward identity \eqref{naive SUSY Ward id}. The fermionic operator $\t j$ is precisely the operator that couples to the gaugino $\t\lambda$ in the background vector multiplet. Then, in the presence of any non-trivial contribution to $\delta_\zeta W$, we have:
\be\label{naive SUSY Ward id ammended}
{\d \ov \d\b \nu} W(\nu, \tau) =\langle  \CJ \rangle = \langle \{ \CQ, \t j\} \rangle = {\d \ov \d\t \lambda}(\delta_\zeta W) \neq 0~,
\ee
schematically. This ``holomorphy anomaly'' equation determines the non-holomorphic dependence of the  partition function on the flavor parameters $\nu, \b\nu$.

For instance, for the $S^3 \times S^1$ partition function, the contribution \eqref{susy anomaly Aqqq intro} from the cubic $U(1)$ anomaly to $\delta_\zeta W$ implies:
\be\label{holo anomaly intro}
{\d W\ov \d \b \nu}  =-{\pi i  \CA_{qqq}  \ov 6\tau_2}  \nu(\nu-\b\nu)~.
\ee
We should then revisit the supersymmetric localization computation of the $S^3 \times S^1$ partition function for $\CN=1$ gauge theories~\cite{Assel:2014paa} in light of the supersymmetric Ward identity \eqref{holo anomaly intro}.

Importantly, the above discussion does not invalidate the localization argument in any way---it is still true that one can localize supersymmetric gauge theories and compute at the weakly-coupled UV fixed point.%
\footnote{This is because the supersymmetric localization argument relies on the decoupling of $\CQ$-exact terms in the gauge sector---in terms of dynamical gauge fields---, and the relevant $\CQ$-exact operators do decouple because the gauge anomalies must vanish.}  The only subtle point is that various one-loop determinants have to be computed carefully, with appropriate regulators, depending on which scheme we decide to work in. 
To compute the partition function $Z(\nu, \tau)$, we must work in a scheme consistent with the anomalous variations~\eqref{U1 anomaly 4d intro} and \eqref{susy anomaly Aqqq intro}, and with diff-invariance. We then obtain results for the one-loop determinants that differ from the  fully holomorphic results obtained in \cite{Closset:2013sxa, Assel:2014paa}. The non-holomorphic dependence is determined by \eqref{holo anomaly intro}.

We show that the absolute value squared of the supersymmetric partition function $Z_{S^3 \times S^1}$ takes the explicit form:
\be\label{Zabs intro}
\big| Z_{S^3 \times S^1}(\nu, \tau)\big|^2 = e^{- {2\pi \tau_2\ov 3} \left(\CA_{qqq}  a_\psi^3 - {\CA_q \ov 2}a_\psi \right)}\, \big| {\bf I}_{S^3}(\nu, \tau) \big|^2~, 
\ee
with ${\bf I}_{S^3}$ the ordinary letter-counting index, and $a_\psi$ related to $\nu$ as in \eqref{nu to apsi intro}. This is to be compared with the expression \eqref{CI S3 intro} for the holomorphic partition function $\CI_{S^3 \times S^1}(\nu, \tau)$. Note that the result \eqref{Zabs intro} does not admit an ``holomorphic square root,'' therefore the supersymmetric partition function $Z_{S^3 \times S^1}(\nu, \tau)$ cannot be holomorphic. The expression \eqref{Zabs intro} is completely gauge-invariant, by construction. Similar formulae hold for other supersymmetric partition functions.

While the result \eqref{Zabs intro} determines the absolute value of the ``gauge-invariant'' partition function, we will further argue that the phase is also fully determined, thanks to supersymmetry, essentially because the anomalous Ward identity \eqref{holo anomaly intro} relates the real and imaginary parts of the supersymmetric effective action.

\subsection*{The small-$\beta$ limit of supersymmetric partition functions}
We also  obtain new general constraints on supersymmetric partition functions in the case of any half-BPS manifold $\CM_d = \CM_{d-1} \times S^1$ (for $d=2$, this is the elliptic genus; for $d=4$ we have a generalized index on $\CM_3$).%
\footnote{In fact, our results will also be valid if the $S^1$ is fibered non-trivially over $\CM_{d-1}$.} In that context, we consider the ``high-temperature limit,'' sending the circle radius $\beta$ to zero, up to order $\beta$ in the small-$\beta$ expansion:
\be\label{small beta exp intro}
W\;  \underset{\beta\to \,0}{\sim} \;  {1\ov \beta} W_{\rm d{-}1}^{(-1)} +  W_{\rm d{-}1}^{(0)}  + \beta  W_{\rm d{-}1}^{(1)}+ \dots~,
\ee
Crucially, the effective action is expected to be {\it local} in the background gauge fields, at each order in $\beta$, essentially because the $d$-dimensional theory compactified on the $S^1$ (and with appropriate chemical potentials) is generically gapped \cite{Banerjee:2012iz, DiPietro:2014bca}. The term of order  ${1/\beta}$ gives the Cardy-like contribution discussed in \cite{DiPietro:2014bca},%
\footnote{Recently, there was a renewed interest in similar Cardy-like limits in the context of black-hole microstate counting \protect\cite{Choi:2018hmj, Honda:2019cio, ArabiArdehali:2019tdm, Kim:2019yrz, Cabo-Bizet:2019osg} (see also \cite{Benini:2018ywd}). We remark that our small-$\beta$ limit is distinct from the one considered in those recent references (while it  is the same as in \protect\cite{DiPietro:2014bca}). 
} while the finite term essentially gives the partition function of the dimensionally-reduced $(d-1)$-dimensional theory on $\CM_{d-1}$---see {\it e.g.} \cite{Aharony:2013dha, Hwang:2018riu}. 
We will study in detail the order-$\beta$ term, $W_{\rm d{-}1}^{(1)}$. That term must reproduce all the consistent flavor anomalies \eqref{delta W intro 1}  of the $d$-dimensional theory, as well as their consequences \eqref{susy anomaly intro}, because the anomalies themselves are local $d$-dimensional functionals, and therefore only contribute at order $\beta$ \cite{Banerjee:2012iz}. 

There indeed exists $(d-1)$-dimensional local terms that reproduce the dimensional reduction of the flavor 't Hooft anomalies, and are consistent with supersymmetry. For instance, in 4d $\CN=1$ theories on $\CM_3 \times S^1$, the cubic 't Hooft anomaly \eqref{U1 anomaly 4d intro}, together with the supersymmetry variation \eqref{susy anomaly Aqqq intro},  uniquely determine the 3d local term:
\be\label{W3d intro}
W_{\rm  3d}^{(1)}={\CA_{qqq}  \ov 12 \pi} \int_{\CM_3} d^3 x \sqrt{g} \Big(-i  \sigma \epsilon^{\mu\nu\rho} A_\mu F_{\nu\rho} + 3  \sigma^2 D - 6 i  \sigma \t\lambda\lambda +\sigma^2 A_\mu \bV^\mu +\sigma^3 \bH\Big)~.
\ee
Here, the fields $(\sigma, A_\mu, \lambda, \t\lambda, D)$ are part of a 3d $\CN=2$ vector multiplet (which descends from the 4d $\CN=1$ vector multiplet), and $\bV_\mu$ and $\bH$ are background supergravity fields necessary to preserve supersymmetry~\cite{Closset:2012ru}. Analogous 3d local terms are necessary to match any other 't Hooft anomalies. 

Plugging the supersymmetric values for the background supergravity and vector multiplet fields into \eqref{W3d intro}---and its analogues for other 't Hooft anomalies---, we determine {\it a priori} the form of the supersymmetric partition function $Z_{\CM_{3} \times S^1}$ at order $\beta$. The result agrees with our explicit computation of the absolute  value of the supersymmetric partition function, and also agrees---as it should, by construction---with the anomalous Ward identity \eqref{holo anomaly intro}.

These considerations lead us to propose a general formula for the supersymmetric partition function $Z_{\CM_{d-1} \times S^1}(\nu, \tau)$, which is fully consistent with gauge invariance and supersymmetry. This is based on the observation, true in a number of examples, that the holomorphic partition function $\CI_{\CM_{d-1}\times S^1}(\nu, \tau)$, {\it does not have any order-$\beta$ term in its small-$\beta$ expansion}---for the $S^3$ index, this was proven in~\cite{Ardehali:2015bla}. Ref. \cite{Ardehali:2015bla} proves this result under the assumption that the $S^3$ partition function of the dimensionally reduced theory is finite, and moreover it shows that counterexamples exist when this assumption is not satisfied. In our paper we will always assume the finiteness of the $\CM_{3}$ partition function. 

Then, we propose that the ``gauge-invariant'' supersymmetric partition function, compatible with all anomalous Ward identities, takes the form:
\be\label{Z to I intro}
Z_{\CM_{d-1}\times S^1}(\nu,\b\nu, \tau) = e^{- \beta  W_{\rm d{-}1}^{(1)}(\nu, \b\nu, \tau)}  \, \CI_{\CM_{d-1}}(\nu, \tau)~.
\ee 
Here, $W_{\rm d{-}1}^{(1)}(\nu, \b\nu, \tau)$ denotes a function computed from the $(d{-}1)$-dimensional functional $W_{\rm d{-}1}^{(1)}$---in particular, we have the contribution \eqref{W3d intro} for the 4d cubic 't Hooft anomaly contribution---upon evaluation onto the supersymmetric locus for the background vector multiplet.

Note also that, if we interpret the 3d local term $\beta  W_{\rm 3d}^{(1)}(\nu, \b\nu, \tau)$ as a 4d local term, by a trivial uplift which breaks 4d diff invariance explicitly, one can view the relation \eqref{Z to I intro} as {\it a change of scheme} from the holomorphic to the ``gauge invariant'' form of the supersymmetric partition function. As expected, the holomorphic partition function corresponds to a scheme that breaks diff-invariance explicitly. We shall leave a more systematic understanding of the allowed counterterms for future work.

\subsection*{Outlook and summary}
 In this paper, we focussed our attention on general constraints on supersymmetric partition functions in the presence of 't Hooft anomalies for the flavor symmetry, for 2d $\CN=(0,2)$ and 4d $\CN=1$ supersymmetric theories with an $R$-symmetry, $U(1)_R$. This clarifies important aspects of the dependence of these partition functions on the {\it flavor parameters.}

 Supersymmetric partition functions can also depend non-trivially on {\it geometric parameters}---in particular, on the modular parameter $\tau$ discussed in this paper, but also on more subtle geometric data. It was first pointed out in \cite{Papadimitriou:2017kzw} that the 't Hooft anomalies for the $U(1)_R$ symmetry---and, more generally, 't Hooft anomalies in the ``gravity sector''---can introduce an anomalous dependence of the partition function on the geometry, thus resolving puzzles that arose from detailed holographic computations \cite{Genolini:2016ecx}. This interesting problem could also be addressed by combining the approach of this paper with the general solution to the WZ consistency conditions for the $\CR$-multiplet given recently in \cite{Papadimitriou:2019yug}.  %
It would also be interesting to generalize this work to include 6d $\CN=(1,0)$ theories, and to discuss theories in 2d and 4d with higher supersymmetry.

Finally, let us mention that the non-holomorphy of partition functions in the flavor parameters, as discussed here, is distinct from (and much simpler than) several other examples of holomorphic anomalies in the literature,  such as the holomorphic anomaly of the topological string partition function \cite{Bershadsky:1993ta}, or of the elliptic genus of theories with continuous spectra \cite{Murthy:2013mya, Troost:2017fpk}, or else in Donaldson-Witten theory \cite{Moore:1997pc}. In our case, the non-holomorphic term appears entirely as a ``supersymmetric Casimir-energy term,'' and not as part of the ``spectrum;'' the interpretation being that it essentially corresponds to a choice of scheme. In all cases, however, it is expected that the non-decoupling of the relevant $\CQ$-exact operators can be understood as a boundary term in the path integral; it would be interesting to understand our results in that  language.

 \medskip
 \noindent This paper is organized as follows. In section~\ref{sec: 2d anomaly}, we discuss the flavor 't Hooft  anomaly in 2d $\CN=(0,2)$ supersymmetric theories, and its consequences for the $T^2$ partition function. In section~\ref{sec: flavorsusy 4d},  after reviewing some necessary formalism, we discuss the analogous problem for 4d $\CN=1$ supersymmetric theories on a fixed half-BPS new-minimal supergravity background, and we consider the small-$\beta$ expansion for theories on $\CM_3 \times S^1$. In section~\ref{sec: MgpS1}, we use those results to revisit the computation of supersymmetric partition functions on a large class of four-dimensional backgrounds called $\CM_{g,p} \times S^1$, which include $S^3 \times S^1$ as a special case. Additional computations and hopefully useful discussions are contained in several appendices.

%%%%%%%%%%%%%%%%%%%%%%%%%%%%%%%%%%%%
%%%%%%%%%%%%%%%%%%%%%%%%%%%%%%%%%%%%
\section{'t Hooft anomalies and the 2d $\CN=(0,2)$ elliptic genus}\label{sec: 2d anomaly}
As an interesting warm-up to the four-dimensional case, let us study the partition function of a 2d $\CN=(0,2)$ supersymmetric theory on the torus, in the presence of flat connections for background gauge fields coupling to the flavor symmetry---this is also known as the flavored $\CN=(0,2)$ elliptic genus.

\subsection{The free fermion and the Quillen anomaly}\label{subsec: quillen anomaly 2d}
The simplest example of a two-dimensional $\CN=(0,2)$ supersymmetric theory is a free massless Fermi multiplet, which contains a single chiral fermion, $\lambda_-$, of positive chirality. Consider the partition function on this free theory on a Riemann surface $\Sigma$, with a background gauge field $a_\mu$ coupled to the $U(1)$ global symmetry:
\be
\SL=  -2 i \b\lambda_- D_\bw \lambda_-~.
\ee
The Dirac operator can be written as:
\be
\quad D_\bw = \nabla_\bw - i a_\bw~,
\ee
with $w$ some local complex coordinate. The naive formula for the partition function:
\be
Z_{\Sigma} = {\rm det}( D_\bw)~, 
\ee
 seems to define a locally holomorphic function of the connection, {\it i.e.} it only depends on $a_\bw$ and not on $a_w$. On the other hand, since the operator $D_\bw$ flips the chirality, its determinant is ill-defined. The standard procedure is to consider ${\rm det}(D_w D_\bw)$ instead. If the latter determinant factorized holomorphically---that is, if it were a product of a holomorphic times an anti-holomorphic factor---, one would naturally obtain a holomorphic definition of the original partition function, simply by taking the ``holomorphic square-root'' of ${\rm det}(D_w D_\bw)$. However, an explicit calculation  by means of a gauge-invariant regularization, such as Pauli-Villars or zeta-function regularization, reveals that \cite{Quillen1985}:
\be\label{eq:DzDzbar}
{\rm det}(D_w D_\bw) = e^{-q} |F(a_\bw)|^2~,
\ee
where $F$ is a holomorphic function, and $q$ is the ``Quillen counterterm:"
\begin{equation}
q = \frac{i}{2\pi} \int_\Sigma dw \wedge d \bw\, a_w a_\bw~,
\end{equation}
schematically. Due to the prefactor $e^{-q}$, the holomorphic factorization fails, and in particular the original partition function cannot be defined as a holomorphic function of the connection. This is known as Quillen's holomorphic anomaly. As we will see, it is also the simplest example of the non-trivial interplay between 't Hooft anomalies and supersymmetry.%
\footnote{Quillen's anomaly can be stated much more generally in terms of Dirac operators that depend holomorphically on some moduli; presumably, the 4d $\CN=1$ case to be discussed in the next section could also be understood in that language.}

A nice geometric interpretation of \eqref{eq:DzDzbar} was provided in  \cite{AlvarezGaume:1986es}. The freedom in choosing the connection $a_\mu$ can be parametrized by the space of flat connection on the Riemann surface, {\it i.e.} the Jacobian $J(\Sigma)$. The latter is a $2g$-dimensional torus for $\Sigma$ a Riemann surface of genus $g$, and it inherits a complex structure from that of $\Sigma$. In the presence of a $U(1)$ 't Hooft anomaly, the partition function of the chiral fermion is not a holomorphic function on $J(\Sigma)$ but rather a section of a holomorphic line bundle over it:
\be
\SL \longrightarrow J(\Sigma)~,
\ee
usually called the \emph{determinant line bundle} \cite{Quillen1985}. The expression \eqref{eq:DzDzbar} is then interpreted as defining a Hermitian norm on  $\SL$ (known as the Quillen metric), and its failure to factorize is due to its curvature. The first Chern class $c_1(\SL)$ is fixed precisely by the 't Hooft anomaly coefficient.

\paragraph{The case $\Sigma=T^2$.}  In anticipation to the discussion of the elliptic genus, let us consider more explicitly the case of a chiral fermion on a torus. We choose coordinates $x\sim x+2\pi$ and $y\sim y+2\pi$, with the complex coordinate:
\be\label{def w T2}
w= x+ \tau y~, \qquad \tau\equiv \tau_1 + i \tau_2~, \qquad \tau_2={ \beta_2 \ov \beta_1}~,
\ee
for some fixed complex structure modulus $\tau$. The metric on the torus reads:
\be\label{eq:metricT2}
ds^2(T^2) = \beta_1^2 dw d\bw = \beta_1^2 (dx+ \tau_1 dy)^2+ \beta_2^2 dy^2~,
\ee
In this case, the Jacobian is itself a torus, parametrized by the complex variable:
\be\label{nu def awb}
\nu \equiv a_x \tau - a_y=2 i \tau_2 \, a_{\bw}~, 
\ee
subject to the identifications:
\be
\nu\sim \nu+1~, \qquad\qquad \nu\sim \nu+\tau~,
\ee
under large gauge transformations. In Appendix~\ref{app:chiralfer 2d}, for review the calculation of the determinant ${\rm det}(D_w D_\bw)$ via zeta-function regularization. From that calculation, we obtain the following result for the partition function of a chiral fermion:
\be\label{eq:Zferm}
Z_{T^2}^\lambda \propto e^{-2\pi  \tau_2 \left({a_x^2\ov 2}- {a_x\ov 2}+ {1\ov 12} \right)}\, \theta_0(\nu; \tau)~
\ee
where the coefficient of proportionality is a phase factor (possibly a function of the real parameters $a_x, a_y, \tau_1, \tau_2$). Here, $\theta_0(\nu; \tau)$ is the ``reduced'' theta-function; its definition and properties can be found in Appendix~\ref{app:theta}. The theta-function is an holomorphic section of the determinant line bundle on $J(T^2)$, in agreement with the discussion above; in particular, it transforms non-trivially under large gauge transformations. The non-holomorphic prefactor in $Z_\lambda$ takes the form of a Casimir energy, being the dominant contribution in the limit $\tau_2\to\infty$. By construction, the absolute value of the partition function, $|Z_\lambda|$, is invariant under the large gauge transformations $\nu\to \nu+1$ and $\nu\to \nu+\tau$, in agreement with the requirement of gauge-invariance described in the introduction.

\subsection{The elliptic genus and the torus partition function}\label{sec:elliptic genus}
Let us now consider a two-dimensional field theory with $\CN=(0,2)$ supersymmetry. 
We are interested in its elliptic genus, which may be defined as the supersymmetric index:
\be
\CI(\nu, \tau)= \Tr\left((-1)^F q^H \b q^{\b H} y^{Q_F} \right)~, \qquad \qquad q=e^{2\pi i \tau}~, \quad y= e^{2\pi i \nu}~.
\ee
Here, $Q_F$ denotes the generator of a $U(1)$ global symmetry, and $y$ is the associated fugacity---more generally, one introduces one fugacity $y_\alpha$ for each $U(1)_\alpha$ in the maximal torus of the global symmetry group $G_F$.

The elliptic genus can also be understood as the supersymmetric partition function on $T^2$ with complex structure modulus $\tau$, as in \eqref{def w T2}. The chemical potential $\nu$ corresponds to a flat background gauge field for the $U(1)$ global symmetry, $a_\mu dx^\mu = a_w dw+ a_\bw d\bw$, with:
\be
\nu \equiv a_x \tau - a_y~,\qquad \qquad a_x\equiv {1\ov 2\pi}\int_{\gamma_x} a~, \qquad  a_y\equiv {1\ov 2\pi}\int_{\gamma_y} a~,
\ee
with $\gamma_x$ and $\gamma_y$ the one-cycles along $x$ and $y$, respectively. As we will review below, a simple argument shows that the derivatives with respect to $\bar{\nu}$ and $\bar{\tau}$ are $Q$-exact and therefore the partition function $Z_{T^2}(\nu, \tau)$ is expected to be holomorphic in $\nu$; a similar argument shows that it should be holomorphic in the geometric parameter $\tau$, as well.

\paragraph{Explicit expression for `Lagrangian' theories.}
The elliptic genus can easily be computed for free theories. For a free chiral multiplet  $\Phi$  or for a free Fermi multiplet $\Lambda$, of charge $q=1$ under the $U(1)$ background gauge field, we have:
\be
 \CI_\Phi(\nu, \tau) \equiv {i\ov \theta(\nu; \tau)}~, 
\quad \qquad
\CI_\Lambda(\nu, \tau) \equiv i \theta(\nu; \tau)~,
\ee
respectively. Here, $\theta(\nu, \tau)$ is the ordinary Jacobi theta function---see Appendix~\ref{app:theta}.

Thanks to supersymmetric localization, we also an have explicit expression for supersymmetric gauged linear $\sigma$-models (GLSM) \cite{Benini:2013nda, Benini:2013xpa, Gadde:2013ftv}. 
Consider an $\CN=(0,2)$ vector multiplet for the gauge group $\GG$, coupled to chiral multiplets $\Phi_i$ and Fermi multiplets $\Lambda_I$ in representations $\FR_i$ and $\FR_I$ of $\GG$, respectively.
The localization formula for the elliptic genus takes the form:
\be\label{ZT2 full}
\CI(\nu, \tau) = \oint_{\rm JK} \prod_{a=1}^\rk {du_a\ov 2\pi i} \; I(u, \nu, \tau)
\ee
where the contour integral denotes the Jeffrey-Kirwan residue \cite{JK1995}. The integrand reads:
\bea\label{def Integrand EG}
&I(u, \nu, \tau)\;& =&\; \prod_i \prod_{\rho_i \in \FR_i}  \CI_\Phi(\rho_i(u)+ \omega_i(\nu), \tau) \prod_I \prod_{\rho_I \in \FR_I}  \CI_\Lambda(\rho_I(u)+ \omega_I(\nu), \tau) \cr
&\;& &\;\qquad\times \left(-2 \pi i \; \eta(\tau)^2\right)^\rk \; \prod_{\alpha \in \Fg} \CI_\Lambda(\alpha(u), \tau)~.
\eea
The products over the indices $i$ and $I$, on the first line, run over the chiral and Fermi multiplets, respectively. The fugacities associated to the gauge group are denoted by $u_a$, and are integrated over; the flavor fugacities are denoted by $\nu$. 
Here, $\rho= (\rho^a)$ denote the weights of the representations $\FR$ under the gauge symmetry. We will also uses indices $\alpha,\beta,\dots$ for the Cartan subalgebra $\Fh_F$ of the flavor symmetry, so that {\it e.g.} $\omega(\nu) = \omega^\alpha \nu_\alpha$, with $\omega= (\omega^\alpha)$ a flavor weight and $\nu= (\nu_\alpha)\in (\Fh_F)_\C$. The second line in \eqref{def Integrand EG} is the contribution from the vector multiplet, with the product over the non-zero roots $\alpha$ of the gauge algebra $\Fg$.

\paragraph{Behavior under large gauge transformations.}
It is instructive to consider the behavior of the elliptic genus under large gauge transformations for the flavor fugacities:
\be
(a_x, a_y) \sim (a_x-m, a_y +n) \qquad \Leftrightarrow\qquad \nu \sim \nu+ n + m \tau~,
\ee
for any $n, m\in \Z$. For instance, for a free Fermi multiplet of charge $q\in \Z$ under the $U(1)$ flavor symmetry, we have:
\bea
&\CI_{\Lambda, q}(\nu, \tau)= \CI_\Lambda(q \nu, \tau)~, \cr
&\CI_{\Lambda, q}(\nu+ n + m\tau, \tau)= (-1)^{q(n+m)}\, e^{-\pi i q^2 \left(2 m \nu + m^2 \tau\right)} \,\CI_{\Lambda, q}(\nu, \tau)~.
\eea
The behavior under large gauge transformations is governed by $\CA_{qq}=q^2$, the 't Hooft anomaly coefficient for a free Fermi multiplet.
More generally, for any GLSM, we have the quadratic $U(1)_\alpha$ 't Hooft anomaly coefficients:
\be\label{def CA 2d}
\CA_{qq}^{\alpha\beta}=-\sum_i \omega_i^\alpha \omega^\beta_i + \sum_{I} \omega_I^\alpha \omega_I^\beta~,
\ee
where the integers $\omega^\alpha_i$ and $\omega^\alpha_I$ denotes the $U(1)_\alpha$ flavor charges of the chiral and Fermi multiplets, respectively. In a well-defined GLSM, all the gauge and gauge-flavor anomalies should vanish:
\be
\CA_{qq}^{ab}=- \sum_i \rho_i^a \rho^b_i + \sum_{I} \rho_I^a \rho_I^b=0~, \qquad
\CA_{qq}^{a\beta}=- \sum_i \rho_i^a \omega^\beta_i + \sum_{I} \rho_I^a \omega_I^\beta=0~.
\ee
This implies that the integrand \eqref{def Integrand EG} is single-valued under $u\sim u+1 \sim u+ \tau$. Under an arbitrary large gauge transformation:
\be
\nu_\alpha \rightarrow \nu_\alpha + n_\alpha+ m_\alpha \tau~,\qquad n_\alpha, m_\alpha\in \Z~,
\ee
for the flavor parameters, the elliptic genus transforms as:%
\footnote{Here we also introduced the ``linear anomaly'' coefficient: $$\CA^\alpha_q=- \sum_i \omega_i^\alpha+ \sum_{I} \omega_I^\alpha~,$$ 
which governs the sign on the right-hand-side of \protect\eqref{shift I gen EG}. This sign is a more subtle effect, related to the parity anomaly in 1d, which we will not discuss in this paper. \label{footnote pseudoA}}
\be\label{shift I gen EG}
\CI(\nu + n + m \tau, \tau)= (-1)^{\CA^\alpha_q (n_\alpha+ m_\alpha)} \, e^{-\pi i \CA^{\alpha\beta} \left(m_\alpha \nu_\beta+ \nu_\alpha m_\beta + m_\alpha m_\beta \tau\right)} \, \CI(\nu, \tau)~,
\ee
where we sum over repeated indices. The fact that the behavior under large gauge transformations is governed by the 't Hooft anomalies can be understood on general ground; we discuss some aspects of this in Appendix~\ref{large gauge 2d}.

\paragraph{Holomorphy versus gauge invariance.} As emphasized in the introduction, the transformation property \eqref{shift I gen EG} is somewhat puzzling when $\CI(\nu, \tau)$ is viewed as a partition function on $T^2$, because the real part of $\log {\CI}$ transforms non-trivially under large gauge transformations. On the other hand, we expect that there exists a scheme in which the absolute value of the partition function, $Z_{T^2}(\nu, \tau)$, is fully gauge invariant:
\be\label{Z abs gauge inv 2d}
|Z_{T^2}(\nu + n + m \tau, \tau)|= |Z_{T^2}(\nu, \tau)|~.
\ee
Consider again a free Fermi multiplet of unit charge. According to \eqref{eq:Zferm}, we have:
\be\label{Z to I fermi}
|Z_{T^2}^\Lambda(\nu, \tau)|= e^{- \pi \tau_2 a_x^2}\, |\CI_\Lambda(\nu, \tau)|~.
\ee
More generally, we will argue that:
\be\label{ZT2 expectation}
Z_{T^2}(\nu, \tau) = e^{-\bW_{\rm 1d}(\nu, \b\nu,\tau, \b\tau)}\, \CI(\nu, \tau)~,
\ee 
for any $\CN=(0,2)$ supersymmetric theory, with $\bW_{\rm 1d}$ a function to be determined below. Note that $\bW_{\rm 1d}$ cannot be holomorphic in $\nu$ if \eqref{Z abs gauge inv 2d} holds true. In the case of a GLSM, it is easy to check that imposing \eqref{Z abs gauge inv 2d} {\it a priori} completely determines the real part of $\bW_{\rm 1d}$ to be:%
\footnote{Recall the definitions $\nu = \tau a_x - a_y$ and $\tau= \tau_1 + i \tau_2$.}
\be\label{ReW1d}
{\rm Re}\Big[\bW_{\rm 1d}(\nu, \b\nu,\tau, \b\tau)\Big]= \pi \tau_2\, \CA^{\alpha\beta} a_{x, \alpha} a_{x, \beta}~,
\ee
 up to a pure constant. The explicit result \eqref{Z to I fermi} is a special case of the relation~\eqref{ReW1d}. 

We see that gauge invariance imposes that the $T^2$ partition function is not fully holomorphic in the flavor parameter $\nu$ (and possibly in $\tau$, as well). Naively, this is in contradiction with supersymmetry. As we will review below, the anti-holomorphic parameter $\b\nu$ couples to a $Q$-exact operator, and therefore we would expect to have:
\be
{\d\ov \d \b \nu} Z_{T^2}(\nu, \tau)= 0~,\label{eq:naive2d}
\ee
as a supersymmetric Ward identity. The presence of a non-trivial quantum correction to the supersymmetric Ward identity \cite{ Itoyama:1985ni} resolves this puzzle, as we explain in the next subsection.

\paragraph{Behavior under modular transformations.} Let us also briefly discuss the behavior of the elliptic genus under large diffeomorphisms of the torus, which span the modular group $SL(2, \Z)$. 
Let $k_g$ denote the gravitational anomaly coefficient, defined as:
\be
k_g=\sum_i {\rm dim}(\CR_i)- \sum_I {\rm dim}(\CR_I) - \dim(\GG)~,
\ee
for a GLSM. Using the properties of the $\theta$-function summarized in Appendix~\ref{app:theta}, one can check that:
  \be\label{modular transform EG}
 \CI\Big({\nu\ov \tau}, -{1\ov \tau}\Big) = e^{{\pi i \ov 2} k_g} e^{{ \pi i \ov \tau}  \CA^{\alpha\beta}\nu_\alpha\nu_\beta}\, \CI(\nu, \tau)~, \qquad \;
 \CI(\nu, \tau+1)=e^{-{\pi i \ov 6} k_g} \CI(\nu, \tau)~,
\ee
under the $S$ and $T$ generators of $SL(2,\Z)$, respectively. This behavior presents physical puzzles similar to the case of the behavior under large gauge transformations, and their resolutions will be similar.

\subsection{Non-holomorphy of the supersymmetric partition function $Z_{T^2}$}\label{sec: 2d susy anomaly}
We will now derive the correction to \eqref{eq:naive2d} in the ``gauge-invariant'' scheme. To this end, we first review the structure of the conserved current $(0,2)$ multiplet, its coupling to a background vector multiplet, and the supersymmetry algebra in WZ gauge. 

Consider a 2d $\CN=(0,2)$ supersymmetric theory with $G_F= U(1)$, for simplicity. We can preserve the two supercharges, $Q_+$ and $\t Q_+$, on the flat torus. Let us denote the supersymmetry variations by:
\be
\delta_\zeta = - i \zeta_- Q_+~, \qquad \delta_{\t\zeta} = - i\t \zeta_- \t Q_+~, 
\ee
where $\zeta_-$ and $\t\zeta_-$ are constant Weyl spinors,%
\footnote{Our conventions for two-dimensional spinors are summarized in Appendix~\protect\ref{Appendix: 2d}.}
 which we take to be bosonic, so that the supersymmetry variations are fermionic.  

\paragraph{Coupling to a background vector multiplet.} 
The $U(1)$ flavor symmetry current, $j^\mu$, sits in a 2d $\CN=(0,2)$ current multiplet, $\CJ$, with component operators:
\be
\CJ= (J~, \, j_+~, \, \t j_+~, j^\mu)~,\qquad\qquad  \d_\mu j^\mu=0~,
\ee
where $j_+$ and $\t j_+$ are fermionic operators. Let us denote by $(j_w, j_\bw)$ the components of $j_\mu$ in complex coordinates; in particular, we have $\d_w j_{\bw} + \d_\bw j_w=0$. The $\CN=(0,2)$ supersymmetry variations of the current multiplet read:%
\footnote{In this subsection, we are setting $\beta_1=1$ to avoid clutter, so that the torus metric is simply $ds^2 = dw d \bw$; equivalently, we are absorbing $\beta_1$ into the definition of $w, \bw$.}
\bea\label{conserved current susy 2}
&\delta_{\zeta} J= - i \zeta_- j_+~, \qquad
&&\delta_{\t\zeta} J=  -i \t\zeta_- \t j_+~, \cr
&\delta_{\zeta} j_+ = 0~, \qquad
&&\delta_{\t\zeta} j_+ = -2 i \t\zeta_- (j_\bw+i \d_\bw J )~,\cr
&\delta_{\zeta}\b j_+ = 2  i \zeta_- (j_\bw-i \d_\bw J)~, \qquad\qquad
&&\delta_{\t\zeta}\t j_+ = 0~, \cr
&\delta_{\zeta} j_w = -\zeta_- \d_w j_+~,\qquad
&&\delta_{\t\zeta} j_w =  \t\zeta_- \d_w \t j_+~,\cr
&\delta_{\zeta} j_\bw = \zeta_- \d_\bw j_+~,\qquad
&&\delta_{\t\zeta} j_\bw =   - \t\zeta_- \d_\bw \t j_+~.
\eea
We couple the current multiplet to an abelian background vector multiplet in the Wess-Zumino (WZ) gauge, which consists of a gauge field $a_\mu=(a_w, a_\bw)$ and its superpartners:
\be
\CV_F= (a_\mu~, \, \lambda_-~, \, \t\lambda_-~, \, D)~.
\ee
Its supersymmetry variations read:
\bea\label{susy V 2d}
&\delta_{\zeta} a_w = - i \zeta_-\t\lambda_- ~, \qquad
&&\delta_{\t\zeta} a_w =  - i \t\zeta_- \lambda_-~, \cr
&\delta_{\zeta} a_\bw =0~, \qquad
&&\delta_{\t\zeta} a_\bw =0~, \cr
&\delta_{\zeta} \lambda_- = i \zeta_- (D+ 2 i f_{w\bw})~,  \qquad
&&\delta_{\t\zeta} \lambda_- = 0~, \cr
&\delta_{\zeta} \t\lambda_- = 0~,  \qquad
&&\delta_{\t\zeta} \t\lambda_- = -i\t \zeta_- (D- 2 i f_{w\bw})~, \cr
&\delta_{\zeta} D= 2 \zeta_- \d_\bw \t\lambda_-~, \qquad \qquad
&&\delta_{\t\zeta} D=   - 2 \t\zeta_- \d_\bw \lambda_-~,
\eea
where $f_{\mu\nu} = \d_\mu a_\nu- \d_\nu a_\mu$ is the field strength.
The supersymmetry algebra in the WZ gauge takes the form:
\be\label{susy algebra 2d}
(\delta_\zeta)^2 \Phi_{(q)}=0~, \quad 
(\delta_{\t\zeta})^2  \Phi_{(q)}=0~, \quad 
\{\delta_\zeta, \delta_{\t\zeta} \} \Phi_{(q)}= -4 i \zeta_- \t\zeta_- \left(\d_\bw - i q a_\bw\right)  \Phi_{(q)}~,
\ee
on any field $\Phi_{(q)}$ of charge $q$ under the $U(1)$ flavor symmetry; note that it is gauge-covariant.
We can also write the last anti-commutator as:
\be\label{susy alg general 2D}
\{\delta_\zeta, \delta_{\t\zeta} \}= -4 i \zeta_- \t\zeta_-  \d_\bw + \delta_{\alpha(a)}~, \qquad \alpha(a) \equiv 4i \zeta_- \t\zeta_- a_\bw~.
\ee
The second term on the right-hand-side can be understood as an $a_\mu$-dependent gauge transformation, $\delta_\alpha$.  Then, \eqref{susy alg general 2D} gives the supersymmetry algebra acting on any fields, including the gauge field itself (with $\delta_\alpha a_\mu= \d_\mu \alpha$). 

The supersymmetric minimal coupling between the current and the background gauge field takes the form:
\be\label{min coupling 2d}
\SL_{\CV\CJ}= 2 a_w j_\bw + 2 a_\bw j_w + D J + \lambda_- j_+ - \t \lambda_- \t j_+~.
\ee
In particular, we can consider supersymmetric background values for $\CV_F$, which are such that the gauginos $\lambda_-$, $\t \lambda_-$ and their variations vanish---that is:
\be\label{D f cond susy 2d}
D=0~, \qquad f_{w\bw}=0~,
\ee
and the gauge field $a_\mu$ must be flat. For future purpose, we will consider small variations $(a_\mu, \lambda_-, \t\lambda_+, \Delta D)$ around that locus, while keeping the gauge field flat. Then, using \eqref{D f cond susy 2d}, one can write \eqref{min coupling 2d} as:
\be\label{minimal coupling 2d with lambda}
\SL_{\CV\CJ}= 2 a_w (j_\bw- i \d_\bw J) + 2 a_\bw (j_w+ i \d_w J) + \Delta D J + \lambda_- j_+ - \t \lambda_- \t j_+~.
\ee
Looking at \eqref{conserved current susy 2}, we see that $a_w$ couples to a $\delta_\zeta$-exact operator,%
\footnote{One can write it as a $\delta_{\t\zeta}$-exact term as well, by integration by part.} which should not affect supersymmetric observables.
This is why we expect the $T^2$ partition function to be holomorphic in the flat connection $a_\bw$, which is related to the flavor parameter $\nu$ as in \eqref{nu def awb}.

\paragraph{Supersymmetry variation of the effective action.}  Let $W$ denote the effective action on the torus, in the presence of an arbitrary background vector multiplet:
\be
W[a_\mu, \lambda, \t\lambda, D]= - \log Z_{T^2}[a_\mu, \lambda, \t\lambda, D]~.
\ee
The $U(1)$ symmetry generally suffers from a quadratic 't Hooft anomaly, with coefficient $\CA_{qq} = \tr(\gamma^3 Q_F^2)$, as in \eqref{def CA 2d}. Under a gauge transformation $\delta_\alpha a_\mu = \d_\mu \alpha$, the effective action must then transform as:
\begin{equation}\label{eq:u1ano2d}
\delta_\alpha W = \frac{i \CA_{qq}}{4\pi} \int_{T^2} \alpha f~,
\end{equation}
where we used the form notation, with $f\equiv da = \frac{1}{2}f_{\mu\nu}dx^\mu \wedge dx^\nu$. Since the gauge transformations commute with supersymmetry (in the WZ gauge), we should have:
\be\label{eq:WZcc}
[\delta_\zeta, \delta_\alpha] W = 0~,\qquad\qquad
[\delta_{\t\zeta}, \delta_\alpha] W= 0~.
\ee
The supersymmetry variations of \eqref{eq:u1ano2d} give:
\bea
 \delta_\zeta \delta_\alpha W & = \frac{i\,\CA_{qq}}{4\pi}  \int \alpha \, d\left(-i\zeta_-{\t\lambda}_- dw\right)
 =- \frac{\CA_{qq}}{4\pi}  \int d\alpha \wedge \left(\zeta_-\t\lambda_- dw\right)  \\
&= \delta_\alpha\Big(\frac{\CA_{qq}}{4\pi}  \int a_{\bar{w}} \, \zeta_-{\t\lambda}_- \, d w\wedge d\bar{w}\Big)=\delta_\alpha \delta_\zeta W~,
\label{eq:gaugesusy}
\eea
where we used \eqref{eq:WZcc} in the last equality, and similarly:
\be
\delta_\alpha \delta_{\t\zeta} W = \delta_\alpha\Big(\frac{\CA_{qq}}{4\pi}  \int a_{\bar{w}} \,\t\zeta_-\lambda_- \, dw\wedge d\bar{w}\Big)~.\label{eq:gaugesusy2}
\ee
This implies the presence of a non-zero supersymmetry variation, $\delta_\zeta W \neq 0$ \cite{Itoyama:1985ni}. The equations \eqref{eq:gaugesusy}-\eqref{eq:gaugesusy2} determine $\delta_\zeta W$ up to the addition of a gauge-invariant term. 
To fix this ambiguity, we use the WZ condition coming from the anti-commutators \eqref{susy alg general 2D} of two supersymmetry transformations, namely:\footnote{The first term on the right-hand-side of the anti-commutator, $- 4 i \zeta_-\t{\zeta}_-\partial_{\bar{w}}$, acts on fields as a translation; for simplicity, we may assume that the theory is free from gravitational anomalies, so this transformation leaves $W$ invariant. The inclusion of the gravitational anomaly in the present discussion is left for future work.}
\be\label{eq:susysusy}
\{\delta_{\zeta} ,\, \delta_{\t\zeta}\} W = \delta_{\alpha(a)}W~,\qquad 
\alpha(a) \equiv 4 i \zeta_-\t\zeta_- a_{\bar{w}}~.
\ee
Taking into account this additional constraint, the supersymmetry variation of $W$ is uniquely fixed to be:
\be\label{eq:susyano2d} 
\delta_\zeta W = \frac{\CA_{qq}}{4\pi}  \int_{T^2} \zeta_-\t\lambda_-  a_\bw\, dw\wedge d\bar{w}~,\qquad\quad
\delta_{\t\zeta} W =\frac{\CA_{qq}}{4\pi}  \int_{T^2} \t\zeta_-\lambda_-  a_\bw\, dw\wedge d\bar{w}~.
\ee
Let us now discuss how these supersymmetric Ward identities allow us to constrain the form of the partition function, $Z_{T^2}(\nu, \tau)$.

 \paragraph{Holomorphy anomaly of $Z_{T^2}(\nu, \tau)$ in $\nu$.}
On a flat torus, we can take the supersymmetric gauge field components $(a_w, a_{\bar w})$ to be constant, in which case the flavor parameters are given by:
\be
\nu =2 i \tau_2 \, a_{\bar{w}}~, \qquad\quad
\bar{\nu} = -2 i \tau_2 \, a_w~.
\ee
An infinitesimal variation of $\bar{\nu}$ then simply corresponds to an insertion of the component $j_\bw$  of the conserved current, which is itself $Q$-exact:
\be\label{dnuW 2d interm}
{\d\ov \d\b \nu} W= {i\ov 2\tau_2} {\d\ov \d\b a_w} W ={i\ov 2\tau_2}  \left\langle \int_{T^2}  d^2 x \sqrt{g} 2 j_\bw   \right\rangle= {1\ov 2\tau_2  \zeta_-}  \left\langle \int_{T^2}  d^2 x \sqrt{g} \delta_\zeta \t j_+   \right\rangle~. 
\ee
Due to the non-trivial supersymmetry variations \eqref{eq:susyano2d}, the last expression in \eqref{dnuW 2d interm} does not vanish. Instead, since the operator $\t j_+$ couples to the gaugino $\t\lambda_-$ as in \eqref{minimal coupling 2d with lambda}, we obtain:
\be
{\d\ov \d\b \nu} W=-{1\ov 2\tau_2  \zeta_-}   {\d \ov \d\t\lambda_-}(\delta_\zeta W)\Big|_{\lambda=\t\lambda=0}= -{\CA_{qq}\ov 8\pi \tau_2} a_\bw \int_{T^2} dw \wedge d\bw~.
\ee
This gives us a simple anomalous Ward identity for the supersymmetric $T^2$ partition function:
\be\label{anomalous ward id 2d}
{\d\ov \d\b \nu} W(\nu, \b \nu)= {\pi \CA_{qq}\ov 2 \tau_2}  \nu~,
\ee
which entirely fixes its dependence on the anti-holomorphic parameter $\b \nu$---we must have:
\be
Z_{T^2}(\nu) = e^{-{\pi \CA_{qq}\ov 2 \tau_2} |\nu|^2} Z_{\rm holo}(\nu)~,
\ee
with $Z_{\rm holo}(\nu)$ a locally holomorphic function of $\nu$. To relate $Z_{\rm holo}(\nu)$ to the holomorphic partition function $\CI(\nu)$ defined above, we will need additional information.  We can already note, however, that if we had:
\be\label{Zholo to EG}
 Z_{\rm holo}(\nu)= e^{{\pi \CA_{qq}\ov 2 \tau_2 } \nu^2} \CI(\nu)~, 
\ee
then the expectation \eqref{ZT2 expectation}-\eqref{ReW1d} would hold true, and the absolute value of the $T^2$ partition function $Z_{T^2}$ would be properly gauge invariant. In the rest of this section, we will argue that \eqref{Zholo to EG} precisely holds.

%%%%%%%%%%
\subsection{The small-$\beta_1$ limit of $Z_{T^2}(\nu, \tau)$}
We can attain a complementary understanding of the non-holomorphy of the partition function by considering the reduction of the 2d $\CN=(0,2)$ theory to a one-dimensional $\CN=2$ supersymmetric theory (as studied {\it e.g.} in \cite{Hori:2014tda}) in the limit when we take one of the circle of $T^2$ to be very small. Using the metric  \eqref{eq:metricT2}, we consider $\beta_1 \rightarrow 0$. This correspond to the limit: 
\be\label{beta1 zero 2d limit}
\tau \rightarrow i \infty~, \qquad\qquad \nu = {\rm fixed}~,
\ee
on the $T^2$ partition function. The basic idea is that, in this limit, all the anomalies should be reproduced by {\it local terms} in the sources of the dimensionally-reduced theory, because generically the effective one-dimensional theory will be gapped \cite{Banerjee:2012iz, DiPietro:2014bca}. Upon evaluating these local terms, we find that they depend on $\bar{\nu}$ exactly as expected from \eqref{anomalous ward id 2d}. This procedure will also allow us to fix the entire ratio between $Z_{T^2}$ and the holomorphic partition function, $\CI_{T^2}$, including the holomorphic term.

In the small-$\beta_1$ limit \eqref{beta1 zero 2d limit}, we study the expansion of the effective action:
\be\label{eq:expW}
W\;  \underset{\beta_1\to \,0}{\sim} \;  {1\ov \beta_1} W_{\rm 1d}^{(-1)} +  W_{\rm 1d}^{(0)}  + \beta_1  W_{\rm 1d}^{(1)} + \dots~,
\ee
as a functional of background vector multiplets. The divergent term, of order $1\ov \beta_1$, is proportional to the gravitational anomaly coefficient, $k_g$; the functional $W_{\rm 1d}^{(-1)}$ is supersymmetric and gauge-invariant.%
\footnote{This is similar to the supersymmetric Cardy formula in 4d $\CN=1$ theories \protect\cite{DiPietro:2014bca}. This point would deserve further discussion.}  The finite term, $W_{\rm 1d}^{(0)}$, is essentially the 1d index of the dimensionally reduced theory \cite{Hori:2014tda}. Here, we are interested in the term linear in $\beta_1$. Let us define:
\be
\bW_{\rm 1d}[\CV_F]Â \equiv \beta_1  W_{\rm 1d}^{(1)}[\CV_F]~.
\ee
It is a local functional of a one-dimensional background vector multiplet, which is strongly constrained by the two-dimensional anomalies. This is because each of the two-dimensinal anomalies, being local, reduces to a term of order $\beta_1$ \cite{Banerjee:2012iz}, which must then be exactly reproduced by the variation of the local functional $\bW_{\rm 1d}$.

\paragraph{Dimensional reduction and 1d $\CN=2$ background vector multiplet.} Let $X^\mu=(x, y)$ denote the 2d coordinate, with $y$ the 1d coordinate in the $\beta_1 \rightarrow 0$ limit. The torus metric:
\be
ds^2(T^2) = \beta_1^2 (dx+ \tau_1 dy)^2+ \beta_2^2 dy^2~,
\ee
is already in the ``Kaluza-Klein'' form, with $ds^2(S^1)=\beta_2^2 dy^2$ the 1d metric and $\tau_1 dy$ a flat graviphoton. The two-dimensional gauge field, $a_\mu$, reduces to a one-dimensional gauge field $A_{\by}$ and a real scalar $\sigma$, which are related to $a_\mu$ by:
\be
a_\mu dX^\mu = \sigma  \beta_1 (dx+ \tau_1 dy) + A_\by d\by~.
\ee
Here, for convenience, we introduced the one-dimensional coordinate $\by\equiv \beta_2 y$, with $\by \sim \by +2 \pi \beta_2$. The 1d vector multiplet consists of the fields:
\be
\CV_F^{(\rm 1d)}= (A_\by~,\, \sigma~, \, \lambda~, \, \t\lambda~, \, D)~,
\ee
with the supersymmetry variations:
\bea
&\delta_{\zeta} A_\by = \zeta \t\lambda~, \qquad
&&\delta_{\t\zeta} A_\by =   \t\zeta \lambda~, \cr
&\delta_{\zeta} \sigma = -  i\zeta \t\lambda~, \qquad
&&\delta_{\t\zeta} \sigma =  - i \t\zeta \lambda~, \cr
&\delta_{\zeta} \lambda = i \zeta (D+ \d_\by \sigma)~,  \qquad
&&\delta_{\t\zeta} \lambda = 0~, \cr
&\delta_{\zeta} \t\lambda = 0~,  \qquad
&&\delta_{\t\zeta} \t\lambda = -i\t \zeta(D-\d_\by \sigma)~, \cr
&\delta_{\zeta} D= \zeta \d_\by \t\lambda~, \qquad \qquad
&&\delta_{\t\zeta} D=   - 2 \t\zeta \d_\by \lambda~.
\eea
This directly follows from \eqref{susy V 2d}.
%%%%%%%%

\paragraph{One-dimensional local functional.} The functional $\bW_{\rm 1d}$ is constrained by the requirement that its gauge variation (under $\delta_\alpha A_\by = \d_\by \alpha$) should reproduces the dimensional reduction of the $U(1)$ anomaly:
\be
\delta_\alpha \bW_{\rm 1d}= {i\CA_{qq}\beta_1\ov 2} \int d\by \alpha (-\d_\by \sigma)= \delta_\alpha \left( {i\CA_{qq}\beta_1\ov 2} \int d\by A_\by \sigma \right)~.
\ee
It should also reproduce the 1d reduction of the supersymmetry variations \eqref{eq:susyano2d}, namely:
\be
\delta_\zeta \bW_{\rm 1d}= - {i\CA_{qq} \beta_1\ov 2} \int \d\by \zeta\t\lambda (\sigma + i A_\by)~, \quad
\delta_{\t\zeta} \bW_{\rm 1d}= - {i\CA_{qq} \beta_1\ov 2} \int \d\by \t\zeta\lambda (\sigma + i A_\by)~.
\ee
This fixes $\bW_{\rm 1d}$ to be:
\be\label{W1d full answer}
\bW_{\rm 1d}=  {i\CA_{qq}\beta_1\ov 2} \int d\by (A_\by \sigma - i \sigma^2)~,
\ee
up to terms of dimension two that are gauge-invariant and supersymmetric. Such terms vanish on the supersymmetric locus ($D=\d_\by \sigma=0$). We can therefore conclude that, in the limit \eqref{beta1 zero 2d limit}, the order-$\beta_1$ term in the expansion of the $T^2$ supersymmetric partition function is fully determined by \eqref{W1d full answer}.

\subsection{The supersymmetric $T^2$ partition function, revisited}\label{subsec: ZT2 revisit}
Consider the elliptic genus of a GLSM, as given explicitly in \eqref{ZT2 full}-\eqref{def Integrand EG}. We can consider its small-$\beta_1$ limit explicitly. One finds:
\be\label{log I2d}
- \log \CI(\nu, \tau) \;  \underset{\tau\to i \infty}{\sim} \;  \tau {\pi i k_g\ov 6}   + W_{\rm 1d}^{(0)} + \CO(e^{2\pi i \tau})~.
\ee
In other words, there is no order-$\beta_1$ contribution appearing in the small-$\beta_1$ expansion of the ``holomorphic partition function.''  This was also observed recently in \cite{ArabiArdehali:2018mil}. 

By contrast, the order-$\beta_1$ term in the expansion of the partition function $Z_{T^2}$ is fully captured by  \eqref{W1d full answer}, which evaluates to:%
\footnote{Note that $\nu = i \beta_2 (\sigma + i A_\by)$ in the 1d variables.}
\be
\bW_{\rm 1d} = \pi \CA_{qq}\beta_1 \beta_2 \, \sigma  (\sigma + i A_\by)= {\pi  \CA_{qq}} {\nu \b\nu-\nu^2 \ov 2\tau_2}~.
\ee
Of course, this satisfies the quantum Ward identity \eqref{anomalous ward id 2d}. Incidentally, we note that $\bW_{\rm 1d}$ is holomorphic in $\tau$ when expressed in terms of the 2d variables $a_x$ and $a_y$:
\be
\bW_{\rm 1d} =- \pi i \CA_{qq} a_x (\tau a_x- a_y)~.
\ee
Based on these considerations, we make a conjecture for the exact form of the ``gauge invariant''  supersymmetric $T^2$ partition function, for any 2d $\CN=(0,2)$ theory. It is simply given by:
\be\label{def ZT2}
Z_{T^2}(a_x, a_y, \tau) = e^{- \bW_{\rm 1d}(a_x, a_y, \tau)} \, \CI(\nu, \tau)~, 
\ee
with $\CI$ the elliptic genus, which is holomorphic in $\nu= \tau a_x-a_y$ and $\tau=\tau_1+ i \tau_2$, and with the prefactor $\bW_{\rm 1d}$ given by:
\be
\bW_{\rm 1d}(a_x, a_y, \tau)=   - {\pi i\ov 2} \CA^{\alpha\beta}  \left(2\tau a_{x, \alpha}  a_{x, \beta}-  a_{x, \alpha} a_{y, \beta}-a_{y, \alpha} a_{x, \beta}\right)~.
\ee
By construction, this is fully consistent with gauge invariance and supersymmetry, with the 't Hooft anomalies properly taken into account. Note that its real part is given by \eqref{ReW1d}, as expected.

\paragraph{Large gauge transformations.} Under large gauge transformations  of the background gauge fields on $T^2$:
\be
(a_{x, \alpha}, a_{y, \alpha})\sim (a_{x, \alpha}+ m_\alpha, a_{y, \alpha}-n_\alpha)~, \qquad n_\alpha, m_\alpha\in \Z~,
\ee
 we obtain:
\bea\label{Z lgt 2d full}
&Z_{T^2}(a_x+m, a_y-n, \tau) &=&\; (-1)^{\CA^\alpha_q (n_\alpha+ m_\alpha)} \cr
&&& \times \, e^{{\pi i}\CA^{\alpha\beta} \left(n_\alpha a_{x, \beta} +m_\alpha a_{y, \beta} +n_\alpha m_\beta\right)} \, Z_{T^2}(a_x, a_y, \tau)~,
\eea
to be compared with \eqref{shift I gen EG}. In particular, the absolute value of $Z_{T^2}$ is gauge invariant. This behavior agrees with general expectations for gauge anomalies on the torus. As explained in Appendix~\ref{large gauge 2d}, general considerations determine the behavior of any flavored $T^2$ partition function under large gauge transformations in terms of the 't Hooft anomaly coefficients, modulo various ambiguities that depend on the renormalization scheme. These general considerations are consistent with the explicit result \eqref{Z lgt 2d full}.

\paragraph{Modular transformations.} One can also check that the partition function  \eqref{def ZT2} transforms naturally under $SL(2, \Z)$. Using \eqref{modular transform EG}, one finds the simple result:
\be\label{modularT1}
Z_{T^2}(a_x, a_y+ a_x, \tau+1)=e^{-{\pi i \ov 6} k_g}   Z_{T^2}(a_x, a_y, \tau)~,
\ee
for the $T$ transformation, and:
\be \label{modularS1} 
Z_{T^2}\Big(a_y, -a_x,-{1\ov\tau}\Big)=  e^{{\pi i \ov 2} k_g}  \, Z_{T^2}(a_x, a_y, \tau)~,
\ee
for the $S$ transformation.
 We therefore see that the anomalous behavior of $Z_{T^2}$ under large diffeomorphisms is determined by the gravitational anomaly only, as one would have naively expected on physical ground; when $k_g=0$, the supersymmetric partition function is modular invariant. 
This has to be contrasted with the behavior of the holomorphic elliptic genus under $SL(2, \Z)$, which is given by  \eqref{modular transform EG}. In fact, we again find perfect agreement with general constraints on $Z_{T^2}$ imposed by the anomalies, as explained in Appendix~\ref{large gauge 2d}.

 \paragraph{Local counterterm and holomorphy.}   We have seen that the non-holomorphic dependence of the $T^2$ supersymmetric partition function is dictated by the 't Hooft anomalies, as in \eqref{anomalous ward id 2d}. Namely, we have:
 \be\label{anomalous ward id 2d bis}
{\d\ov \d\b \nu_\alpha} W(\nu, \b \nu)= {\pi\ov 2 \tau_2}  \CA_{qq}^{\alpha\beta} \nu_\beta~,
\ee
 for an arbitrary abelian flavor symmetry, $G_F= \prod_\alpha U(1)_\alpha$, with 't Hooft anomaly coefficients $\CA_{qq}^{\alpha\beta}$. Given the explicit form of the superymmetric Ward identity \eqref{eq:susyano2d}, it is clear that one could cancel this anomaly---and, therefore, restore the holomorphy of the supersymmetric partition function in $\nu$---by adding the following local term to the effective action:
 \be\label{Wct in 2d}
W_{\rm ct}=-{\CA_{qq}\ov 8\pi} \int d^2 x \sqrt{g}\,  a_{\mu} a^\mu~.
\ee
 Thus, there certainly exists a scheme where the holomorphy in $\nu$ is manifest, corresponding to the partition function~\eqref{Zholo to EG}.%
 \footnote{In this scheme, we would still not have holomorphy in $\tau$. To recover the fully holomorphic elliptic genus $\CI(\nu, \tau)$, we should add an additional counterterm corresponding to the prefactor in \protect\eqref{Zholo to EG}, which breaks diffeomorphism invariance explicitly. We leave a better understanding of this point for future work.} 
The counterterm  \eqref{Wct in 2d} is not gauge invariant, however. Note also that, since:
\be
\delta_\zeta (W+W_{\rm ct})=0~, \qquad \delta_{\t\zeta} (W+W_{\rm ct})=0~, 
\ee
it follows from \eqref{eq:susysusy} that $\delta_{\alpha(a)} (W+W_{\rm ct})=0$. This does not mean that we have cancelled the gauge anomaly, however, only that this counterterm also cancelled the $U(1)$ anomaly under a {\it particular} gauge transformation with parameter $\alpha= \alpha(a)$.

 In conclusion, the holomorphy of the $T^2$ supersymmetric partition function in $\nu$ can only be maintained, by adding the counterterm \eqref{Wct in 2d}, at the cost of violating gauge invariance for background gauge fields. (Note that \eqref{Wct in 2d} violates explicitly the gauge-invariance of the {\it real part} of $W$.)

\subsection{Anomalous supersymmetry variation from WZ gauge-fixing}
As explained in the introduction, we can understand the anomalous supersymmetry variation, $\delta_\zeta W\neq 0$, as a simple consequence of fixing the WZ gauge. Let us show this in more detail. Here, we will use a superspace notation, for convenience.

\paragraph{WZ gauge-fixing.} The $\CN=(0,2)$ abelian background vector multiplet can be written in terms of two superfields:
\bea
&\CV = C + i\theta^+\chi_+ + i\t\theta^+ \t\chi_+ + 2\theta^+ \t\theta^+ a_{\bar z}~, \\
&\CV_{w} = a_w + i\theta^+ (\t\lambda_- -i\partial_w \chi_+) + i\t\theta^+ (\lambda_- + i\partial_w \t\chi_+) - \theta^+ \t\theta^+ (D + 2\partial_w\partial_{\bar w}C)\ .
\eea
The $U(1)$ gauge symmetry, $\delta_\alpha a_\mu = \partial_\mu \alpha$, can then be supersymmetrized by:
\be
\delta_\Omega \CV = \frac{i}{2}(\Omega - \t\Omega)~, \qquad \qquad
\delta_\Omega\CV_w = \frac12 \partial_w (\Omega + \t\Omega)\ ,
\ee
where the gauge parameters sit in the chiral and anti-chiral superfields:
\bea
&\Omega = \omega + \sqrt2 \theta^+\psi^\Omega_+ -2i\theta^+\t\theta^+ \partial_{\bar w}\omega+ \cdots~, \cr
&\t\Omega = \t \omega -\sqrt 2 \t\theta^+ \t\psi^\Omega_+ + 2i\theta^+\t\theta^+ \partial_{\bar w}\t \omega+ \cdots~.
\eea
Using this larger gauge freedom, one can fix the WZ gauge:
\be\label{WZ gauge 2d v}
C=\chi_+ = \t\chi_+ = 0~.
\ee
The WZ gauge \eqref{WZ gauge 2d v} is not invariant under supersymmetry, because:
\be
\delta \chi_+ = 2i\t\zeta_- a_{\bar z}~, \qquad\qquad \delta\t\chi_+ = -2i \zeta_- a_{\bar z}\ .
\ee 
Nonetheless, one can define consistent supersymmetry transformations, which we denote by $\delta_\zeta$, $\delta_{\t\zeta}$, by performing a compensating gauge transformation that restores the WZ gauge:
\be\label{del zeta in WZ gauge 2d}
\delta_\zeta = \delta_\zeta^{(0)} + \delta_{\Omega(\zeta)}~, \qquad
\delta_{\t\zeta} = \delta_{\t\zeta}^{(0)} + \delta_{\Omega(\t\zeta)}~.
\ee
Here, $\delta_{\zeta,\t\zeta}^{(0)}$ is the ``bare" supersymmetry transformation before gauge fixing, and $\delta_{\Omega(\t\zeta)}$, $\delta_{\t\Omega(\zeta)}$ are compensating gauge transformations with field-dependent gauge parameters:
\bea\label{compensating component}
&\Omega(\zeta)\;:\;\;&&\omega= \psi_+^\Omega =  0~, \qquad &&\t\omega=0~, \quad \t\psi_+^\Omega = 2\sqrt 2 i \zeta_- a_{\bar z}~,\cr
&\Omega(\t\zeta)\; :\; \;&&\omega = 0~, \quad \psi_+^\Omega = -2\sqrt 2 i \t\zeta_- a_{\bar z}~, \qquad &&\t\omega=\t\psi_+^\Omega=0~,
\eea
respectively.

\paragraph{'t Hooft anomaly and  supersymmetry.}
The t 'Hooft anomaly for the background $U(1)$ symmetry \eqref{eq:u1ano2d} can be supersymmetrized as:
\be\label{anomaly superspace 2d}
\delta_{\Omega}W = \frac{\CA_{qq}}{8\pi} \int d^2 x \sqrt g \left[\int d\theta^+ \Omega \CY_-  + \int d\t\theta^+\t\Omega \t\CY_- \right]~,
\ee
where $\CY_-$ and $\t\CY_-$ are the gauge-invariant gaugino multiplets:~\footnote{The SUSY-covariant derivatives  are given by ${\rm D}_+= \d_+ -2 i \t\theta^+ \d_\bw$ and $\t{\rm D}_+=-\t \d_+ +2 i \theta^+ \d_\bw$.}
\be
\CY_- = \t {\rm D}_+ (\partial_w\CV + i \CV_w)~, \qquad \quad \t\CY_- = {\rm D}_+ (\partial_w \CV - i \CV_w)~.
\ee
In components, this reads:
\be\label{delt Om W 2d}
\delta_{\Omega}W = \frac{\CA_{qq}}{8\pi} \int d^2 x \sqrt g \left(2(\omega+\t \omega)f_{w\bar w} -i(\omega-\t \omega) D + \sqrt 2(\psi^\Omega_+\lambda_- - \t\psi^\Omega_+ \t\lambda_-)\right)~,
\ee
where we identify $\alpha = \half (\omega+ \t \omega)$ as the ordinary $U(1)$ gauge parameter.
The expression \eqref{anomaly superspace 2d} is fully supersymmetric with respect to the ``bare'' supersymmetry variation:
\be
\delta_{\zeta}^{(0)}\Big( \delta_{\Omega}W \Big)=0~.
\ee
On the other hand, we have the WZ consistency condition:
\be
\left[\delta_{\zeta}^{(0)},\delta_{\Omega}\right] W =0~.
\ee
From the last two equations, we have $\delta_\Omega\delta_\zeta^{(0)}W=0$, which implies that:
\be
\delta_\zeta^{(0)}W = 0~.
\ee
In other words, the $U(1)$ 't Hooft anomaly is compatible with supersymmetry and there  is no genuine ``supersymmetry anomaly.''  Then, by its definition \eqref{del zeta in WZ gauge 2d}, the supersymmetry {\it in the Wess-Zumino gauge} does have an anomalous variation, namely:
\be
\delta_{\zeta}W = \delta_{\Omega(\zeta)} W~, \qquad \qquad \delta_{\t\zeta}W = \delta_{\Omega(\t\zeta)} W~.
\ee
Plugging the values \eqref{compensating component} inside \eqref{delt Om W 2d},  one finds:
\be
\delta_{\zeta}W =- \frac{i \CA_{qq}}{2\pi} \int d^2 x\, \zeta_- \t\lambda_- a_{\bar w}~,\qquad\qquad
\delta_{\t\zeta}W =- \frac{i\CA_{qq}}{2\pi} \int d^2x\,  \t\zeta_- \lambda_- a_{\bar w}~.
\ee
This indeed reproduces the anomalous supersymmetry variations \eqref{eq:susyano2d}.

%%%%%%%%%%%%%%%%%%%%%%%%%%%%%%%%%%%%%%%%%%%%%%%%%%%%%%%%%
%%%%%%%%%%%%%%%%%%%%%%%%%%%%%%%%%%%%%%%%%%%%%%%%%%%%%%%%%
\section{'t Hooft anomalies and 4d  $\CN=1$ partition functions}\label{sec: flavorsusy 4d}
%%%%%%%%%%%%%%%%%%%%%%%%%%%%%%%%%%%%%
Let us now consider 4d $\CN=1$ supersymmetric theories. In this section, after reviewing the curved-space rigid supersymmetry formalism \cite{Festuccia:2011ws, Dumitrescu:2012ha} and the main result of~\cite{Closset:2013vra} on the holomorphic dependence of supersymmetric partition functions on continuous parameters, we derive the supersymmetric Ward identities on a general half-BPS background. This determines the non-holomorphic dependence of the supersymmetric partition function, in our ``gauge invariant'' scheme for background gauge fields. In the case of half-BPS four-manifolds of topology $\CM_3 \times S^1$, with $S^1$ a circle of radius $\beta_{S^1}$, we also discuss how the 't Hooft anomaly together with supersymmetry determine the order-$\beta_{S^1}$ term in the small-$\beta_{S^1}$ expansion of the partition function. Our four-dimensional conventions are summarized in Appendix~\ref{Appendix: 4d}.

\subsection{Curved-space supersymmetry for 4d $\CN=1$ theories}
Consider a four-dimensional $\CN=1$ supersymmetric theory with a $U(1)_R$ symmetry. By assumption, there exists an $\CR$-multiplet, which contains the $R$-symmetry current $j^\mu_{R}$ together with the energy-momentum tensor $T^{\mu\nu}$, and which can be consistently coupled to four-dimensional new-minimal supergravity \cite{Sohnius:1981tp}---see {\it e.g.} \cite{Gates:1983nr, Dumitrescu:2011iu}. At the linearized level:
\be
\Delta\SL= -\half \Delta g_{\mu\nu} T^{\mu\nu} + A_\mu^{(R)} j^\mu_{(R)}+ \Psi_\mu^\alpha S^\mu_\alpha+ \t \Psi_{\alphadot\mu} \t S^{\alphadot\mu}  +{i\ov 4} \epsilon^{\mu\nu\rho\lambda} B_{\mu\nu} \CF_{\rho\lambda}~.
\ee
We consider a curved-space rigid supersymmetric background on a compact four-manifold $\CM_4$, which can be  obtained as a rigid limit of new-minimal supergravity, with the background fields:
\be\label{sugra fields 4d}
g_{\mu\nu}~, \qquad A_\mu^{(R)}~, \qquad V^\mu = {i\ov 2} \epsilon^{\mu\nu\rho\lambda}  \d_\nu B_{\rho\lambda}~,
\ee
such that the gravitino and its supersymmetry variations vanish  \cite{Festuccia:2011ws}:
\be\label{susy condition}
 \Psi_\mu =\t  \Psi_\mu=0~, \qquad \delta  \Psi_\mu=\delta \t\Psi_\mu=0~.
\ee
This means that there exists spinors $\zeta_\alpha$ and/or $\t\zeta^\alphadot$ on the Riemannian manifold $(\CM_4, g_{\mu\nu})$ that satisfy the generalized Killing equations:
\be
(\nabla_\mu - i A_\mu^{(R)})\zeta = -{i\ov 2} V^\nu \sigma_\mu \t\sigma_\nu \zeta~, \qquad\qquad
(\nabla_\mu + i A_\mu^{(R)})\t\zeta = {i\ov 2} V^\nu \t\sigma_\mu \sigma_\nu \t\zeta~.
\ee
The Killing spinors are globally defined and nowhere-vanishing. Their existence allows us to define curved-space rigid supersymmetries, with the supersymmetry variations on fields written as:
\be
\delta_\zeta= i\zeta \CQ~, \qquad\qquad  \delta_{\t \zeta} = i\t\zeta \t\CQ~,
\ee
where $\CQ_\alpha$ and $\t\CQ^\alphadot$ denote the ``curved-space supercharges'' (one for each Killing spinor). Note that the Killing spinors are bosonic and of $R$-charge $\pm 1$ while the supercharges have $R$-charge $\mp 1$, so that $\delta$ is a fermionic operation of vanishing $R$-charge.

\subsubsection{Complex structure and background supergravity fields}
 Given a Killing spinor $\zeta$, the real bilinear tensor:
\be
{J^\mu}_\nu= - 2i {\zeta^\dagger {\sigma^\mu}_\nuÂ \zeta\ov |\zeta|^2}~,
\ee
defines an integrable complex structure on $\CM_4$  \cite{Klare:2012gn, Dumitrescu:2012ha}. One can preserve a single supercharge on any Hermitian four-manifold. Given an Hermitian manifold, 
\be
(\CM_4,~ g_{\mu\nu},~ {J^\mu}_\nu)~,
\ee
 one can explicitly solve for $\zeta$ and for the remaining supergravity background fields:
\be\label{V and AR 4d explicit}
V_\mu = \half \nabla_\nu {J^\nu}_\mu+ U_\mu~, \qquad A_\mu^{(R)} = \h A_\mu^{(R)} + {i\ov 4} {J_\mu}^\nu \nabla_\rho {J^\rho}_\nu~,
\ee
with $U_\mu$ satisfying $U_\nu {J^\nu}_\mu  = i U_\mu$ and $\nabla_\mu U^\mu=0$. Here, for later reference, we defined:
\be\label{def Ah and Ac}
\h A_\mu^{(R)} =A_\mu^c -\half \nabla_\nu {J^\nu}_\mu~, \qquad\qquad A_\mu^c \equiv  {1\ov 4}(\d_\nu \log\sqrt{g}) {J^\nu}_\mu+ \d_\mu s~.
\ee
The expression for $A_\mu^c$ is only valid in complex coordinates adapted to the complex structure; the function $s$ in \eqref{def Ah and Ac} is a $U(1)_R$ gauge transformation parameter. We will use the notation $X^\mu$ ($\mu=1, \cdots, 4$) for the four-dimensional real coordinates, and:
\be\label{w z def0}
w= w(X)~, \qquad z=z(X)~,
\ee
for the complex coordinates $w, z$ on $\CM_4$. 

\subsubsection{Half-BPS Hermitian geometry}
In the following, our main interest will be in half-BPS supersymmetric backgrounds which preserve two Killing spinors of opposite chiralities, $\zeta$ and $\t \zeta$. The second Killing spinor, $\t \zeta$, defines a second complex structure:
\be
{\t J^\mu}_{\phantom{\mu}\nu}= - 2i {\t\zeta^\dagger {\t\sigma^\mu}_{\phantom{\mu}\nu}Â \t\zeta\ov |\t\zeta|^2}~,
\ee
and a complex Killing vector:
\be
K^\mu = \zeta \sigma^\mu \t \zeta~, \qquad \qquad \nabla_\mu K_\nu+ \nabla_\nu K_\mu=0~.
\ee
We will assume that $K$ commutes with its complex conjugate, $[K, \b K]=0$. Then, the background geometry admits two commuting complex structures, $[J, \t J]=0$, and $K$ is anti-holomorphic with respect to both:
\be
{J^\mu}_\nu K^\nu = - i K^\mu~, \qquad {\t J^\mu}_{\phantom{\mu}\nu} K^\nu = - i K^\mu~,
\ee
 We choose the holomorphic coordinates \eqref{w z def0} such that:
 \be\label{K first def 4d}
 K = {2\ov \beta_1}\,\d_{\b w}~,
 \ee 
 with $\beta_1 >0$ a real positive constant introduced for later convenience. 
The holomorphic coordinates adapted to the complex structure $\t J$ are $(w, \bz)$.
The background geometry is locally a $T^2$ fibration,%
\footnote{This is also true globally for $\CM_4$ as a topological manifold, but the $T^2$ fibers need not coincide with the orbits of $K^\mu$, as the latter need not close. When they do, we have an elliptic fibration over an Riemann surface orbifold. (See {\it e.g.} \protect\cite{Closset:2018ghr} for a thorough discussion of the 3d analogue.)} 
 and the Hermitian metric (with respect to $J$) can be written in the canonical form \cite{Dumitrescu:2012ha}:
\be\label{M4 metric wz gen}
ds^2(\CM_4) =\Omega(z,\bz) (dw + h(z, \bz) dz)(d\bw + \b  h(z, \bz)  d\bz) + 2 g_{z\bz}(z,\bz) dz d\bz~.
\ee
The factor $\Omega(z,\bz)$ is related to the norm of the Killing vector, and of the Killing spinors, as:
\be
|K|^2\equiv  K_\mu \b K^\mu= 2 |\zeta|^2 |\t\zeta|^2= 2   \beta_1^{-2} \Omega^2~.
\ee
For simplicity (and without much loss of generality), we will restrict ourselves to the case $\Omega=\beta_1$, so that $|K|^2=2$.

The supergravity background fields $V_\mu$ and $A_\mu^{(R)}$ are still given by \eqref{V and AR 4d explicit}, with the additional constraint:
\be\label{U as kappa}
U_\mu = \kappa K_\mu~, \qquad K^\mu\d_\mu  \kappa =0~.
\ee
The parameter $\kappa$ is part of the definition of the supersymmetric background, and it can be chosen arbitrarily. In the following, we will mostly be interested in the case of a product four-manifold $\CM_4 \cong \CM_3 \times S^1$, and we will require that the 4d $\CN=1$ supersymmetric background admits a consistent reduction to a 3d $\CN=2$ supersymmetric background on $\CM_3$ \cite{Closset:2012ru}. This will fix $\kappa$ uniquely.

\paragraph{Topological-holomorphic twist.} In  the following adapted complex frame:%
\footnote{Our conventions for spinors are given in Appendix~\ref{Appendix: 4d}. Here and henceforth, we set $\Omega= \beta_1$.}
\bea\label{e1 e2 def 4d}
&e^1 = \beta_1 (dw + h(z, \bz) dz)~, \qquad  && e^2 = \sqrt{2 g_{z\bz}(z,\bz)}  dz~,\cr
&e^{\b 1} = \beta_1 (d\bw + \b h(z, \bz) d\bz)~, \qquad 
&& e^{\b 2} = \sqrt{2 g_{z\bz}(z,\bz)}  d\bz~,
\eea
the Killing spinors take the particularly simple form: 
\be
\zeta_\alpha =e^{is} \mat{1\\ 0}~, \qquad \t\zeta^\alphadot =e^{-is} \mat{-1\\ 0}~.
\ee
Let us define the projectors onto holomorphic vectors in either complex structure:
\be\label{def proj holo}
{\Pi^\mu}_\nu = \half ({\delta^\mu}_\nu - i {J^\mu}_\nu)~, \qquad 
{\t\Pi^\mu}_{\phantom{\mu}\nu}  = \half ({\delta^\mu}_\nu - i {\t J^\mu}_{\phantom{\mu}\nu})~, 
\ee
and similarly for the anti-holomorphic projectors:
\be
{\b\Pi^\mu}_{\phantom{\mu}\nu} = \half ({\delta^\mu}_\nu +i {J^\mu}_\nu)~, \qquad 
\b{\t\Pi}^\mu_{\phantom{\mu}\nu}  = \half ({\delta^\mu}_\nu + i {\t J^\mu}_{\phantom{\mu}\nu})~.
\ee
Since the two complex structure commute, we can project any vector $X^\mu$ into its four components $X^w, X^\bw, X^z, X^\bz$, respectively, and similarly for any tensor.
One can also define an adapted connection, $\h \nabla$, that preserves {\it both} complex structures:
\be
\h \nabla_\mu g_{\nu\rho}=0~, \qquad \h\nabla_\mu {J^\nu}_\rho=0~, \qquad
\h\nabla_\mu  \t J^\nu_{\phantom{\nu}\rho} =0~.
\ee
It is defined by:
\be
\h \Gamma^\mu_{\phantom{\mu}\nu\rho}= \Gamma^\mu_{\phantom{\mu}\nu\rho} + K^\mu_{\phantom{\mu}\nu\rho}
\ee
with $\Gamma^\mu_{\phantom{\mu}\nu\rho}$ the Levi-Civita connection, and:
\be
K_{\mu\nu\rho} = \half {J_\mu}^\alpha {J_\nu}^\beta  {J_\rho}^\gamma (dJ)_{\alpha\beta\gamma}~,
\ee
 the contorsion tensor, with $ (dJ)_{\mu\nu\rho} \equiv \nabla_{\mu} J_{\nu\rho}+ \nabla_{\nu} J_{\rho\mu}+ \nabla_{\rho} J_{\mu\nu}$.%
 \footnote{The expression for $K_{\mu\nu\rho}$ takes the same form when written in terms of ${\t J^\mu}_{\phantom{\mu}\nu}$.} This connection has non-trivial torsion:
\be
 T^\mu_{\phantom{\mu}\nu\rho}= K^\mu_{\phantom{\mu}\nu\rho}- K^\mu_{\phantom{\mu}\rho\nu}= \epsilon^{\mu}_{\phantom{\mu}\nu\rho\lambda} {J^\lambda}_\sigma   \nabla_\kappa J^{\kappa\sigma}~.
\ee
In terms of the adapted connection, the Killing spinor equations take the simple form:
\be
(\h\nabla_\mu- i \h A^{(R)}_\mu)\zeta=0~, \qquad \qquad
(\h\nabla_\mu+ i \h A^{(R)}_\mu)\t\zeta=0~,
\ee
with $\h A_\mu^{(R)}$ defined in \eqref{def Ah and Ac}.
The connection $\h \nabla$ has a $U(1)$ holonomy which can be ``twisted'' by the $U(1)_R$ gauge field. Whenever the orbits of $K^\mu$ close, we have a proper elliptic fibration over a two-dimensional complex space $\Sigma$ and the 4d $\CN=1$ supersymmetric background is precisely the four-dimensional uplift of the topological $A$-twist on $\Sigma$ \cite{Witten:1988xj, Dumitrescu:2012ha}.
The half-BPS supersymmetric background \eqref{V and AR 4d explicit}-\eqref{M4 metric wz gen} can then be understood as a topological-holomorphic twist---that is, holomorphic along the fiber direction \cite{Closset:2014uda} (with coordinate $w$)  Â and topological along the base (with coordinate $z$).

%%%%%%%%%%%%%%%%%
\subsection{Background vector multiplet and minimal coupling}
Consider an $\CN=1$ theory with a flavor symmetry group $G_F$, and let us introduce a background vector multiplet $\CV_F$ for that symmetry. Its component fields are:
\be\label{CV WZ gauge components}
\CV_F= (a_\mu~,\, \lambda_\alpha~,\,  \t \lambda^\alphadot~, \, D)~,
\ee
with $\lambda, \t\lambda$ a ``background'' gaugino. 
Note again that we fixed the Wess-Zumino gauge for $\CV_F$, and work only in terms of the remaining ``physical'' components \eqref{CV WZ gauge components}. The supersymmetry variations in WZ gauge \cite{Wess:1974tw} are famously modified by including a compensating gauge transformation into the supersymmetry transformation, to preserve the WZ gauge. This implies that the supersymmetry algebra itself become gauge-covariant under $G_F$.

Let us then consider the background vector multiplet \eqref{CV WZ gauge components} in addition to the supersymmetric curved-space background on $\CM_4$. Note that, while we require the supergravity background fields to be fixed to definite values so that \eqref{susy condition} holds, as described above, we do not impose any requirement on $\CV_F$ at this point---we will need to consider arbitrary background values of the fields  \eqref{CV WZ gauge components}, including for the gaugino.

Given two Killing spinors $\zeta$ and $\t\zeta$, we have two independent supersymmetries on $\CM_4$, denoted by $\delta_\zeta$ and $\delta_{\t\zeta}$, respectively, satisfying the following curved-space supersymmetry algebra \cite{Closset:2014uda}:
\bea\label{SUSY alg 4d}
& \delta_\zeta^2 \Phi_{(s, r)}=\delta_{\t\zeta}^2\Phi_{(s, r)}=0~, \qquad \cr
&\{ \delta_\zeta, \delta_{\t \zeta}\} \Phi_{(s, r)}= 2 i \Big(\CL_K- i r K^\mu\Big(A_\mu^{(R)}+{3\ov 2} V_\mu\Big) - i  K^\mu a_\mu\Big) \Phi_{(s, r)}~,
\eea
on any field $\Phi_{(s, r)}$ of spin $s$, $R$-charge $r$ and arbitrary flavor charges. Here, $\CL_K$ denotes the Lie derivative along the Killing vector $K^\mu$, and the background gauge field $a_\mu$ acts in the appropriate $G_F$ representation.

The curved-space rigid supersymmetry transformations for the vector multiplet \eqref{CV WZ gauge components} are given by:
\bea\label{susy V 4d}
& \delta_{\zeta} a_\mu = i \zeta \sigma_\mu \t \lambda ~,\qquad
&& \delta_{\t\zeta} a_\mu = i   \t \zeta\, \t \sigma_\mu \lambda~,\cr
&  \delta_{\zeta}  \lambda = i \zeta D +  \sigma^{\mu\nu}\zeta \, f_{\mu\nu}~, 
&& \delta_{\t\zeta}  \lambda = 0~, \cr
& \delta_{\zeta}  \t \lambda =0~,
&& \delta_{\t\zeta}  \t \lambda = - i \t \zeta D +  \t \sigma^{\mu\nu} \t \zeta \,  f_{\mu\nu}~,\cr
& \delta_{\zeta}  D = - D_\mu (\zeta \sigma^\mu \t \lambda) + 2 i V_\mu \zeta \sigma^\mu \t \lambda~,\qquad 
&& \delta_{\t\zeta} D = D_\mu (\t \zeta \, \t \sigma^\mu \lambda) + 2 i V_\mu  \t \zeta \, \t \sigma^\mu \lambda~,
\eea
in the WZ gauge. Here, we defined the field strength
\be
f_{\mu\nu} =  \d_\mu a_\nu - \d_\nu a_\mu - i [a_\mu, a_\nu]~,
\ee
and the covariant derivatives $D_\mu$ are understood to be appropriately gauge covariant, including under $U(1)_R$---for instance, $D_\mu \lambda = \nabla_\mu\lambda - i A^{(R)}_\mu \lambda - i[a_\mu,\lambda]$.

\paragraph{Coupling to an abelian conserved current.} Let us focus on the case of an abelian $G_F$, for simplicity. In a supersymmetric theory, any $U(1)$ conserved current operator  $j^\mu$ (with the conservation equation $\nabla_\mu j^\mu=0$ on $\CM_4$), sits in a linear multiplet:
\be\label{linear mult}
\CJ = (J~, \, j^\mu~, \, j_\alpha~, \, \t j^\alphadot~)~.
\ee
The supersymmetry variations are:
\bea\label{susy JF}
&\delta_{\zeta} J =i\zeta j~, 
&&\;\quad\qquad\delta_{\t\zeta} J =  i \t\zeta\t j~,\cr
&\delta_{\zeta} j_\alpha = 0~,\quad 
&&\;\quad\qquad\delta_{\t\zeta} j_\alpha = (\sigma^\mu\t\zeta)_\alpha \big( \d_\mu J - i j_\mu - 2i V_\mu J \big)~,\cr
& \delta_{\zeta} \t j^{\dot\alpha} =  -(\t\sigma^\mu\zeta)^\alphadot \big( \d_\mu J + i j_\mu +&&\hspace{-0.3cm} 2i V_\mu J \big)~, 
\;\; \delta_{\t\zeta} \t j^{\dot\alpha} = 0~, \cr
& \delta_{\zeta} j_\mu =-2 \nabla^\nu(\zeta \sigma_{\mu\nu} j)~,
&&\;\quad\qquad \delta_{\t\zeta} j_\mu =2 \nabla^\nu(\t\zeta \t\sigma_{\mu\nu} \t j)~.
\eea
We then have the  supersymmetric minimal coupling:
\be\label{minimal coupling 4d}
S_{\CV \CJ}[\CV_F]= \int d^4 x \sqrt{g}~ \SL_{\CV \CJ}[\CV_F]~, \qquad \SL_{\CV \CJ}[\CV_F]\equiv  a_\mu j^\mu + D J - \lambda j - \t\lambda  \t j~.
\ee
In the following, we will consider small variations of the background vector multiplet around a fixed value, $\CV_F^{(0)}$, denoted by $\Delta \CV_F$:
\be
\CV= \CV_F^{(0)}+ \Delta \CV_F~.
\ee
On any supersymmetric background $\CM_4$, the first-order correction to the effective action upon a small change of the sources is captured by the expectation value of the minimal coupling \eqref{minimal coupling 4d}.

\paragraph{Supersymmetric background.}
A supersymmetric background for the vector multiplet is a bosonic configuration $(a_\mu, D)$ such that the gaugino and its variations vanish:
\be
\lambda= \t \lambda=0~, \qquad \delta_\zeta \lambda=0~, \qquad  \delta_{\t \zeta}\t\lambda =0~.
\ee
On our half-BPS geometry, this corresponds to:
\be\label{susy halfBPS 4d V}
f_{w\bw}=0~, \quad f_{\b w \b z}=0~, \quad f_{\b w z}=0~, \quad
D= -\half J^{\mu\nu} f_{\mu\nu}~.
\ee
We will consider {\it real} background gauge fields $a_\mu$ on $\CM_{g,p} \times S^1$, in which case the only non-vanishing gauge-field curvature component is $f_{z\bz}$. The background gauge field is the connection of a holomorphic line bundle over $\CM_4$ \cite{Closset:2013vra}, which we denote by ${\bf L}_F$.

%%%%%%%%%%%%%%%%%
\subsection{$\CM_3\times S^1$ supersymmetric partition functions and holomorphy}
Consider any 4d $\CN=1$ supersymmetric theory with a $U(1)_R$ symmetry.
Its supersymmetric partition function on $\CM_4$, denoted by $Z_{\CM_4}$, is the path integral on a fixed supersymmetric background for the geometry and for the vector multiplet $\CV_F$:
\be\label{background fields}
{\bf L}_F \rightarrow \CM_4 \; : \; (g_{\mu\nu}~,\, A_\mu^{(R)}~, \,V_\mu~;\;   \, a_\mu~,\, D)~.
\ee
Naively, the partition function may depend on all the background fields shown in~\eqref{background fields}. 
However, it was argued in~\cite{Closset:2013vra} that, given an Hermitian four-manifold $\CM_4$ and a choice of flavor line bundle ${\bf L}_F$, the partition function $Z_{\CM_4}$ only depends on two types of {\it continuous} complex parameters:
\be
Z_{\CM_4}(\nu, \tau)~.
\ee
Here, we have:%
\footnote{Here we are glossing over some of the needed geometric data \protect\cite{Closset:2013vra}, for simplicity. } 
\begin{itemize}
\item[(i)] the ``geometric'' parameters $\tau$, which are {\it complex structure moduli} of the complex manifold $\CM_4$;
\item[(ii)] the flavor parameters $\nu$, which are {\it holomorphic line bundle moduli} of the line bundles ${\bf L}_F$. (That is, for abelian symmetries; they are holomorphic vector bundle moduli in general.)
\end{itemize}
In particular, for any fixed complex structure on $\CM_4$, the supersymmetric partition function is independent of the choice of Hermitian metric. Moreover, the partition function was found to be {\it holomorphic} in all of its continuous parameters:
\be
{\d\ov \d \b\nu} Z_{\CM_4}(\nu, \tau)=0~, \qquad 
{\d\ov \d \b\tau} Z_{\CM_4}(\nu, \tau)=0~.
\ee
These results seemingly holds for any supersymmetric background \eqref{background fields} preserving at least one supercharge. For half-BPS geometries, we have further restrictions, and the theory becomes fully topological along the $z, \bz$ directions \cite{Closset:2013vra}.

These results were derived by  a standard argument. Consider the effective action $W[b]$, defined as minus the logarithm of the partition function in the presence of the supersymmetric background fields $b_n$:
\be
W[b]= - \log Z_{\CM_4}[b]~,
\ee
and consider an infinitesimal variation:
\be
b_n = b^{(0)}_n + \Delta b_n~,
\ee
of the background fields $b_n$ around some particular supersymmetric value, $b^{(0)}_n$, {\it while keeping $b_n$ supersymmetric.} Each independent variation $ \Delta b_n$ couples to a particular bosonic operator $\CO_n$ contained in the $\CR$-multiplet or in the linear multiplet $\CJ$ of the quantum field theory coupled to $\CM_4$, so that:
\be\label{selection rule gen}
{\d \ov \d \Delta b_n} W[b^{(0)}]=  \Big\langle \int d^4x \sqrt{g} \CO_n \Big\rangle_{\CM_4}~,
\ee
at first order in $\Delta b_n$.
Whenever the insertion is $Q$-exact:
\be
 \int d^4x \sqrt{g} \,\CO_n =  \Big\{ Q, \int d^4x \sqrt{g} \, \psi_n \Big\}~,
\ee
for $\psi_n$ some fermionic operator, the expression \eqref{selection rule gen} vanishes and the partition function is independent of that particular parameter $b_n$. This result relies on the fact that the expectation value of any $Q$-exact operator vanishes:
\be
\big\langle \CO \big\rangle_{\CM_4}= \big\langle \{Q, \psi \} \big\rangle_{\CM_4}=0~.
\ee
As in 2d, the existence of flavor 't Hooft anomalies will amend this simple picture, and imply an anomalous dependence of the partition function on some $Q$-exact couplings~\cite{Itoyama:1985qi, Papadimitriou:2019yug} .

\subsection{Supersymmetry variation of the effective action}
Let $W=W[\CV_F]$ denote the effective action in the presence of {\it arbitrary} flavor-symmetry sources \eqref{CV WZ gauge components}, including the background gauginos:
\be
W[\CV_F]=W[a_\mu, \lambda, \t\lambda, D]~,
\ee
on a fixed background $\CM_4$; we now consider $G_F= U(1)$, for simplicity.
Since the theory is supersymmetric, we would expect that:
\be
\delta_\zeta W\stackrel{?}{=} 0~, \qquad\qquad \delta_{\t\zeta} W \stackrel{?}{=} 0~,
\ee
under the supersymmetry transformations \eqref{susy V 4d} of the background fields. This is inconsistent with $U(1)$ gauge invariance and diff-invariance whenever the $U(1)$ symmetry suffers from 't Hooft anomalies \cite{Itoyama:1985qi}, in the following sense. Let $\delta_\alpha$ denote a gauge transformation with gauge parameter $\alpha$, with:
\be
\delta_\alpha a_\mu = \d_\mu \alpha~, \qquad \delta_\alpha \lambda =\delta_\alpha \t \lambda =\delta_\alpha D=0~.
\ee
The consistent anomalous variation of the effective action takes the form:
\be\label{Anomaly U1 4d}
\delta_\alpha W = {i \CA_{qqq} \ov 96\pi^2} \int d^4x \sqrt{g} \, \alpha \epsilon^{\mu\nu\rho\sigma} f_{\mu\nu} f_{\rho\sigma} +  {i \CA_{q} \ov 384\pi^2} \int d^4x \sqrt{g} \, \alpha \CP+ \cdots~,
\ee
with:
\be
 f_{\mu\nu}= \d_\mu a_\nu - \d_\nu a_\mu~, \qquad\qquad  \CP= \half \epsilon^{\mu\nu\rho\sigma} R_{\mu\nu\lambda\kappa}{R_{\rho\sigma}}^{\lambda \kappa}~,
 \ee
 the $U(1)$ field strength and the Pontryagin density, respectively, with $R_{\mu\nu\rho\lambda}$ the Riemann tensor of the metric $g_{\mu\nu}$. The ellipsis in \eqref{Anomaly U1 4d} denotes additional contributions from mixed $U(1)_R$-$U(1)$ anomalies, to be discussed in subsection~\ref{subsec: U1R contrib 4d} below.
The cubic and mixed $U(1)$-gravitational 't Hooft anomalies coefficients are normalized such that $\CA_{qqq}=q^3$ and $\CA_q= q$ for a single chiral fermion $\psi_\alpha$ of charge $q$. In a Lagrangian theory, we have:
\be
\CA_{qqq}= \tr(\gamma^5 Q_F^3)~, \qquad\qquad \CA_{q} =  \tr(\gamma^5 Q_F)~,
\ee
where the trace is a sum over all the chiral fermions charged under $U(1)$. 
 The supersymmetry variations of the $U(1)$ anomaly \eqref{Anomaly U1 4d} are non-trivial. We have:
\be\label{del zeta alpha W}
\delta_\zeta( \delta_\alpha W) =  { \CA_{qqq} \ov 24\pi^2} \int d^4x \sqrt{g} \, \d_\mu \alpha \, \epsilon^{\mu\nu\rho\lambda} \zeta\sigma_\nu \t\lambda f_{\rho\lambda}~,
\ee
and similarly for $\delta_{\t\zeta}( \delta_\alpha W)$.
Since the supersymmetry and gauge transformations commute,  we necessarily have
$\delta_\alpha( \delta_{\zeta} W) \neq 0$, and therefore there must be a non-trivial supersymmetry variation of the quantum effective action, $\delta_{\zeta} W \neq 0$. More generally, the supersymmetry variations must satisfy the following Wess-Zumino consistency conditions \cite{Itoyama:1985qi, Papadimitriou:2019yug}:
\bea\label{WZ consistency conditons 4d}
&[\delta_\alpha, \delta_\zeta]W= [\delta_\alpha, \delta_{\t\zeta}]W=0~, & \cr
& \{ \delta_\zeta,  \delta_{\t\zeta}\}W = \delta_{\alpha(a)}W~,& \quad {\rm with}\quad \alpha(a)= -2 i K^\mu a_\mu~.
\eea
Note the $a_\mu$-dependent gauge transformation on the second line, which is a consequence of working in the WZ gauge for the vector multiplet. 
The consistency conditions \eqref{WZ consistency conditons 4d} determines the anomalous supersymmetry variations, $\delta_\zeta W$ and $\delta_{\t\zeta} W$, on a fixed curved-space supersymmetric background.  

At this point, an important comment is in order. The curved-space supersymmetry algebra \eqref{SUSY alg 4d} takes the form:
\be
 \{ \delta_\zeta,  \delta_{\t\zeta}\} = \delta_{\xi(K)}+ \delta_{\alpha(a)}~,
\ee
with $\delta_{\xi(K)}$ a diffeomorphism along the Killing vector $K$. In four dimensions, we can always choose a scheme which preserves diffeomorphism invariance, and therefore $\delta_\xi W=0$ for any $\xi$. This leads to the WZ conditions \eqref{WZ consistency conditons 4d}. The results of this paper are then only valid in such a diff-invariant scheme. The condition \eqref{WZ consistency conditons 4d} has the important implication that, in any diff-invariant scheme, the variations $\delta_\zeta W$ and $\delta_{\tilde{\zeta}} W$ cannot vanish, because their commutator gives a gauge-transformation, and the latter is non-zero in any scheme by the assumption that a flavor 't Hooft anomaly exists \cite{Papadimitriou:2017kzw}.~\footnote{Note, however, that this argument holds only when the background supergravity fields are allowed to vary arbitrarily. On a fixed half-BPS geometry---that is, for {\it fixed Killing spinors}---, there might well exist a diff-invariant local term that cancels $\delta_\zeta W$, and therefore cancels the gauge anomaly for the particular gauge parameter $a(\alpha)$ in \protect\eqref{WZ consistency conditons 4d}, as we discussed at the end of section~\protect\ref{subsec: ZT2 revisit}. In any case, this appears not to be the case in 4d \protect\cite{Itoyama:1985qi}.}

Let us decompose the supersymmetry variations into contributions from the cubic and linear flavor 't Hooft anomalies:
\be\label{flavor susy 4d full}
\delta_\zeta W=\delta_\zeta W|_{q^3}+ \delta_\zeta W|_{q}+ \cdots~, \qquad\qquad
\delta_{\t\zeta} W=\delta_{\t\zeta} W|_{q^3}+ \delta_{\t\zeta} W|_{q}+ \cdots~,
\ee
where the ellipsis denotes the contribution from other potential anomalies, to be discussed later on. 
One finds:
\bea\label{Anomaly SUSY qqq 4d}
&\delta_\zeta W|_{q^3}= -{ \CA_{qqq} \ov 24\pi^2} \int d^4x \sqrt{g} \, \left(\epsilon^{\mu\nu\rho\sigma} \zeta \sigma_\mu \t\lambda \, a_\nu f_{\rho\sigma} -  3i \zeta \lambdaÂ \, \t\lambda\t\lambda \right)~,\cr
&\delta_{\t\zeta} W|_{q^3}= -{ \CA_{qqq} \ov 24\pi^2} \int d^4x \sqrt{g} \, \left(\epsilon^{\mu\nu\rho\sigma} \t\zeta\t \sigma_\mu \lambda \, a_\nu f_{\rho\sigma} + 3 i\t \zeta \t\lambdaÂ \, \lambda\lambda \right)~,
\eea
for the cubic-anomaly contribution. The term linear in the gaugino obviously follows from \eqref{del zeta alpha W}; the second term, which is gauge-invariant and cubic in the gaugino, is fixed by the second line of \eqref{WZ consistency conditons 4d}. This result was first obtained in flat space in~\cite{Itoyama:1985qi}, and much more recently on curved space but at first order in the gauginos in~\cite{Papadimitriou:2019yug}; here we give the full expression valid on any curved-space $\CN=1$ new-minimal rigid-supersymmetric background.

The mixed $U(1)$-gravitational anomaly contribution to $\delta_\zeta W$ is harder to study on a fixed supergravity background. On any half-BPS four-manifold, the Pontraygin density turns out to be a total derivative:
\be\label{CP at tot der}
\CP = -\nabla_\mu \CP^\mu~,
\ee
with  the vector $\CP^\mu$ a local expression of dimension $3$ in the supergravity fields \cite{Assel:2014tba}. Then, one can easily check that the following terms solve the Wess-Zumino consistency conditions:
\bea\nn
&\delta_\zeta W|_{q}=- { \CA_q \ov 384\pi^2} \int d^4 x\sqrt{g} \, \zeta \sigma_\mu \t\lambda \, \CP^\mu~,\quad\delta_{\t\zeta} W|_{q}= -{ \CA_q \ov 384\pi^2} \int d^4 x\sqrt{g} \,\t \zeta \t\sigma_\mu \lambda \, \CP^\mu~.
\eea
This solves  \eqref{WZ consistency conditons 4d}, as one can see using \eqref{CP at tot der} and the fact that $\CP^\mu$ is invariant under the isometry $K$ of the background.
Note that the vector $\CP^\mu$ is only defined up to a divergenceless quantities; to completely fix it, one should consider the full background new-minimal supergravity away from the rigid-supersymmetry limit. One actually finds \cite{Papadimitriou:2019yug}:
\be\label{del W Aq zero}
\delta_{\zeta} W|_{q}=0~, \qquad \delta_{\t\zeta } W|_{q}=0~,
\ee
on any half-BPS supersymmetric background. We will give another argument to that effect in section~\ref{subsec: susy an from wz 4d}.

\subsection{Non-holomorphy of the supersymmetric partition function $Z_{\CM_4}$}
Even though the supersymmetry variation \eqref{flavor susy 4d full} vanishes when we set the background gauginos to zero, it has some important consequence for supersymmetric partition functions, as we now explain.

On curved space with two Killing spinors $\zeta$ and $\t\zeta$, it will be convenient to introduce the following twisted notations for the gauginos in the background vector multiplet:
\be\label{twisted lambdas def}
\t\Lambda_\mu = \zeta \sigma_\mu \t\lambda~, \qquad \qquad \Lambda_\mu = \t\zeta \t\sigma_\mu \lambda~.
\ee
The minimal coupling \eqref{minimal coupling 4d} to the linear multiplet then reads:
\be\label{min coupling Lambda}
\SL_{\CV \CJ}=  a_\mu j^\mu + D J + \Lambda_\mu {\bf j}^\mu + \t\Lambda_\mu\t{\bf j}^\mu~,
\ee
where we introduced the twisted fermionic operators:
\be
\t{\bf j}^\mu = \half  {\zeta^\dagger \sigma^\mu \t j \ov|\zeta|^2}~, \qquad \qquad
{\bf j}^\mu= \half  {\t\zeta^\dagger\t \sigma^\mu  j \ov|\t\zeta|^2}~,
\ee
It follows from their definition that $\t\Lambda$ and $\t{\bf j}$ are holomorphic one-forms and vectors, respectively, with respect to the complex structure ${J^\mu}_\nu$; similarly, $\Lambda$ and ${\bf j}$ are holomorphic with respect to $\t J^\mu_{\phantom{\mu}\nu}$.

\paragraph{The case of a single Killing spinor $\zeta$.} Let us first look at the case of a single curved-space supercharge, with Killing spinor $\zeta$. We consider small variations of the background gauge field near the supersymmetric locus:
\be
 f_{\bw \bz}=0~,\qquad D= -\half J^{\mu\nu} f_{\mu\nu}~.
\ee
In particular, we allow a small variation $\Delta D$ of the background field $D$ around its supersymmetric value. We also turn on background gauginos:%
\footnote{When considering the effect of a single supercharge, we cannot use $\t\zeta$ to ``twist'' $\lambda_\alpha$ into a one-form as in \protect\eqref{twisted lambdas def}; instead, we could write it in terms of two scalars $\gamma= \zeta\lambda$ and $\eta= {\zeta^\dagger \lambda\ov |\zeta|^2}$, but this will not be necessary for our discussion.}
\be
\lambda_\alpha~,\qquad \qquad\t\Lambda_\mu = \zeta \sigma_\mu \t\lambda~.
\ee
Then, the effective action $W= W[a_\mu, \lambda_\alpha, \t\Lambda_\mu, \Delta D]$  has a nilpotent supersymmetry:
\be
\delta_\zeta a_\mu = i \t\Lambda_\mu~,
\qquad \quad
\delta_\zeta \lambda_\alpha=i \zeta_\alpha \Delta D~,\qquad\quad
\delta_\zeta \t \Lambda_\mu=0~, \qquad \quad \delta_\zeta \Delta D = 0~,
\ee
with $\delta_\zeta W$ determined by the 't Hooft anomalies as discussed above.
Let us define the operator:
\be\label{def CJ}
\CJ^\mu \equiv j^\mu - i \d^\mu J + 2 V^\mu J~,
\ee
in terms of the linear multiplet operators \eqref{linear mult}. It follows from \eqref{susy JF} that:
\be
\delta_\zeta \t{\bf j}^\mu = i  {\Pi^\mu}_\nu\CJ^\nu~,
\ee
where ${\Pi^\mu}_\nu$ is the projector on the holomorphic indices, as defined in \eqref{def proj holo}.
Using the complex coordinate indices $X^\mu= (z^i, \bz^{\b i})$, we can write the minimal coupling as:
\be\label{lin coupling 1Q 4d}
\SL_{\CV\CJ} = a_i \CJ^i+ a_{\b i}(j^{\b i} + i \d^{\b i} J +2 V^{\b i} - 2 U^{\b i}) + J \h D - \lambda j + \t\Lambda_\mu \t{\bf j}^\mu~.
\ee
We see that the {\it holomorphic} gauge field components, $a_i$, couple to a $Q$-exact operator. We have:
\be\label{delta W ai}
{\delta W\ov \delta a_i}= \left\langle \CJ^i \right\rangle_{\CM_4} = -i  \left\langle \delta_\zeta \t {\bf j}^i \right\rangle_{\CM_4}~. 
\ee
Assuming that $Q$-exact operators decouple, we would conclude that the partition function is independent of $a_i$, and only depends on the background gauge field $a_\mu$ through $a_{\b i}$; in fact, one can then easily show that the dependence is only through the {\it holomorphic moduli} of the line bundle ${\bf L}_F$; they are in one-to-one correspondence with the Dolbeault cohomology classes:
\be
[\Delta a_{\b i}] \in H^{0,1}(\CM_4)~.
\ee
Due to the quantum correction to the supersymmetric Ward identity, however, the right-hand-side of \eqref{delta W ai} does not vanish. Instead, since $\t{\bf j}^i$ couples to $\t\Lambda_i$ as shown in \eqref{delta W ai}, we have:
\be\label{holomorphy anomaly gen 4d}
{\delta W\ov \delta a_i} = - i {\delta \ov \delta \t \Lambda_i}  (\delta_\zeta W)\Big|_{\lambda=\t\lambda=0}~.
\ee
This anomalous Ward identity gives rise to a {\it holomorphy anomaly} in the flavor-parameter dependence of supersymmetric partition functions. Since the supersymmetry variation \eqref{Anomaly SUSY qqq 4d} has a term linear in $\t\Lambda$, the holomorphy anomaly \eqref{holomorphy anomaly gen 4d} will generally be non-vanishing.

\paragraph{The case of a two Killing spinors, $\zeta$ and $\t \zeta$.}
Consider now the case of a half-BPS background. We define the linear-multiplet operators:
\be\label{def CJt}
\t\CJ^\mu \equiv j^\mu + i \d^\mu J + 2 V^\mu J~,
\ee
in addition to \eqref{def CJ}. We now have the relations:
\be
\delta_\zeta \t{\bf j}^\mu = i  {\Pi^\mu}_\nu\CJ^\nu~, \qquad \qquad 
\delta_{\t\zeta} {\bf j}^\mu = i  {\t\Pi^\mu}_\nu\t\CJ^\nu~, 
\ee
in terms of the projectors \eqref{def proj holo}. This is equivalent to:
\be
\delta_\zeta \t{\bf j}^w = i  \CJ^w~, \qquad
\delta_\zeta \t{\bf j}^z = i  \CJ^z~, \qquad
\delta_{\t\zeta} {\bf j}^w = i \t\CJ^w~, \qquad
\delta_{\t\zeta} {\bf j}^\bz = i \t\CJ^\bz~.
\ee
Expanding around the supersymmetry locus \eqref{susy halfBPS 4d V}, the linear coupling \eqref{min coupling Lambda} reads:
\bea\label{lin S full 2Qs}
&\SL_{\CV\CJ} &=&\; a_w \CJ^w + a_\bw (\t \CJ^\bw - \kappa K^\bw)+ a_z \CJ^z+ a_\bz \CJ^\bz + J \Delta D\cr
&&&\;+ \Lambda_w {\bf j}^w+\Lambda_z {\bf j}^z + \t\Lambda_w\t{\bf j}^w + \t\Lambda_\bz\t{\bf j}^\bz~. 
\eea
In this case, only $a_{\bw}$ does not couple to a $\CQ$-exact operator, and therefore the partition function is expected to depend only on the corresponding line bundle modulus, which we denote by $\nu$.
Note also that the coupling $a_w \CJ^w$ in \eqref{lin S full 2Qs} is equivalent to $a_w \t\CJ^w$ by integration by part. We then find the anomalous Ward identities:
\bea\label{full holomorphy anomaly gen 4d}
&{\delta W\ov \delta a_w} = - i {\delta \ov \delta \t \Lambda_w}  (\delta_\zeta W)\Big|_{\lambda=\t\lambda=0}=  - i {\delta \ov \delta \Lambda_w}  (\delta_{\t\zeta} W)\Big|_{\lambda=\t\lambda=0}~,\cr
&{\delta W\ov \delta a_z} =- i {\delta \ov \delta \t \Lambda_z}  (\delta_\zeta W)\Big|_{\lambda=\t\lambda=0}~,\cr
&{\delta W\ov \delta a_\bz} =- i {\delta \ov \delta \Lambda_\bz}  (\delta_{\t\zeta} W)\Big|_{\lambda=\t\lambda=0}~.
\eea
In section~\ref{sec: MgpS1}, we will evaluate these identities explicitly, for a large class of half-BPS supersymmetric partition funtions.

\subsection{The small-$\beta_{S^1}$ limit of $Z_{\CM_3 \times S^1}(\nu, \tau)$}
Whenever the four-dimensional background is of the form:
\be
\CM_4 \cong \CM_3 \times S^1~,
\ee
it is interesting to consider the small-circle limit, $\beta_{S^1}\rightarrow 0$, where $\beta_{S^1}$ is the radius of the $S^1$. Formally, this is a ``high-temperature'' limit of the supersymmetric partition function. At finite temperature---and also with supersymmetry-preserving boundary conditions on the Euclidean time circle, provided that generic fugacities are turned on---, it is expected that the effective action is {\it local} in the background gauge field and metric, order by order in $\beta_{S^1}$ \cite{Banerjee:2012iz, DiPietro:2014bca}.
For any set of background fields $b$, we have:
\be\label{Wb expansion}
W[b]\;  \underset{\beta_{S_1}\to \,0}{\sim}\; {1\ov \beta_{S^1}} W^{(-1)}_{\rm 3d}[b] +  W^{(0)}_{\rm 3d}[b]  + \beta_{S^1} W^{(1)}_{\rm 3d}[b]  + \cdots~.
\ee
 The $1/\beta_{S^1}$ term was studied in \cite{DiPietro:2014bca}; it is fully gauge invariant and supersymmetric. The finite term $W^{(0)}_{\rm 3d}$ roughly corresponds to the supersymmetric partition function of the 3d $\CN=2$ theory on $\CM_3$ obtained from the 4d $\CN=1$ theory by dimensional reduction, up to important subtleties that will not affect our discussion---see \cite{Ardehali:2015hya, DiPietro:2016ond, Hwang:2017nop, Hwang:2018riu}.  
 Here, we are interested in {\it the order-$\beta_{S^1}$ functional} in \eqref{Wb expansion}. 
Let us introduce the notation:
 \be\label{def W3d}
 \bW_{\rm 3d}[b] \equiv   \beta_{S^1} W^{(1)}_{\rm 3d}[b]~.
 \ee
This term should capture all the anomalies of the four-dimensional theory, including the anomalous variation $\delta_\zeta W$, which allows us to fix it entirely---modulo fully gauge-invariant and supersymmetric terms, which do not contribute on the supersymmetric locus.

 \subsubsection{Dimensional reduction and 3d $\CN=2$ supersymmetry}\label{subsubsec:M3 background}
 Before discussing the anomalies, we need to recall some formalism to deal with supersymmetry on $\CM_3$. Here, we assume that the background is half-BPS, with two Killing spinors $\zeta$ and $\t\zeta$.
 
 \paragraph{3d $\CN=2$ supergravity background fields.} We couple the 4d $\CN=1$ theory to a fixed $\CN=1$ new-minimal supergravity background on $\CM_3 \times S^1$, and similarly the three-dimensional reduced theory couples to a 3d $\CN=2$ supersymmetric background on $\CM_3$, of the type studied in \cite{Closset:2012ru}. The three-dimensional bosonic supergravity fields are denoted by:
 \be
 g_{\mu\nu}^{(\rm 3d)}~, \qquad \bA_\mu^{(R)}~, \qquad \bH~, \qquad \bV_\mu~.
 \ee
 The three-dimensional Killing spinor equations read:
 \bea
& (\nabla_\mu- i \bA_\mu^{(R)}) \zeta = - \half \bH \gamma_\mu \zeta +{ i\ov 2} \bV^\nu \gamma_\mu \gamma_\nu \zeta~, \cr
&  (\nabla_\mu+ i \bA_\mu^{(R)})\t \zeta = - \half \bH \gamma_\mu \t\zeta -{ i\ov 2} \bV^\nu \gamma_\mu \gamma_\nu \t\zeta~.
\eea
Our conventions for the 3d $\gamma$-matrices are given in Appendix~\ref{def gamma 3d}.
We choose the following Kaluza-Klein reduction ansatz for the metric:
 \be\label{dsM4 KK}
 ds^2(\CM_4) =  (\beta_{S^1} dt + c_\mu dx^\mu)^2+ds^2(\CM_3)~.
 \ee
We will focus on the case of a topologically-trivial circle bundle over $\CM_3$, in which case the graviphoton $c_\mu$ is actually a well-defined one-form on $\CM_3$.
 The three-dimensional background field $\bV_\mu$ should be given in terms of the 3d graviphoton, as:
 \be\label{bV def}
 \bV^\mu = -i \epsilon^{\mu\nu\rho}\d_\nu c_\rho~.
 \ee 
 Let us also define the unit-norm one-form:
 \be
 \h e^0 = \beta_{S^1} dt + c_\mu dx^\mu~.
 \ee
Then, the 3d and 4d supergravity fields are related as \cite{Closset:2012ru}:
 \bea\label{V and A 4d to 3d}
 V_\mu dX^\mu = \bH  \h e^0  +\half \bV_\mu dx^\mu~, \qquad 
  A_\mu^{(R)} dX^\mu = -\half \bH  \h e^0  + \left(\bA_\mu^{(R)} + {1\ov 4} \bV_\mu\right) dx^\mu~,
 \eea
 where $X^\mu$ and $x^\mu$ denote the coordinates on $\CM_4$ and $\CM_3$, respectively. 
 
 \paragraph{Transversely holomorphic structure (THF) on $\CM_3$.}  The 4d Killing spinors descend to 3d Killing spinors. In terms of the latter, one can define the following real one-form on $\CM_3$:
 \be
 \eta_\mu \equiv {\zeta^\dagger \gamma_\mu \zeta\ov |\zeta|^2} = - {\t\zeta^\dagger \gamma_\mu \t\zeta\ov |\t\zeta|^2}~,
 \ee
 with the last equality being a property of the half-BPS backgrounds under consideration \cite{Closset:2012ru}. We also define the three-dimensional tensor:
 \be
{ \Phi^\mu}_\nu = - \epsilon^{\mu}_{\phantom{\mu}\nu\rho}\eta^\rho~.
 \ee
 These objects define a metric-compatible THF, which is obtained from the complex structure on $\CM_4$ by dimensional reduction. In the conventions of Appendix~\ref{Appendix: 3d}, we have:
 \be
 \eta_\mu = {J^{0}}_\mu= \h e^0_\nu {J^\nu}_\mu~, \qquad { \Phi^\mu}_\nu= { J^\mu}_\nu\big|_{\rm 3d}~.
 \ee
 The one-form $\eta= \eta_\mu dx^\mu$ has unit norm and defines the foliation. The THF-adapted metric on $\CM_3$ takes the canonical form:
 \be
 ds^2(\CM_3) = \eta^2 + 2 g_{z\bz} dz d\bz~,
 \ee
 with $z, \bz$ the transverse holomorphic coordinates  \cite{Closset:2012ru}.
 
 \paragraph{Dimensional reduction of the background gauge field.} The 4d $\CN=1$ vector multiplet reduces to a 3d $\CN=2$ vector multiplet:%
 \footnote{The scalar $D$ is defined with a shift with respect to its 4d counterpart, $D_{\rm 3d}= D_{\rm 4d} - \sigma H$ \protect\cite{Closset:2012ru}.}
 \be
 \CV_F^{(\rm 3d)}= (A_\mu~, \, \sigma~, \, \lambda_\alpha~, \, \t\lambda_\alpha~, \, D)~.
 \ee
 The three-dimensional gauge field $A_\mu$ and scalar $\sigma$ are related to the four-dimensional gauge field $a_\mu$ by:
 \be\label{amu 4d to 3d}
 a_\mu dX^\mu = \sigma \h e^0+ A_\mu dx^\mu~.
 \ee
 The three-dimensional supersymmetry transformations on $\CM_3$ are given by:
\bea\label{susy V 3d i}
 & \delta_{\zeta} A_\mu = - i \zeta \gamma_\mu \t \lambda~,\cr
& \delta_{\zeta} \sigma = - \zeta \t \lambda~,\cr
& \delta_{\zeta} \lambda_\alpha = i \zeta_\alpha \left(D + \sigma \bH\right)
			  -i  (\gamma^\mu \zeta)_\alpha \Big(\half \eps_{\mu\nu\rho} F^{\nu\rho}+  \d_\mu \sigma + i \bV_\mu \sigma\Big)~,\cr
&\delta_{\zeta} \t \lambda_\alpha =0~,\cr
&\delta_{\zeta} D = \nabla_\mu \big(\zeta \gamma^\mu \t \lambda\big)  - i \bV_\mu \zeta \gamma^\mu \t \lambda-  \bH \zeta \t \lambda~,
\eea
and:
\bea\label{susy V 3d ii}
 & \delta_{\t\zeta} A_\mu = - i  \t \zeta \gamma_\mu \lambda~,\cr
& \delta_{\t\zeta} \sigma =  \t \zeta \lambda~,\cr
& \delta_{\t\zeta} \lambda_\alpha = 0~,\cr
&\delta_{\t\zeta} \t \lambda_\alpha = - i \t \zeta \left(D + \sigma \bH\right) 
			 -i  (\gamma^\mu \t\zeta)_\alpha \Big(\half \eps_{\mu\nu\rho} F^{\nu\rho} -  \d_\mu \sigma + i \bV_\mu \sigma\Big)~,\cr
&\delta_{\t\zeta} D = -\nabla_\mu \big(\t \zeta \gamma^\mu \lambda \big)    - i \bV_\mu  \t \zeta \gamma^\mu \lambda +  \bH \t \zeta \lambda~,
\eea
with $F_{\mu\nu}$ the field strength of $A_\mu$. Here, we again focus on the case of an abelian vector multiplet, for simplicity. On the supersymmetric locus, we have:
\be\label{halfBPS vec 3d}
D= - \sigma \bH+ \half \eps^{\mu\nu\rho} \eta_\mu F_{\nu\rho}+ i \eta^\mu \bV_\mu \sigma~,\qquad \qquad \d_\mu \sigma=0~.
\ee
Note also that the combination $\eta^\mu A_\mu + i \sigma$ is invariant under supersymmetry.

\subsubsection{'t Hooft anomalies and three-dimensional local functional}  \label{subsec: introduce W3d}
Let us now consider the reduction of the 4d anomalies to three dimensions. Since the 3d effective action should be local, in the small-$\beta_{S^1}$ expansion, the 3d reduction of the anomalies should be the variation of a 3d local term of order $\beta_{S^1}$, which we denoted by $\bW_{\rm 3d}$ in \eqref{def W3d}.

\paragraph{Reduction of the $U(1)$ anomaly.}
Consider first the $U(1)$ 't Hooft anomaly \eqref{Anomaly U1 4d}. Its three-dimensional reduction is given by:
\bea\label{red of Aqqq to 3d}
&\delta_\alpha W \big|_{{\rm 3d}, \, q^3} &=&\;- {i\CA_{qqq} \ov 12 \pi} \beta_{S^1} \int d^3 x \sqrt{g} \alpha \Big(-\epsilon^{\mu\nu\rho}\d_\mu \sigma F_{\nu\rho}- 2 i \bV^\mu  \sigma \d_\mu \sigma\Big)\cr
&&=&\;-{i\CA_{qqq} \beta_{S^1} \ov 12 \pi} \int d^3 x \sqrt{g}\, \d_\mu \alpha \left(  \sigma\epsilon^{\mu\nu\rho} F_{\nu\rho}+ i \bV^\mu \sigma^2 \right)~,
\eea
for the cubic anomaly, and:
\be\label{red of Aq to 3d}
\delta_\alpha W \big|_{{\rm 3d}, \, q} ={i\CA_q\beta_{S^1}\ov 192 \pi} \int d^3 x\sqrt{g} \,\d_\mu \alpha \, {\bf N}^\mu~,
\ee
for the mixed $U(1)$-gravitational anomaly, formally.~\footnote{Here, we defined the vector:
\be\nn
{\bf N}^\mu = 4 i {\bf G}^{\mu\nu} \bV_\nu - 2 i \bV^\mu \bV^\nu \bV_\nu~, \qquad\qquad  {\bf G}_{\mu\nu} \equiv {\bf R}_{\mu\nu} - \half g_{\mu\nu} {\bf R}~,
\ee
with $ {\bf G}_{\mu\nu}$ the Einstein tensor of the three-dimensional metric, for $\CM_4$ a circle bundle over $\CM_3$, with $\bV^\mu$ defined as in \protect\eqref{bV def}. Then, 
the Pontryagin density of the 4d metric \protect\eqref{dsM4 KK} can always be written as
$\CP = - \nabla_\mu {\bf N}^\mu$. Whenever the $S^1$ is trivially fibered, $\CP=0$.
 } 
On any four-manifold of the form $\CM_3 \times S^1$, the  Pontryagin density vanishes, and therefore so does \eqref{red of Aq to 3d}, namely:
\be
\delta_\alpha W \big|_{{\rm 3d}, \, q} =0~.
\ee

\paragraph{Reduction of the supersymmetry variations.}
Similarly, the three-dimensional reduction of $\delta_\zeta W$ and $\delta_{\t\zeta} W$ is given by:
\bea
&\delta_\zeta W \big|_{{\rm 3d}, \, q^3} &=&\; {\CA_{qqq} \beta \ov 12 \pi}\int d^3 x \sqrt{g}  \Big\{ \epsilon^{\mu\nu\rho} \left(i \zeta\t\lambda \, A_\mu F_{\nu\rho} +2 \zeta\gamma_\mu\t\lambda\, A_\nu \d_\rho \sigma  \right)   \cr
&&& \;- 2 \zeta\t\lambda \,\sigma A_\mu \bV^\mu +\epsilon^{\mu\nu\rho}  \zeta\gamma_\mu\t\lambda\,  \sigma F_{\nu\rho} +2 i  \zeta\gamma_\mu\t\lambda\,  \sigma^2 \bV^\mu - 3 i \zeta \lambda \t\lambda\t\lambda \Big\}~,\cr
&\delta_{\t\zeta} W \big|_{{\rm 3d}, \, q^3} &=&\; {\CA_{qqq} \beta \ov 12 \pi}  \int d^3 x \sqrt{g}  \Big\{ \epsilon^{\mu\nu\rho} \left(-i \t\zeta\lambda \, A_\mu F_{\nu\rho} +2 \t\zeta\gamma_\mu\lambda\, A_\nu \d_\rho \sigma  \right)  \cr
&&& \;  +2 \t\zeta\lambda \,\sigma A_\mu \bV^\mu +\epsilon^{\mu\nu\rho}\t\zeta\gamma_\mu\lambda\,  \sigma F_{\nu\rho} +2 i  \t\zeta\gamma_\mu\lambda\,  \sigma^2 \bV^\mu + 3 i\t \zeta \t\lambda \lambda\lambda \Big\}~,
\eea
for the cubic anomaly contribution \eqref{Anomaly SUSY qqq 4d}, and by: 
\be\label{susy grav 3d}
\delta_\zeta W \big|_{{\rm 3d}, \, q}= \delta_{\t\zeta} W \big|_{{\rm 3d}, \, q} =0~,
\ee
 for the mixed $U(1)$-gravitational anomaly contribution, according to \eqref{del W Aq zero}.

\paragraph{Explicit formula for $\bW_{\rm 3d}$.}  As we explained, there should exist a three-dimensional local functional, $\bW_{\rm 3d}$, of the vector multiplet, in the fixed 3d $\CN=2$ supergravity background on $\CM_3$, such that:
\be
\delta_\alpha \bW_{\rm 3d} = \delta_\alpha W \big|_{\rm 3d}~, \qquad
\delta_\zeta \bW_{\rm 3d} =\delta_\zeta W \big|_{\rm 3d}~, \qquad
\delta_{\t\zeta} \bW_{\rm 3d} =\delta_{\t\zeta} W \big|_{\rm 3d}~.
\ee
Up to any three-dimensional action that is gauge-invariant and supersymmetric under \eqref{susy V 3d i}-\eqref{susy V 3d ii}, the functional $\bW_{\rm 3d}$ must take the form:%
\footnote{Here we take only into account the contribution from the $U(1)$ flavor anomaly. There can be additional contributions from mixed $U(1)$-$U(1)_R$ anomalies, which we will consider in section~\ref{subsec: U1R contrib 4d} below.}
\be
\bW_{\rm 3d}=\bW_{\rm 3d}|_{q^3} + \bW_{\rm 3d}|_{q}~.
\ee
The mixed $U(1)$-gravitational anomaly contribution must vanish, $\bW_{\rm 3d}|_{q}=0$.
By a straightforward computation, one can check that there is a unique possibility for the cubic-anomaly contribution, given by:
\bea\label{W3d full Aqqq}
&\bW_{\rm 3d}|_{q^3} =\; {\CA_{qqq}  \beta_{S^1}\ov 12 \pi} \int d^3 x \sqrt{g} \hspace{-0.3cm}&&\Big(-i  \sigma \epsilon^{\mu\nu\rho} A_\mu F_{\nu\rho} + 3  \sigma^2 D - 6 i  \sigma \t\lambda\lambda \cr
&&&\;\; +\sigma^2 A_\mu \bV^\mu +\sigma^3 \bH\Big)~.
\eea
The first line in \eqref{W3d full Aqqq} is the flat-space answer, while the second line adds additional couplings to the curved-space supersymmetric background.

%%%%%%%%%%%%%%
\subsection{Generalization to any abelian flavor symmetry}
The above discussion can be generalized straightfowardly to any abelian flavor group:
\be\label{max torus 4d}
G_F =\prod_{\alpha} U(1)_\alpha~.
\ee
(For a non-abelian $G_F$, we focus on the maximal torus, which is what is generally done when discussing supersymmetric partition functions.) The $G_F$ anomaly can be obtained from the anomaly polynomial:
\be
P_6= {1\ov 24 \pi^2} \sum_{\alpha, \beta, \gamma} \CA_{qqq}^{\alpha\beta\gamma} f_\alpha \wedge f_\beta \wedge f_\gamma- {1\ov 384 \pi^2}\sum_\alpha \CA^\alpha f_\alpha \wedge \tr( R \wedge R)~,
\ee
by the descent relations, $P_6 = dQ_5$ and $\delta_{\boldsymbol{\alpha}} Q_5 = d(\boldsymbol{\alpha}\omega_4)$. 
Then, the cubic 't Hooft anomaly reads:
\be\label{cubic anom 4d gen}
\delta_{\boldsymbol{\alpha}} W\big|_{q^3} =\sum_{\alpha, \beta, \gamma} {i \CA_{qqq}^{\alpha\beta\gamma} \ov 24\pi^2} \int_{\CM_4} \, \boldsymbol{\alpha}_\alpha \, f_{\beta}\wedge f_\gamma~,
\ee
with $\boldsymbol{\alpha}$ denoting an arbitrary abelian gauge parameter along \eqref{max torus 4d}, and with $f_\alpha = da_{\alpha}$ for the background gauge fields $a_{\alpha, \mu}$. The sum over repeated flavor indices will be left implicit, in the following. Note that the 't Hooft anomaly coefficient $\CA^{\alpha\beta\gamma}_{qqq}$ is fully symmetric in the three flavor indices $\alpha, \beta, \gamma$. 

The expression \eqref{cubic anom 4d gen} is valid up to ambiguities that correspond to the gauge-variations of local terms---those, in turn, correspond to ambiguities in the descent relations. For instance, given two abelian symmetries $U(1)_\alpha$ and $U(1)_\beta$ with $\alpha\neq \beta$, we could always add the local term:~\footnote{In this equation, there is no sum over repeated flavor indices.} 
\be
W_{\rm loc} = {i \ov 24 \pi^2} \int_{\CM_4} \left( s \CA_{qqq}^{\alpha\alpha\beta} a_\beta \wedge a_\alpha \wedge f_\alpha + t \CA_{qqq}^{\alpha\beta\beta}  a_\alpha \wedge a_\beta \wedge f_\beta \right)~,
\ee
with $s, t \in \R$ some arbitrary coefficients. The addition of such counterterms to the effective action can ``move around'' the mixed anomalies; for instance, setting $s=-2$ and $t=1$ would ensure that the variation $\delta_{\boldsymbol{\alpha}_\alpha} W$ only depends on the gauge field $a_\alpha$, at the price of pushing all the mixed anomalies into $\delta_{\boldsymbol{\alpha}_\beta} W$. Here and in the following, we always work in the ``symmetric'' scheme which treats all the $U(1)_\alpha$ factors equally, in which case \eqref{cubic anom 4d gen} holds. 

The supersymmetry variations \eqref{Anomaly SUSY qqq 4d} are then generalized similarly, in the fully symmetric scheme:%
\footnote{Here, the indices $\alpha, \beta, \cdots$ are flavor indices, while the spinor indices are left implicit.}
\bea\label{Anomaly SUSY qqq 4d gen}
&\delta_\zeta W|_{q^3}= -{ \CA_{qqq}^{\alpha\beta\gamma}  \ov 24\pi^2} \int d^4x \sqrt{g} \, \left(\epsilon^{\mu\nu\rho\sigma} \zeta \sigma_\mu \t\lambda_\alpha \, a_{\beta,\nu} f_{\gamma,\rho\sigma} -  3i \zeta \lambda_\alphaÂ \, \t\lambda_\beta\t\lambda_\gamma \right)~,\cr
&\delta_{\t\zeta} W|_{q^3}= -{ \CA_{qqq}^{\alpha\beta\gamma}  \ov 24\pi^2} \int d^4x \sqrt{g} \, \left(\epsilon^{\mu\nu\rho\sigma} \t\zeta\t \sigma_\mu \lambda_\alpha \, a_{\beta,\nu} f_{\gamma,\rho\sigma} + 3 i\t \zeta \t\lambda_\alphaÂ \, \lambda_\beta\lambda_\gamma \right)~,
\eea
The 3d local term that reproduces the 3d reduction of the anomalies \eqref{cubic anom 4d gen} and \eqref{Anomaly SUSY qqq 4d gen} takes the form:
\bea\label{W3d full Aqqq gen}
&\bW_{\rm 3d}|_{q^3} =\; {\CA_{qqq}^{\alpha\beta\gamma}  \beta_{S^1}\ov 12 \pi} \int d^3 x \sqrt{g} \hspace{-0.3cm}&&\Big(-i   \epsilon^{\mu\nu\rho} \sigma_\alpha A_{\beta,\mu} F_{\gamma,\nu\rho} + 3  \sigma_\alpha \sigma_\beta D_\gamma  \cr
&&& +\sigma_\alpha \sigma_\beta A_{\gamma,\mu} \bV^\mu +\sigma_\alpha\sigma_\beta\sigma_\gamma \bH  - 6 i  \sigma_\alpha \t\lambda_\beta\lambda_\gamma\Big)~,
\eea
generalizing \eqref{W3d full Aqqq}.

\subsection{R-symmetry contributions to $\delta_\zeta W$}\label{subsec: U1R contrib 4d}
Finally, we should discuss the effect of the $R$-symmetry background gauge field. Consider again the case $G_F= U(1)$, to avoid clutter. The anomalous $U(1)$ variation of the effective action has contributions from mixed 't Hooft anomalies $U(1)^2$-$U(1)_R$ and $U(1)$-$U(1)_R^2$, which read:
\be\label{anomaly R 4d}
\delta_\alpha W\big|_{R} =  {i \CA_{qqR} \ov 48\pi^2} \int d^4x \sqrt{g} \, \alpha \epsilon^{\mu\nu\rho\sigma} f_{\mu\nu} f_{\rho\sigma}^{(R)}+ {i \CA_{qRR} \ov 96\pi^2} \int d^4x \sqrt{g} \, \alpha \epsilon^{\mu\nu\rho\sigma} f_{\mu\nu}^{(R)} f_{\rho\sigma}^{(R)}~.
\ee
This is the contribution that was denoted by an ellipsis in \eqref{Anomaly U1 4d}.
Here, $f^{(R)}=d a^{(R)}$ is the field strength of the $U(1)_R$ gauge field:
\be\label{amu R def}
a_\mu^{(R)} \equiv A_\mu^{(R)}+ {3\ov 2 } V_\mu~.
\ee
The second term in \eqref{anomaly R 4d} vanishes on the half-BPS background, because:
\be
\epsilon^{\mu\nu\rho\sigma} f_{\mu\nu}^{(R)} f_{\rho\sigma}^{(R)}= 0~,
\ee
in that case. The mixed 't Hooft anomalies \eqref{anomaly R 4d} then contribute to right-hand-side of the supersymmetry variation of $W$ as:
\bea\label{Anomaly SUSY qR 4d}
&\delta_\zeta W|_{R}&=&\; -{ \CA_{qqR} \ov 24\pi^2} \int d^4x \sqrt{g} \, \epsilon^{\mu\nu\rho\sigma} \zeta \sigma_\mu \t\lambda \left(a_\nu f_{\rho\sigma}^{(R)} + a_\nu^{(R)} f_{\rho\sigma}\right)\cr
&&&\; -{ \CA_{qRR} \ov 24\pi^2} \int d^4x \sqrt{g} \, \epsilon^{\mu\nu\rho\sigma} \zeta \sigma_\mu \t\lambda \, a_\nu^{(R)} f_{\rho\sigma}^{(R)}~,
\eea
and similarly for $\delta_{\t\zeta} W$. 
Note again that,  in this paper, we focus on the flavor anomaly on a fixed half-BPS background. The more general form of the mixed anomalies on arbitrary supergravity backgrounds, including the contributions from the gravitino, were recently discussed in \cite{Papadimitriou:2019yug}.

\paragraph{4d $U(1)_R$ vector multiplet.}  An important property of the 4d $\CN=1$ supergravity multiplet is that it contains an ordinary {\it vector multiplet for the $R$-symmetry,} $\CV_R$, as a sub-multiplet of supersymmetry \cite{Sohnius:1982fw}. The gaugino $\lambda^{(R)}$ inside $\CV_R$ is a contraction of the gravitino $\Psi_\mu$. On the half-BPS background $\CM_4$, with $\Psi_\mu=\delta_\zeta \Psi_\mu= \delta_{\b\zeta}\Psi_\mu=0$, we are left with the bosonic fields:
\be
\CV_R= \left( a_\mu^{(R)}~, \, D^{(R)}_{\rm 4d}\right)~,
\ee
with $a_\mu^{(R)}$ given in terms of the supergravity fields by \eqref{amu R def}, and:
\be
D^{(R)}_{\rm 4d}= {1\ov 4} \left(R + 6 V_\mu V^\mu\right)~,
\ee
with $R$ the four-dimensional Ricci scalar. One can verify that these combinations of background fields satisfy the conditions \eqref{susy halfBPS 4d V} for an half-BPS vector multiplet, namely:
\be
f_{w\bw}^{(R)}= f_{\b w \b z}^{(R)}= f_{\b w z}^{(R)}=0~, \qquad
D^{(R)}_{\rm 4d}= -\half J^{\mu\nu} f_{\mu\nu}^{(R)}~.
\ee
Note that the background gauge field $a_\mu^{(R)}$ is not real, unlike the background flavor gauge fields that we consider in this paper; supersymmetry generally imposes that $a_\mu^{(R)}$ be complex.

The $U(1)_R$ vector multiplet controls the $R$-charge dependence of supersymmetric partition functions  \cite{Closset:2014uda}. Consider any mixing of the $R$-charge with abelian flavor symmetries $U(1)_\alpha$:
\be
R\rightarrow R+\sum_\alpha t_\alpha\, Q_{F}^\alpha~,
\ee
with $t_\alpha \in \R$ some mixing parameters. The effect of this shift on the supersymmetric partition function, computed with background vector multiplets $\CV_{F, \alpha}$, is simply to shift the latter by the $U(1)_R$ vector multiplet itself:
\be\label{shift F to FR}
\CV_{F, \alpha} \rightarrow \CV_{F, \alpha}+ t_\alpha \CV_R~.
\ee
This allows us to treat the $R$-charge dependence systematically once we understand the ``flavor sector.''  For instance, as a typical example of this approach, let us consider a free chiral multiplet. The mixed flavor-$R$ 't Hooft anomaly coefficients for a chiral multiplet of $U(1)$ charge $q$ and $R$-charge $r$ are:
\be
\CA_{qqR}= q^2 (r-1)~, \qquad \CA_{qRR}= q (r-1)^2~,
\ee
and therefore we can ignore the effect of those anomalies if we choose $r=1$. Then, the general case can be obtained by starting from the ``reference'' $R$-charge $R_0$ with $r=1$ and shifting the partition function of the free chiral to $R= R_0 +(r-1) Q$ using \eqref{shift F to FR}.

Similarly, the anomalous variations \eqref{anomaly R 4d}-\eqref{Anomaly SUSY qR 4d} can be obtained from \eqref{cubic anom 4d gen} and \eqref{Anomaly SUSY qqq 4d gen}, simply by replacing one flavor index ``$\alpha$'' by ``$R$.'' This again corresponds to a fully symmetric scheme, treating $a_\mu^{(R)}$ and $a_{\alpha, \mu}$ on the same footing. 

\paragraph{Dimensional reduction and 3d $U(1)_R$ vector multiplet.} Finally, let us briefly discuss the 3d reduction of the $U(1)_R$ vector multiplet:
\be
\CV_R^{(\rm 3d)} = (\CA_\mu^{(R)}~,\, \sigma^{(R)}~, \, D_{\rm 3d}^{(R)})~, 
\ee
 Using \eqref{V and A 4d to 3d} and \eqref{amu 4d to 3d}, one can check that  \cite{Closset:2014uda}:
 \be
 \CA_\mu^{(R)} = \bA_\mu + \bV_\mu~, \qquad  \sigma^{(R)}= \bH~, \qquad D_{\rm 3d}^{(R)}= {1\ov 4}({\bf R}+ 2 \bH^2 + 2 \bV_\mu \bV^\mu)~, 
 \ee
where ${\bf R}$ is the 3d Ricci scalar, which is related to the 4d Ricci by $R= {\bf R}+ \half \bV_\mu \bV^\mu$. On any half-BPS background $\CM_3$, we have the analogue of the condition \eqref{halfBPS vec 3d} for a vector multiplet on its supersymmetric locus, namely:
\be
\epsilon^{\mu\nu\rho} \eta_\mu \d_\nu (\bA_\rho + \bV_\rho) + i \eta_\mu \bV^\mu \bH= {1\ov 4} {\bf R} + {3\ov 2 } \bH^2 + \half \bV_\mu \bV^\mu~, \quad \quad \d_\mu \bH=0~.
\ee
One can then find 3d local terms that reproduce the mixed $U(1)$-$U(1)_R$ 't Hooft anomalies and are consistent with supersymmetry, by an appropriate generalization of \eqref{W3d full Aqqq}.

\subsection{Anomalous supersymmetry variation from WZ gauge-fixing}\label{subsec: susy an from wz 4d}
As in 2d, we can understand the 4d ``anomalous supersymmetry variation,'' $\delta_\zeta W \neq 0$, as a direct consequence of the Wess-Zumino gauge-fixing procedure. This provides a conceptually and  technically  simpler derivation of those anomalous variations, and demonstrates that they are simple consequences of the ordinary 't Hooft anomalies, when working in the WZ gauge.

For simplicity of presentation, in this subsection, let us work in flat-space, $\R^4$, so that we can use the 4d $\CN=1$ superfield notation. (The generalization to rigid supersymmetry on $\CM_4$ is completely straigthforward.) 

\paragraph{WZ gauge-fixing.} Let us consider a theory coupled to a $U(1)$ background vector multiplet,
\be
\CV_F=\left( C~, \, \chi~, \, \t \chi~, \, M~, \, \t M~, \, a_\mu~, \, \lambda~, \, \t\lambda~, \, D\right)~.
\ee
This corresponds to the real superfield:
\bea\label{CS expans V}
&\CV_F= C + i \theta\chi - i \t\theta\t\chi + {i\ov2} \theta\theta M - {i\ov2} \t\theta\t\theta \t M - \theta \sigma^\mu \t\theta a_\mu(x) \cr
&\quad+ i \theta\theta \t\theta \big(\t\lambda+ {i\ov 2 }\t\sigma^\mu\d_\mu \chi\big)
 - i \t\theta\t\theta \theta \big(\lambda +{i\ov 2 }\sigma^\mu\d_\mu \t\chi \big) 
+ \half \theta\theta \t\theta\t\theta \, \big(D+ \half \d_\mu \d^\mu C\big)~.
\eea
The gauge freedom of the gauge field, $\delta_\alpha a_\mu = \d_\mu \alpha$, is supersymmetrized to:
\be\label{delta Omega def}
\delta_\Omega \CV_F=  {i\ov2 }\big(\Omega - \t \Omega \big)~,
\ee
 with the gauge parameters valued in a pair of chiral and anti-chiral multiplets, $\Omega$ and $\t \Omega$---in terms of superfields, we have:
\bea
&\Omega = \omega + \sqrt{2} \theta \psi^\Omega + \theta\theta F^\Omega + \cdots~, \qquad
&\t\Omega=\t \omega + \sqrt{2} \t\theta \t\psi^\Omega + \t\theta\t\theta \t F^\Omega + \cdots~.
\eea
 Let us define the gauge parameters $\alpha$ and $\gamma$ by:
\be
\omega = \alpha + i \gamma~, \qquad\qquad \t\omega = \alpha - i \gamma~.
\ee
In components,  \eqref{delta Omega def} is equivalent to:
\bea\label{deltaOmega compo}
&\delta_\Omega C= -\gamma~, \qquad &&\delta_\Omega \chi = {1\ov \sqrt2} \psi^\Omega~, \qquad 
&& \delta_\Omega\t\chi = {1\ov \sqrt{2} } \t\psi^\Omega~, \cr
&\delta_\Omega a_\mu= \d_\mu \alpha~, \qquad &&\delta_\Omega  M = F^\Omega~, \qquad 
&& \delta_\Omega\t M =\t F^\Omega~, 
\eea
while $\lambda, \t\lambda$ and $D$ are gauge-invariant. Using this gauge-freedom, we can fix the so-called WZ gauge \cite{Wess:1974tw}:
\be
C= \chi = \t \chi= M =\t M=0~.
\ee
This leaves $\delta_\alpha \CV_F$ as a residual $U(1)$ gauge transformation.
Obviously, the WZ gauge does not commute with supersymmetry. Nonetheless, one can define a consistent set of supersymmetry transformations by accompanying any ``bare'' supersymmetry transformation, denoted by $\delta_\zeta^{(0)}$ or $\delta_{\t\zeta}^{(0)}$, with a gauge transformation to restore the WZ gauge:
\be
\delta_\zeta = \delta_\zeta^{(0)} + \delta_{\Omega(\zeta)}~, \qquad 
\delta_{\t\zeta} = \delta_{\t\zeta}^{(0)} + \delta_{\Omega(\t\zeta)}~.
\ee
The compensating gauge transformations are given by specializing \eqref{deltaOmega compo} to the field-dependent gauge parameters:
\be\label{omega(zeta)}
\Omega(\zeta)\; \; : \; \; \omega =\psi^\Omega_\alpha= F^\Omega=\t\omega=0~,  \quad \t\psi^\Omega_\alphadot = i \sqrt2 (\zeta \sigma^\mu)_\alphadot\, a_\mu~, \quad \t F^\Omega = - 2 \zeta \lambda~,
\ee
or:
\be\label{omega(t zeta)}
\Omega(\t\zeta)\; \; : \; \; 
\psi^\Omega_\alpha =- i \sqrt2 (\sigma^\mu \t \zeta)_\alpha\, a_\mu~, \quad F^\Omega = - 2 \t\zeta \t\lambda~,\quad
 \omega =\t\omega=\t\psi^\Omega_\alphadot=\t F^\Omega=0~, 
\ee
respectively.~\footnote{In flat space, $\zeta$ and $\t\zeta$ are arbitrary constant spinors. They become our non-trivial Killing spinors $\zeta$ and $\t\zeta$ on $\CM_4$ once we generalize this discussion to curved space.}

\paragraph{'t Hooft anomaly and supersymmetry.}  Let us first consider the cubic 't Hooft anomaly:
\be\label{del W cubic}
\delta_\alpha W= {i \CA_{qqq} \ov 96\pi^2} \int d^4 x \, \alpha \epsilon^{\mu\nu\rho\sigma} f_{\mu\nu} f_{\rho\sigma}~.
\ee
Its supersymmetric completion takes the form \cite{Piguet:1984aa}:
\be\label{susy hooft Aqqq superspace}
\delta_\Omega W = -{i\CA_{qqq} \ov 48\pi^2} \int d^4 x \left[\int d^2 \theta\, \Omega \CW^\alpha \CW_\alpha- \int d^2 \t\theta\, \t\Omega \t\CW_\alphadot \t\CW^\alphadot \right]~,
\ee
possibly up to the $\delta_\Omega$-variation of a local term. Here, we defined the standard field-strength chiral  and anti-chiral multiplets:~\footnote{With the usual definitions: ${\rm D}_\alpha=\d_\alpha + i (\sigma^\mu\t\theta)_\alpha \d_\mu$ and $\t{\rm D}_\alphadot = - \t\d_\alphadot - i (\theta \sigma^\mu)_\alphadot \d_\mu$.}
\be
\CW_\alpha=-{1 \ov4 } \t {\rm D} \t {\rm D}  {\rm D}_\alpha V~, \qquad \t\CW_\alphadot = -{1\ov 4} {\rm D} {\rm D} \t{\rm D}_\alphadot V~.
\ee
Expanding in components, \eqref{susy hooft Aqqq superspace} gives:
 \bea\label{susy hooft Aqqq comp}
&\delta_\Omega W &&=\;{i \CA_{qqq} \ov 48\pi^2} \int d^4 x \Bigg\{  \alpha\left(\half \epsilon^{\mu\nu\rho\sigma} f_{\mu\nu} f_{\rho\sigma} + 2 i \d_\mu\left(\lambda \sigma^\mu \t\lambda\right) \right)\cr
&&&+i \gamma \left( f_{\mu\nu}f^{\mu\nu}- 2 D^2 +2 i\lambda \sigma^\mu \d_\mu \t\lambda +2 i \t\lambda \t\sigma^\mu \d_\mu \lambda\right) + F^\Omega \lambda \lambda - \t F^\Omega \t\lambda \t \lambda\cr
&&&+\sqrt2 \left(\psi^\Omega \sigma^{\mu\nu} \lambda - \t \psi^\Omega \t\sigma^{\mu\nu}  \t \lambda\right)f_{\mu\nu} - i \sqrt2 \left(\psi^\Omega \lambda + \t\psi^\Omega \t\lambda \right) D \Bigg\}~.
\eea
Specializing to the WZ gauge, we see that \eqref{susy hooft Aqqq comp} reproduces \eqref{del W cubic} up to the variation of a local term---namely, the cubic 't Hooft anomaly $\delta_\alpha W$ in \eqref{del W cubic} is given by:
\be\label{del alpha W with Wloc}
\delta_\alpha W = \delta_\Omega W\Big|_{\rm WZ} + \delta_\alpha W_{\rm loc}~,
\ee
with the simple local term:
\be\label{Wloc 4d def for thooft an}
 W_{\rm loc} \equiv - {\CA_{qqq} \ov 24 \pi^2} \int d^4 x\, a_\mu \lambda \sigma^\mu \t\lambda~.
\ee
The gauge variation of \eqref{Wloc 4d def for thooft an} in \eqref{del alpha W with Wloc} cancels the second term on the first line of \eqref{susy hooft Aqqq comp}. 

\paragraph{Anomalous supersymmetry variation in WZ gauge.} Before fixing the Wess-Zumino gauge, the ``bare'' supersymmetry transformations are non-anomalous \cite{Piguet:1980fa}:
\be\label{susy nonanom 4d}
\delta_\zeta^{(0)} W=0~, \qquad\qquad \delta_{\t\zeta}^{(0)} W=0~.
\ee
This is the statement that, the one hand, there is no genuine supersymmetry anomaly; and, on the other hand, the 't Hooft anomaly \eqref{susy hooft Aqqq superspace} does not contribute any quantum correction to \eqref{susy nonanom 4d}. The latter statement directly follows from the WZ consistency condition:
\be
[\delta_\zeta^{(0)}, \delta_\Omega]W= 0~,
\ee
and the fact that  \eqref{susy hooft Aqqq superspace} is supersymmetric, $\delta_\zeta^{(0)} \delta_\Omega W=0$. The WZ gauge fixing modifies the supersymmetry algebra, introducing an explicit dependence on the gauge fields due to \eqref{omega(zeta)}-\eqref{omega(t zeta)}. The compensating gauge transformation also introduces non-vanishing WZ-gauge supersymmetry variations, since:
\bea\label{delta zeta W from wz}
&\delta_\zeta W= \Big(\delta_\zeta^{(0)} + \delta_{\Omega(\zeta)}\Big)W= \delta_{\Omega(\zeta)} W~, \cr
&\delta_{\t\zeta} W= \Big(\delta_{\t\zeta}^{(0)} + \delta_{\Omega(\t\zeta)}\Big)W= \delta_{\Omega(\t\zeta)}W~,
\eea
and $\delta_\Omega W \neq 0$ due to the $U(1)$ 't Hooft anomalies. 
This gives an alternative way to compute the anomalous variations $\delta_\zeta W$ in WZ gauge. To compare to the results from the previous subsections, we should take into account the fact that we are working in a scheme in which the cubic anomaly is given by \eqref{del W cubic} exactly. This corresponds to \eqref{del alpha W with Wloc}, including the contribution from the local term \eqref{Wloc 4d def for thooft an}. We thus have:
\bea\label{del zeta from gauge fixing}
\delta_\zeta W =  \delta_{\Omega(\zeta)} W+ \delta_\zeta W_{\rm loc}~, \cr
\delta_{\t\zeta} W =  \delta_{\Omega(\t\zeta)} W+ \delta_{\t\zeta} W_{\rm loc}~.
\eea 
Here, the gauge variation \eqref{susy hooft Aqqq comp} is specialized to $\Omega= \Omega(\zeta)$ or $\Omega= \Omega(\t\zeta)$, and the second term is the (WZ-gauge) supersymmetry variation of \eqref{Wloc 4d def for thooft an}. One finds:
\bea\label{label del Omega of zeta}
&\delta_{\Omega(\zeta)} W= {\CA_{qqq} \ov 24 \pi^2} \int d^4 x\, \left(\zeta \sigma^\mu \t\sigma^{\nu\rho} \t\lambda a_\mu f_{\nu\rho} + i \zeta\sigma^\mu \t\lambda a_\mu D + i  \zeta \lambda \t\lambda\t\lambda \right)~, \cr
&\delta_{\Omega(\t\zeta)} W= {\CA_{qqq} \ov 24 \pi^2} \int d^4 x\, \left(-\t\zeta \t\sigma^\mu \sigma^{\nu\rho} \lambda a_\mu f_{\nu\rho} + i\t\zeta\t\sigma^\mu \lambda a_\mu D - i  \t\zeta \t\lambda \lambda\lambda \right)~, 
\eea
and:
\bea
&\delta_\zeta W_{\rm loc}= {\CA_{qqq} \ov 24 \pi^2} \int d^4 x\, \left(\zeta \sigma^{\nu\rho} \sigma^\mu\t\lambda a_\mu f_{\nu\rho} - i \zeta\sigma^\mu \t\lambda a_\mu D +2 i  \zeta \lambda \t\lambda\t\lambda \right)~, \cr
&\delta_{\t\zeta} W_{\rm loc}= {\CA_{qqq} \ov 24 \pi^2} \int d^4 x\, \left(-\t\zeta \t\sigma^{\nu\rho}\t\sigma^\mu  \lambda a_\mu f_{\nu\rho} - i\t\zeta\t\sigma^\mu \lambda a_\mu D -2 i  \t\zeta \t\lambda \lambda\lambda \right)~.
\eea
It follows that \eqref{del zeta from gauge fixing} exactly reproduces the result \eqref{Anomaly SUSY qqq 4d}, which was obtained by solving the WZ consistency conditions for supersymmetry and gauge transformations directly in WZ gauge.

Note also that, even if we worked in WZ gauge in the scheme defined by \eqref{susy hooft Aqqq comp} with $\Omega=\t \Omega= \alpha$, in which case the anomalous supersymmetry variation takes the more cumbersome form \eqref{label del Omega of zeta}, we would still obtain the same results for supersymmetric partition functions, because the local term \eqref{Wloc 4d def for thooft an} vanishes on the supersymmetric locus for the background vector multiplet.

One can similarly discuss the mixed $U(1)$-gravitational anomaly in this language, generalized to supergravity. The supersymmetrization of the anomaly is similar to \eqref{susy hooft Aqqq superspace}, with $\CW_\alpha$ replaced by a chiral multiplet, $\CT_{\mu\nu\alpha}$, that contains the Riemann tensor in its $\theta$-component \cite{Assel:2014tba}. The bottom and $\theta\theta$ components of $\CT_{\mu\nu\alpha}$ vanish once we set the gravitino to zero and therefore, using \eqref{delta zeta W from wz}, we find that the mixed $U(1)$-gravitational anomaly does not contribute to $\delta_\zeta W$ on a rigid-supersymmetric background.

%%%%%%%%%%%%%%%%%%%%%%%%%%%%%%%%%%%%%%%%%%%%%%%%%%%%%%%%%
\section{4d $\CN=1$ supersymmetric partition functions, revisited}\label{sec: MgpS1}
%%%%%%%%%%%%%%%%%%%%%%%%%%%%%%%%%%%%%%%%%%%%%%%%%%%%%%%%%
In this section, as an application of the above formalism, we study how the supersymmetric Ward identities constrain supersymmetric partition functions on some specific half-BPS backgrounds. This leads us to discuss the explicit form of the partition functions $Z(\nu, \tau)$ on $\CM_3 \times S^1$, which differ from the well-known holomorphic results, $\CI(\nu, \tau)$, in the Casimir-like prefactor.

\subsection{The $\CM_{g,p} \times S^1$ background}
We will focus on the special case of $\CM_4$ a principal elliptic fiber bundle over a closed genus-$g$ Riemann surface:
\be
T^2  \longrightarrow \CM_4 \stackrel{\pi}\longrightarrow \Sigma_g~.
\ee
In that case, the four-manifold necessarily takes the form:
\be
\CM_4 \cong \CM_{g, p} \times S^1~, \qquad \qquad S^1  \stackrel{p}\longrightarrow \CM_{g,p}  \stackrel{\pi}\longrightarrow \Sigma_g~,
\ee
where the three-manifold $\CM_{g,p}$ is a degree-$p$ principal circle bundle over $\Sigma_g$. As a special case, this includes $S^3\times S^1$ (for $p=1, g=0$) and $\Sigma_g \times T^2$ (for $p=0$). The corresponding supersymmetric backgrounds were studied  {\it e.g.} in \cite{Festuccia:2011ws, Dumitrescu:2012ha,Klare:2012gn, Assel:2014paa, Nishioka:2014zpa, Closset:2017bse}.%
\footnote{We closely follow the discussion in \protect\cite{Closset:2017bse}, with only  slightly different conventions; in particular, the sign of the complex structure here differs to the one in \protect\cite{Closset:2017bse} by a sign.}
The metric on $\CM_{g, p}\times S^1$ is chosen to be:
\be\label{MgpS1 metric psi t}
ds^2(\Mgp\times S^1) =  \beta_1^2 \big(d\psi + \tau_1 dt + \CC(z, \bz)\big)^2 +\beta_2^2 dt^2+ 2 g_{z\bz}\, dz d\bz~,
\ee
with the angular coordinates $t\in [0,2\pi)$ and $\psi\in [0, 2\pi)$, with $t$ the coordinate along the circle direction. In \eqref{MgpS1 metric psi t}, $g_{z\bz}$ is the Hermitian metric on $\Sigma_g$, which we normalize such that ${\rm vol}(\Sigma_g)=\pi$;  the one-form $\CC$ is the connection of a principal circle bundle over $\Sigma$, with first Chern class $p$, which satisfies:
\be
\d_z \CC_\bz- \d_\bz \CC_z =  p\, 2i g_{z\bz}~, \qquad \quad \CC= \pi^\ast(\CC_\Sigma)~, \qquad\quad {1\ov 2 \pi} \int_{\Sigma_g} d \CC_\Sigma = p \in \Z~,
\ee
with $\CC_\Sigma$ a two-dimensional gauge field on $\Sigma_g$ which pulls back to $\CC$.
The complex coordinates $w$ along the $T^2$ fiber is given explicitly by:
\be\label{w def gen}
w=  \psi + \tau t + f(z, \bz)~, \qquad \qquad 
\CC_z = \d_z \b f~, \qquad \CC_\bz = \d_\bz f~.
\ee
where the complex function $f(z, \bz)$ is related to the one-form $\CC$ as shown, and we introduced the complex  parameter:
\be
\tau= \tau_1+ i \tau_2~, \qquad\qquad  \tau_2 \equiv {\beta_2\ov \beta_1}~.
\ee
This $\tau$ is a {\it complex structure modulus} of the Hermitian manifold $\CM_4 \cong \CM_{g, p} \times S^1$---of course, it is also the complex structure of the $T^2$ fiber itself; note that it is kept constant over the base, $\Sigma_g$.
The Killing vector $K^\mu$ is given by:
\be
K= {2 \beta_1^{-1}} \d_{\bw} =-i \beta_2^{-1} \big(\tau \d_\psi - \d_t\big)~.
\ee
The metric \eqref{MgpS1 metric psi t} is equivalent to the canonical expression \eqref{M4 metric wz gen}-\eqref{e1 e2 def 4d}, with:
\be\label{real frame MgpS1}
e^{1}\equiv e^{(1)}+ i e^{(2)}~, \qquad e^{(1)}=  \beta_1 \big(d\psi + \tau_1 dt + \CC(z, \bz)\big)~, \qquad e^{(2)}=\beta_2 dt~.
\ee 
and $e^2\equiv  e^{(3)}+ i e^{(4)}$ as given in \eqref{e1 e2 def 4d}. The other new-minimal supergravity background fields, $V_\mu$ and $A_\mu^{(R)}$, can be computed from the general formula \eqref{V and AR 4d explicit} with \eqref{U as kappa}. This gives:
\be\label{V A Mgp 4d}
V_\mu dX^\mu =p \beta_1 e^{(1)}+ \kappa K_\mu dX^\mu~, \qquad A^{(R)}_\mu dX^\mu =-p \beta_1 e^{(1)}+ {i p \beta_1 \ov 2} e^{(2)}+ A_\mu^{c} dX^\mu~, 
\ee
with $A_\mu^c$ defined in \eqref{def Ah and Ac}. The gauge field $A_\mu^c$ is the pull-back of a non-trivial two-dimensional $R$-symmetry gauge field with $1-g$ units of flux on $\Sigma_g$:
\be\label{def Ac Sigmag}
A^c= \pi^\ast(A_\Sigma)~, \qquad \qquad {1\ov 2\pi} \int_{\Sigma_g} dA_\Sigma=1- g~.
\ee
The free parameter $\kappa$ in \eqref{V A Mgp 4d} can be fixed by requiring the existence of a consistent dimensional reduction of this 4d $\CN=1$ supergravity background to a corresponding three-dimensional $\CN=2$ supergravity background on $\CM_{g,p}$  \cite{Closset:2012ru}. This imposes:
\be\label{kappa 4d fixed}
\kappa=p \beta_1 {\b \tau\ov \tau}~.
\ee
This is an important point for the computations that follow, since only with this particular choice, \eqref{kappa 4d fixed}, do we have a consistent reduction of the 4d $\CN=1$ half-BPS background to a 3d $\CN=2$ half-BPS background on the three-manifold $\CM_{g,p}$.

\paragraph{Kaluza-Klein form of the metric.} In order to discuss the 3d reduction, it is useful to work with a slightly different frame, obtained from \eqref{real frame MgpS1} by an $SO(2)$ rotation:
\be\label{rot frame 4d}
\mat{\eta \\ \h e^{0} } \equiv {1\ov |\tau|} \mat{\tau_2 & - \tau_1 \\ \tau_1 & \tau_2} \mat{e^{(1)}\\ e^{(2)}}~.
\ee
Let us define the radii:
\be
\beta_{S^1} \equiv \beta_1 |\tau|~, \qquad \qquad\quad\beta_{\rm 3d} \equiv {\beta_2 \ov |\tau|}~.
\ee
In term of these, \eqref{rot frame 4d} Â gives us the one-forms:
\be\label{def eta e0h}
\eta = \beta_{\rm 3d} (d\psi + \CC)~, \qquad \qquad \quad    \h e^{0}= \beta_{S^1} dt + c_\mu dx^\mu~,
\ee
with $c_\mu$, the so-called graviphoton, given by:
\be
c_\mu= {\tau_1\ov \tau_2} \eta_\mu~.
\ee
This brings the metric into the Kaluza-Klein (KK) form:
\be\label{MgpS1 metric KK}
ds^2(\Mgp\times S^1) =  (\beta_{S^1} dt + c_\mu dx^\mu)^2+ds^2(\CM_{g,p})~.
\ee
Note that, in our conventions, we have the volumes:
\be
{\rm vol}(\CM_{g,p})= 2\pi^2 \beta_{\rm 3d}~, \qquad {\rm vol}(\CM_{g,p}\times S^1)=  4\pi^3\beta_{S^1}  \beta_{\rm 3d}=4\pi^3\beta_1  \beta_2~.
\ee
%%%%

\paragraph{Supersymmetric background on $\CM_{g,p}$.}
In the following, we will also need the explicit form of the corresponding 3d $\CN=2$ supersymmetric background on $\CM_{g,p}$, as discussed in general terms in section~\ref{subsubsec:M3 background}.
The three-dimensional metric on $\CM_{g,p}$ takes the simple form:
\be\label{3d sugra fields 0}
ds^2(\CM_{g,p})=\eta^2+ ds^2(\Sigma_g)~, 
\ee
in the coordinates $(\psi, z, \bz)$ adapted to the THF. Here, $\eta$ is defined as in \eqref{def eta e0h}, and $ds^2(\Sigma_g)= 2 g_{z\bz}\, dz d\bz$ is the metric on $\Sigma_g$. This manifold is also a Seifert manifold,%
\footnote{See \protect\cite{Closset:2018ghr} for a systematic study of 3d $\CN=2$ supersymmetric backgrounds on Seifert  manifolds.}  
with $\psi$ the coordinate on the Seifert fiber of constant radius $\beta_{\rm 3d}$.  The other three-dimensional supergravity fields are given by:
\be\label{3d sugra fields 1}
{\bf H}= -i p \beta_{\rm 3d} +2 p \beta_{\rm 3d} {\tau_1\ov \tau_2}~,\qquad 
{\bf V}_\mu = - 2 i p \beta_{\rm 3d} {\tau_1\ov \tau_2} \eta_\mu~, \qquad
{\bf A}_\mu^{(R)}= - p  \beta_{\rm 3d}  \eta_\mu + A_\mu^c~,
\ee
with $A_\mu^c$ defined as in \eqref{def Ac Sigmag}. The three-dimensional background field ${\bf V}_\mu$ is the dual field strength of a {\it real} graviphoton $c_\mu$, with:%
\footnote{We have the useful relation: $\epsilon^{\mu\nu\rho}\d_\nu \eta_\rho= 2 p \beta_{\rm 3d} \eta^\mu$ in this geometry.}
\be
{\bf V}^\mu = -i \epsilon^{\mu\nu\rho} \d_\nu c_\rho~, \qquad\qquad c_\mu = {\tau_1\ov \tau_2} \eta_\mu~.
\ee
This is as expected for a 3d background obtained by dimensional reduction from 4d, with the KK metric \eqref{MgpS1 metric KK}. Requiring the existence of this supersymmetric reduction to 3d uniquely fixes the  parameter $\kappa$ in the 4d background to \eqref{kappa 4d fixed}.

\subsection{Supersymmetric Ward identity on $\CM_{g,p}\times S^1$}\label{subsec: flavor susy 4d Mgp}

Let us evaluate $\delta_\zeta W$ on the $\CM_{g,p}\times S^1$ supergravity background (focussing on the cubic 't Hooft anomaly). We may write the anomaly in terms of the  gaugino one-forms defined in \eqref{twisted lambdas def}. In the complex frame basis, they are given by:
\be
\t\Lambda = \t\Lambda_1 e^1 +  \t\Lambda_2 e^2~, \qquad 
\Lambda = \Lambda_1 e^1 +  \Lambda_{\b2} e^{\b2}~.
\ee
We will choose $ \t\Lambda_1,  \t\Lambda_2$ constant, for our purposes.
Then, the cubic-anomaly contribution \eqref{Anomaly SUSY qqq 4d} takes the simple form:
\be\label{susy anomaly 4d form}
\delta_\zeta W \big|_{q^3} = - {\CA_{qqq}\ov 12 \pi^2} \int \t\Lambda \wedge a\wedge da~.
\ee
at first order in the gaugino.%
\footnote{The higher-order terms in $\lambda$ can be ignored for our purpose, since they do not affect the anomalous Ward identities \protect\eqref{full holomorphy anomaly gen 4d}.} The 4d background $U(1)$ gauge field takes the form:
\be\label{a 4d def}
a= \h a_1 e^1 +\h a_{\b1} e^{\b1} + {\bf a}^{(\m)}~, \qquad \h a_1 = {\h a_w \ov \beta_1}~, \qquad  \h a_{\b1} = {\h a_\bw \ov \beta_1}~,
\ee
in terms of the complex frame, with $\h a_\bw= (\h a_w)^\ast \in \C$ an arbitrary complex parameter.
The term ${\bf a}^{(\m)}$ in \eqref{a 4d def} denotes a topologically non-trivial gauge field with flux $\m\in \Z$. If $p=0$, we have $\CM_4 \cong \Sigma_g\times T^2$ and $-\m$ units of flux through the Riemann surface $\Sigma_g$:%
\footnote{The choice of $-\m$ instead of $+ \m$ here is to conform to the definition of the ``flux operator'' in \protect\cite{Closset:2017bse}, where the orientation convention for $\Sigma$ was the opposite as it is here.}
\be\label{def am p0}
{1\ov 2\pi} \int_{\Sigma_g} d {\bf a}^{(\m)} =- \m~, \qquad {\rm if} \quad p=0~,
\ee
If $p\neq 0$, on the other hand, the flux $\m$ is a torsion flux, $\m\in \Z_p$, and ${\bf a}^{(\m)}$ is a flat connection with non-trivial holonomy along the Seifert fiber of the three-manifold $\CM_{g,p}$:
\be\label{def am pn0}
e^{i \int_\gamma  {\bf a}^{(\m)}}= e^{2 \pi i {\m\ov p}}~, \qquad {\rm if} \quad p\neq 0~.
\ee
While the two cases, $p=0$ or $p\neq 0$, must be treated separately, the final result is valid for every $\CM_{g,p}\times S^1$ background. The gauge field component $a_\bw$ are related to the above parameters as:
\be
a_{\bw} =\begin{cases}  \h a_\bw \qquad & {\rm if}\; p=0 \\   \h a_\bw - {i \tau \ov 2 \tau_2} {\m \ov p}~. \qquad & {\rm if }\; p\neq 0\end{cases}~,
\ee
and similarly for the complex conjugate $a_w$. The flavor parameters $\nu, \b\nu$ are then defined as:
\be\label{nu as aw def}
\nu = 2 i \tau_2\, a_\bw \equiv \tau a_\psi- a_t~, \qquad \quad
\b\nu =- 2 i \tau_2\, a_w = \b\tau a_\psi- a_t~.
\ee
Here, for future reference, we also gave these expressions in terms of the real gauge-field components $a_\psi$ and $a_t$.

\paragraph{The case $p\neq 0$, $\m=0$.}  Consider first the case without flux. When $\m=0$,  the gauge-field \eqref{a 4d def} is a well-defined one-form, and the integral \eqref{susy anomaly 4d form} can be evaluated straightforwardly. One finds:
\be
\delta_\zeta W \big|_{q^3, \, \m=0} = -{\pi i  p \CA_{qqq} \ov 3 \tau_2}   \t\Lambda_w \nu(\nu-\b\nu)~.
\ee

\paragraph{The general case.} In the general case with $\m \neq 0$, the integral \eqref{susy anomaly 4d form} is not well-defined, strictly speaking. Nonetheless, we claim that the general result should be:
\be\label{susy anomaly 4d qqq gen res}
\delta_\zeta W \big|_{q^3} = - \pi  \CA_{qqq} \t\Lambda_w\left( {i p \ov 3\tau_2} \nu(\nu-\b\nu) + \m \nu\right)~.
\ee
The term linear in $\m$ can be deduced by requiring consistency with explicit results for one-loop determinants and with the 3d reduction, as we will discuss below. A completely rigorous derivation of \eqref{susy anomaly 4d qqq gen res} would require an {\it a-priori} definition of \eqref{susy anomaly 4d form} for topologically non-trivial gauge fields, possibly akin to the definition of the 3d Chern-Simons functional in terms of a bounding four-manifold. We leave this as an interesting question for future work.

\subsubsection{The holomorphy anomaly on $Z_{\CM_{g,p}\times S^1}$}
Given the above discussion, the holomorphy anomaly of the supersymmetric partition function (in the flavor sector) follows from \eqref{full holomorphy anomaly gen 4d}. We have:
\be\label{holo anomaly MgpS1}
{\d W\ov \d \b\nu} = {1\ov 2 \tau_2} {\d \ov \d \t\Lambda_w} \delta_\zeta W~.
\ee
We also see that, since the supersymmetry variation \eqref{susy anomaly 4d qqq gen res} is independent of $\t\Lambda_z$ (and, similarly, $\delta_{\t\zeta} W$ is independent of $\Lambda_{\bz}$), there is no dependence of the partition function on background flat connections on the Riemann surface base, $\Sigma_g$. 

The supersymmetry variation \eqref{susy anomaly 4d qqq gen res} contributes the following holomorphy anomaly in the flavor parameters:
\be\label{holo anomaly 4d qqq gen res}
{\d W\ov \d \b \nu}  \Big|_{q^3}= -{ \pi  \CA_{qqq}}\left( {i p  \ov 6\tau_2^2}  \nu(\nu-\b\nu) + {\m\ov 2\tau_2} \nu\right)~.
\ee
The generalization to any abelian symmetry is straightforward.

\subsubsection{The small-$\beta_{S^1}$ limit and the 3d functional $\bW_{\rm 3d}$.}\label{subsec: W3d Mgp}
Let us now compute $\bW_{\rm 3d}$, as defined in section~\ref{subsec: introduce W3d}. It is obtained in the limit $\tau \rightarrow 0$ with $\nu/\tau$ kept finite. We again focus on the cubic 't Hooft anomaly contribution. We should then evaluate the local action:
\be\label{3d W1 q3 action explicit}
\bW_{\rm 3d}|_{q^3} = {\CA_{qqq}  \beta_{S^1}\ov 12 \pi} \int d^3 x \sqrt{g}\Big(-i  \sigma \epsilon^{\mu\nu\rho} A_\mu F_{\nu\rho} + 3  \sigma^2 D+\sigma^2 A_\mu \bV^\mu +\sigma^3 \bH\Big)~,
\ee
on the  3d $\CN=2$ supergravity background \eqref{3d sugra fields 0}-\eqref{3d sugra fields 1}. 
We also consider the background vector multiplet on its supersymmetric locus, \eqref{halfBPS vec 3d}. Then, the three-dimensional gauge field takes the form:
\be
A_\mu = \h a_0 \eta_\mu + {\bf A}^{(\bf m)}~,
\ee
with $\h a_0 \in \R$ a constant, and with the non-trivial gauge field ${\bf A}^{(\bf m)}$ defined exactly as ${\bf a}^{(\bf m)}$  in \eqref{def am p0} or \eqref{def am pn0}. Let us also introduce the real parameter:
\be
a_0 \equiv  A_\mu \eta^\mu = \begin{cases}  \h a_0 \qquad & {\rm if}\; p=0 \\  \h a_0 +{1\ov \beta_{\rm 3d}}{\m \ov p} \qquad & {\rm if }\; p\neq 0\end{cases}~.
\ee
The three-dimensional flavor parameters $\sigma$ and $a_0$ are related to the 4d flavor parameters \eqref{nu as aw def} by:
\be\label{nu to sigma}
{\nu\ov \tau}= \beta_{\rm 3d} (a_0 + i \sigma)~, \qquad \quad
{\b\nu\ov \b\tau}= \beta_{\rm 3d} (a_0 - i \sigma)~.
\ee
Equivalently, we have:
\be
  \beta_{\rm 3d} \sigma= {\tau_2 a_t\ov |\tau|^2}~, \qquad\qquad
     \beta_{\rm 3d} a_0=  a_\psi - {\tau_1 a_t \ov |\tau|^2}~,
\ee
in terms of the 4d gauge-field components $a_t$ and $a_\psi$. Similarly to the 4d expression \eqref{susy anomaly 4d form} for $\delta_\zeta W$, the action \eqref{3d W1 q3 action explicit} is only well-defined for $\m=0$. We claim that the complete expression for \eqref{3d W1 q3 action explicit} on this supersymmetric background is:
\bea\label{W3d full MgpS1}
&\bW_{\rm 3d}|_{q^3} &=&\; \pi \CA_{qqq} \beta_{S^1} \beta_{\rm 3d} \Bigg\{ p \beta_{\rm 3d} \left(-{2i\ov 3}\sigma a_0^2 + \big(1-i {\tau_1\ov 3\tau_2} \big) \sigma^2 a_0 + {\tau_1+ i \tau_2\ov 3\tau_2}\sigma^3\right) \cr
&&&\;\qquad\qquad  \qquad+ \m \left(i \sigma a_0 - \sigma^2\right) \Bigg\}~.
\eea
For $\m=0$, this is obtained by a straightforward computation using the supergravity and flavor background fields. In the general case with $\m\neq 0$, this is essentially a conjecture---assuming there exists a first-principle definition of the 3d action for non-trivial gauge fields. 
Using the relation \eqref{nu to sigma}, one can write \eqref{W3d full MgpS1} as:
\be\label{W3d full MgpS1 nunub}
\bW_{\rm 3d}|_{q^3} = {i \pi p \CA_{qqq} \ov \tau_2^2}  \left({\nu\b\nu^2\ov 12} - {\nu^2\b\nu \ov 6} + {\b\tau (2 \tau- \b \tau) \nu^3\ov 12 \tau^2} \right) -  {\pi \m \CA_{qqq} \ov 2 \tau_2} \left(\nu\b\nu - {\b\tau \nu^2\ov \tau}\right)~.
\ee
In particular, we see that $\bW_{\rm 3d}$ reproduces the holomorphy anomaly \eqref{holo anomaly 4d qqq gen res}---as it should, by construction. It will also be useful to give \eqref{W3d full MgpS1 nunub} in terms of the real parameters $a_\psi$ and $a_t$; we have:
\be\label{W3d ax ay}
\bW_{\rm 3d}(a_\psi, a_t, \tau)\big|_{q^3} = {i \pi \CA_{qqq}\ov 3 \tau^2}\, a_t (\tau a_\psi -a_t) \big( p (a_t - 2 \tau a_\psi) + 3 \m \tau\big)~. 
\ee
Incidentally, $\bW_{\rm 3d}|_{q^3}$ is holomorphic in $\tau$ when viewed as a function of $(a_\psi, a_t)$ and $\tau$. Finally, for future reference, let us also give the explicit expression for the real part:
\be\label{Re W3d}
{\rm Re}\left[\bW_{\rm 3d}\big|_{q^3} \right]= {\pi \CA_{qqq}\ov |\tau|^2} \left( \tau_2 (p a_\psi - \m) a_t^2 - {2p \tau_1\tau_2 \ov 3 |\tau|^2}  a_t^3 \right)~.
\ee

\subsection{The holomorphic supersymmetric partition function}
In the rest of this section, we would like to discuss the ``gauge invariant'' supersymmetric partition functions  $Z_{\CM_{g,p}\times S^1}$, in light of our general results for the supersymmetric Ward identities in WZ gauge. Before proceeding, let us first review the standard expressions for the holomorphic partition functions.

\subsubsection{The $S^3 \times S^1$ partition function}
The most studied 4d $\CN=1$ supersymmetric partition function is certainly the $S^3\times S^1$ partition function, which computes the 4d $\CN=1$ supersymmetric index:
\be\label{index pq}
\CI_{S^3\times S^1}({\bf p}, {\bf q}, y)=  \Tr_{S^3}\Big( (-1)^F {\bf p}^{J_3+J_3' +\half R} {\bf q}^{J_3-J_3' +\half R}  \prod_\alpha y_\alpha^{Q_F^\alpha}\Big)~.
\ee 
The parameters ${\bf p}$ and ${\bf q}$ are complex structure parameters on $S^3\times S^1$ seen as a Hopf surface \cite{Closset:2013vra}, and $y_\alpha= e^{2\pi i \nu_\alpha}$ are the flavor parameters.
In this work, we will only discuss the special case:
\be
{\bf p}= {\bf q}= q= e^{2\pi i \tau}~,
\ee
which corresponds to $(g,p)=(0,1)$ for the $\CM_{g, p} \times S^1$ background defined above \cite{Closset:2017bse}.  

The $S^3$ index is easily computed for a free theory. For a free chiral multiplet, $\Phi$, of unit charge under a background $U(1)$ and of $R$-charge $r\in \R$, we have:
\be\label{CIPhi def}
\CI^\Phi_{S^3\times S^1}(\nu, \tau)= \CF_1^\Phi(\nu + \tau(r-1), \tau)~, 
\ee
in terms of the function \cite{Closset:2017bse}:
\be\label{CF1 Phi def}
 \CF_1^\Phi(\nu, \tau) \equiv e^{2\pi i \tau \left({\nu^3\ov 6 \tau^3}- {\nu\ov 12\tau}\right)} \, \Gamma_0(\nu, \tau)~.
\ee
Here, the ``restricted elliptic $\Gamma$-function,'' $\Gamma_0$, is the ${\bf p}= {\bf q}$ limit of the elliptic-$\Gamma$ function, defined as:%
\footnote{The definition of $\Gamma_0$ corresponds to the limit ${\bf q}={\bf p}$ of the elliptic $\Gamma$-function with a shifted argument $\nu+\tau$. See Appendix~\ref{app:theta}.}
\be\label{def Gamma0}
\Gamma_0(\nu, \tau) = \prod_{k=0}^\infty \left({1- e^{2\pi i (-\nu + (k+1)\tau)}\ov 1- e^{2\pi i (\nu + (k+1) \tau)}}\right)^{k+1}~,
\ee
for $\tau$ on the upper-half plane. The prefactor in \eqref{CIPhi def} is the contribution from the supersymmetric Casimir energy for a free chiral multiplet  \cite{Assel:2014paa, Lorenzen:2014pna}. More generally, there exists an explicit formula for any 4d $\CN=1$ gauge theory, with gauge group $\GG$ and chiral multiplets $\Phi_i$ in representations $\FR$ of the gauge algebra $\Fg$. It is given by \cite{Assel:2014paa, Assel:2015nca}:
\be\label{susy S3 index}
\CI_{S^3\times S^1}(\nu, \tau) = e^{2\pi i \tau \CE(\nu, \tau)} \, {\bf I}_{S^3}(\nu, \tau)~.
\ee
Here, ${\bf I}_{S^3}$ denotes the ``ordinary'' 4d $\CN=1$ index; for SCFTs, it starts at $1$ when expanded in powers of $q$ \cite{Romelsberger:2005eg, Kinney:2005ej, Romelsberger:2007ec, Dolan:2008qi}:
\be\label{S3 index integral expl}
{\bf I}_{S^3} \equiv \frac{(q; q)^{2\rk}_\infty}{|W_\GG|}  \oint_{\prod_a \mathbb{T}_{x_a}}  \prod_{a=1}^{\rk} \frac{dx_a}{2\pi i  x_a} \,   {\prod_i\prod_{\rho_i}  \Gamma_0\big(\rho_i(u)+ \nu_i+\tau(r_i-1), \tau\big)\ov \prod_{\alpha\in \Fg} \Gamma_0\big(\alpha(u)-\tau, \tau\big)}~,
\ee
with $x_a\equiv e^{2\pi i u_a}$. The supersymmetric Casimir energy contribution in \eqref{susy S3 index} is given entirely in terms of the 't Hooft anomalies \cite{Assel:2015nca, Bobev:2015kza, Martelli:2015kuk}:
\be\label{Susy Casimir flavor}
\CE(\nu, \tau) = {1\ov 6 \tau^3}\CA^{\alpha\beta\gamma}\nu_\alpha\nu_\beta \nu_\gamma
- {1\ov 12\tau} \CA^\alpha \nu_\alpha + \cdots~,
\ee
where the ellipsis denotes contributions from the 't Hooft anomaly coefficients involving the $R$-symmetry.

\paragraph{Behavior under large gauge transformations.} On $S^3\times S^1$, one has the equivalence $\nu \sim \nu+1$ for the flavor parameter, corresponding to the large gauge transformation $a_t \sim a_t-1$ along the $S^1$. While the ``pure index'' \eqref{S3 index integral expl} is clearly gauge invariant, the holomorphic partition function \eqref{susy S3 index} is not. Due to the supersymmetric Casimir energy contribution, we have:%
\footnote{Here we ignored the effect of the mixed flavor-$U(1)_R$ anomalies, for simplicity.}
\be\label{CI S3 lgt}
\CI_{S^3\times S^1}(\nu+ n, \tau) = e^{-\pi i \CA_{q}^{\alpha} n_\beta}\, e^{{\pi i \ov \tau^2} \CA_{qqq}^{\alpha\beta\gamma}\left(n_\alpha u_\beta u_\gamma + n_\alpha n_\beta u_\gamma + {1\ov 3} n_\alpha n_\beta n_\gamma \right)}\, \CI_{S^3\times S^1}(\nu, \tau)~,
\ee
under the transformation $\nu_\alpha \rightarrow \nu_\alpha+ n_\alpha$ for the flavor parameters. Similarly to the case of the elliptic genus discussed in section~\ref{sec: 2d anomaly}, the puzzling aspect of the transformation \eqref{CI S3 lgt} is that the absolute value of $\CI_{S^3\times S^1}$ is not gauge-invariant. As we discussed, it cannot possibly be gauge invariant whenever the partition function is holomorphic.

\paragraph{The three-dimensional limit.}  Consider the small-$\beta_{S^1}$ limit of the holomorphic partition function, namely:
\be\label{Mgp 3d limit}
\tau \rightarrow 0~,\qquad \nu \rightarrow 0~, \qquad {\nu\ov \tau}= {\rm finite}.
\ee
In this limit, we claim that:
\be\label{log CIS3 expanded}
- \log \CI_{S^3\times S^1}(\nu, \tau)\;  \underset{\beta_{S_1}\to \,0}{\sim}\; {1\ov \beta_{S^1}} W^{(-1)}_{\rm 3d} +  W^{(0)}_{\rm 3d} + \CO\big(e^{-{1\ov \beta_{S^1}}}\big)~.
\ee
 Here, $W^{(-1)}_{\rm 3d}$ is given by the Cardy-like formula discussed in \cite{DiPietro:2014bca, DiPietro:2016ond}, and $W^{(0)}_{\rm 3d}$ is essentially the partition function of the dimensionally-reduced theory. The important point, for our purpose, is that there is no order-$\beta_{S^1}$ term in the expansion~\eqref{log CIS3 expanded}. This is analogous to the expansion~\eqref{log I2d} for the 2d elliptic genus.

 Note that the supersymmetric Casimir energy in~\eqref{Susy Casimir flavor} precisely survives in the limit  \eqref{Mgp 3d limit}. Therefore, the property~\eqref{log CIS3 expanded} is equivalent to the statement that the order-$\beta_{S^1}$ term in the expansion of the ordinary index is governed by the supersymmetric Casimir energy:
\be\label{IS3 expansion}
 -\log {\bf I}_{S^3}(\nu, \tau)\;  \underset{\beta_{S_1}\to \,0}{\sim}\; {1\ov \beta_{S^1}} W^{(-1)}_{\rm 3d} +  W^{(0)}_{\rm 3d} + 2 \pi i \tau \CE(\nu, \tau)+  \CO\big(e^{-{1\ov \beta_{S^1}}}\big)~.
\ee
For general gauge theories, using the explicit localisation formula \eqref{S3 index integral expl}, this expansion follows from the estimate:
\be\label{gamma0 estimate}
-\log \Gamma_0(\tau \t\nu, \tau) \;  \underset{\tau\to \,0}{\sim}\; {2 \pi i\ov \tau} {\t\nu \ov 12} +{\bf w}^{(0)}(\t\nu) + 2 \pi i \tau \left({\t\nu^3\ov 6}- {\t\nu \ov 12}\right) +\CO\big(e^{-{1\ov \beta_{S^1}}}\big)~,
\ee
 for the restricted elliptic $\Gamma$-function \eqref{def Gamma0}, with $\t \nu$ held fixed (and modulo $2\pi i$, corresponding to a choice of branch cut for the $\log$). Here, the finite term is given by:
  \bea
& {\bf w}^{(0)}(\t\nu)= -\log\left(e^{-{\pi i\ov 2} \t \nu^2 + {\pi i \ov 12}} f_\Phi(\t\nu)\right)~, \cr 
&f_\Phi(\t\nu) \equiv  \exp{\left({1\ov 2 \pi i} \dilog\left(e^{2\pi i \t\nu}\right)+ \t\nu \log\left(1-e^{2\pi i \t\nu}\right)\right)}~.
 \eea
corresponding to a free chiral multiplet on the round $S^3$ \cite{Jafferis:2010un}, with $\t\nu \equiv {\nu/\tau}$ the 3d flavor parameter \eqref{nu to sigma}. The expansion \eqref{gamma0 estimate} was proven in \cite{Ardehali:2015bla}.

\subsubsection{The $\CM_{g,p} \times S^1$ partition function}
The $\CM_{g,p} \times S^1$ holomorphic partition functions were computed in \cite{Closset:2017bse}. They can be written as generalized indices:
\be\label{CI Mgp}
\CI_{\CM_{g,p}\times S^1}(\tau, \nu)_\m =\Tr_{\CM_{g,p}}\Big((-1)^F q^{2J+ R}  \prod_\alpha y_\alpha^{Q_F^\alpha}\Big)~,
\ee
with $q=e^{2\pi i \tau}$ and $y= e^{2\pi i \nu}$ as above. For $p=0$, we have the topologically-twisted index on $\Sigma_g \times S^1$ \cite{Nekrasov:2014xaa, Benini:2015noa, Honda:2015yha}; for $p=1, g=0$, we recover the $S^3$ index discussed above.  The parameter $\m$ in \eqref{CI Mgp} denotes a background flux for the flavor symmetry on $\CM_{g,p}$, as discussed in subsection~\ref{subsec: flavor susy 4d Mgp} above.

For our purpose, we only need to know that the $\CM_{g,p}\times S^1$ holomorphic partition function for a 4d $\CN=1$ gauge theory may be written as:
\be
\CI_{\CM_{g,p}\times S^1}(\tau, \nu)_\m  = \sum_{\h u \in \CS_{\rm BE}} \CF_1(\h u, \nu, \tau)^p \, \CH(\h u, \nu, \tau)^{g-1} \, \prod_\alpha \Pi_\alpha(\h u, \nu, \tau)^{\m_\alpha}~,
\ee
where the sum is over the solutions to the ``4d Bethe equations,'' with $\h u$ a gauge parameter evaluated onto the so-called Bethe vacua  \cite{Nekrasov:2014xaa}. The objects ${\CF}_1$, $\CH$ and $\Pi_\alpha$ are called the fibering operator, the handle-gluing operator, and the flavor flux operators, respectively. 

For a free chiral multiplet, the (holomorphic) fibering operator is given by \eqref{CF1 Phi def}. The flavor flux operator, on the other hand, takes the form:
\be
\Pi^\Phi(\nu, \tau)= {e^{-{\pi i \nu^2\ov \tau}}\ov \theta(\nu, \tau)}~,
\ee
with $\theta(\nu, \tau)$ defined in Appendix~\ref{app:theta}. The handle-gluing operator of a free chiral multiplet of $R$-charge $r$ (with $r\in \Z$, on a generic $\CM_{g,p} \times S^1$) takes the form $\CH^\Phi= (\Pi^\Phi)^{r-1}$ \cite{Closset:2017bse}. 

\paragraph{Large gauge transformations.} On $\CM_{g,p} \times S^1$, there are two types of large gauge transformations for the gauge parameters.  In addition to $\nu \sim \nu +1$, corresponding to the large gauge transformation $a_t \sim a_t-1$ along the circle, we also have $\nu \sim \nu + \tau$ together with $\m \sim \m +p$, a large gauge transformation along the $S^1$ fiber of $\CM_{g,p}$.

Consider a 4d $\CN=1$ gauge theory whose partition function is computed as in \eqref{CI Mgp}. For generic values of $g,p$, the $R$-charges of the matter fields should be integer-quantized on $\CM_{g,p}\times S^1$. Then, the entire $R$-charge dependence of $\CI_{\CM_{g,p}}$ is through the handle-gluing operator, $\CH$, which drops out for $g=1$.  Here we focus on the contributions from the fibering and flux operator---schematically:
\be\label{CI Mgp schem}
\CI(\nu,Â \tau)_\m = \CF_1(\nu, \tau)^p\, \Pi(\nu, \tau)^\m~.
\ee
Under the large gauge transformation:
\be\label{lgt Mgp I gen}
(\nu_\alpha, \m_\alpha) \sim  (\nu_\alpha+ n_\alpha + m_\alpha \tau, \m_\alpha+ m_\alpha p)~,  \qquad n_\alpha, m_\alpha\in \Z~,
\ee
 for the flavor parameters, the fibering and the flux operator transform as:
 \bea
& \CF_1(\nu+ n+ m \tau, \tau) &=&\; e^{\varphi^{\CF_1}_{n, m}} \, \Pi(\nu, \tau)^{-m} \, \CF_1(\nu, \tau)~,\cr
&\Pi_\alpha(\nu+n + m \tau, \tau)&=&\; e^{\varphi^{\Pi_\alpha}_{n, m}}\, \Pi_\alpha(\nu,Â \tau)~,
 \eea
with the prefactors:%
\footnote{Here, $ \CA_{qq}^{\alpha\beta}$ denotes some quadratic ``pseudo-anomalies'' in 4d, similar to the ones discussed in footnote~\protect\ref{footnote pseudoA}. We leave a proper understanding of those terms for future work.}
\bea\nn
& e^{\varphi^{\CF_1}_{n, m}}\equiv  e^{-{\pi i \ov 2} \CA_{qq}^{\alpha\beta} m_\alpha m_\beta} e^{-{\pi i \ov 6} \CA_q^{\alpha}(n_\alpha + 3 m_\alpha)} e^{\pi i \CA^{\alpha\beta\gamma}_{qqq} \left(n_\alpha m_\beta m_\gamma + {n_\alpha m_\beta\ov \tau} (n_\gamma + 2 \nu_\gamma)+ {n_\alpha \ov 3 \tau^2}(n_\beta n_\gamma + 3 n_\beta \nu_\gamma + 3 \nu_\beta \nu_\gamma) \right)}~,\cr
&e^{\varphi^{\Pi_\alpha}_{n, m}} \equiv (-1)^{\CA_{qq}^{\alpha\beta} (n_\beta + m_\beta)} \, e^{-{\pi i \ov \tau}  \CA^{\alpha\beta\gamma}_{qqq}(n_\beta n_\gamma + 2 m_\beta \nu_\gamma)}~.
\eea
 Then, the holomorphic partition function~\eqref{CI Mgp schem} transforms as:
\be
\CI(\nu+ n + m \tau,Â \tau)_{\m+ m p} = e^{\varphi^\CI_{n, m}} \,\CI(\nu,Â \tau)_\m~, \qquad 
e^{\varphi^\CI_{n, m}} \equiv e^{p \varphi^{\CF_1}_{n, m}} \,  \prod_\alpha e^{(\m_\alpha + m_\alpha p) \varphi^{\Pi_\alpha}_{n, m}}~.
\ee
 One can easily generalize this expression to include the effect of the handle-gluing operators~\cite{Closset:2017bse}. 
 
 \paragraph{The three-dimensional limit.} As in the special case of the $S^3$ index, we again claim that there is no order-$\beta_{S^1}$ term to the supersymmetric index~\eqref{CI Mgp}. More precisely, one can again write the index as a ``pure index'' contribution times the supersymmetric Casimir energy contribution:
 \be
 \CI_{\CM_{g,p}\times S^1}(\nu, \tau)_\m = e^{2\pi i \tau \CE_{\CM_{g,p}}(\nu, \tau)_\m} \, {\bf I}_{\CM_{g,p}}(\nu, \tau)_\m~.
 \ee
 Explicit expressions for the Casimir energy were given in \cite{Closset:2017bse}.
 Then, we claim that the analogue of \eqref{IS3 expansion} holds for any $\CM_{g,p}$.
 In gauge-theories, this can be shown easily for the handle-gluing operator and for the flux operators, which are given in terms of $\theta$-functions. For the fibering operator, that small-$\beta_{S^1}$ property again follows from the estimate \eqref{gamma0 estimate}.

\subsection{The ``gauge-invariant'' supersymmetric partition function}\label{subsec: 4d Z revisit}
As we have reviewed, there are various exact results for 4d $\CN=1$ supersymmetric indices in the literature, and those formulae are locally holomorphic in all the parameters. While this holomorphy agrees with the known classical supersymmetric Ward identities on curved space~\cite{Closset:2013vra}, it clashes with simpler requirements that stem from gauge invariance for real background gauge fields coupling to the flavor symmetry. For instance, there should exists a scheme in which the absolute value of the supersymmetric partition function is gauge invariant.

The presence of a non-zero contribution from the flavor 't Hooft anomaly to the supersymmetric Ward identity, in the ``gauge invariant scheme,''  resolves this apparent puzzle---it is then possible to have a supersymmetric partition function which respects all the quantum Ward identities at the same time. The only price to pay is that the partition function is not fully holomorphic in the flavor parameters.

Given this understanding, it is now straightforward to revisit previous localisation computations to derive the ``gauge invariant'' partition functions. As explained before, the localisation argument still goes through. The only modification comes from a more careful computation of the various one-loop determinants on the supersymmetric locus. Since the new non-holomorphic terms depend on 't Hooft anomalies, they factor out of various ``localization integrals'' in gauge theories, and only contribute to a Casimir-like overall pre-factor. 

\paragraph{One-loop determinant and holomorphy.}
Let us focus on the contribution from a free chiral multiplet, $\Phi$. This is a free theory, whose path integral can be computed on any $\CM_{g, p} \times S^1$. In fact, we can factorize this one-loop determinant in terms of  the fibering and flux operators \cite{Closset:2017bse}:%
\footnote{Here, without loss of generality, we pick the $R$-charge $r=1$.}
\be
Z_{\CM_{g,p}\times S^1}^\Phi(\nu, \tau) = \h \CF_1^\Phi(\nu, \tau)^p \, \h \Pi^\Phi(\nu, \tau)^\m~.
\ee
Here, the hatted symbols denote the ``gauge invariant'' objects, by opposition to the holomorphic fibering and flux operators discussed above. In particular, the $S^3\times S^1$ partition function is given by:
\be
Z_{S^3\times S^1}^\Phi(\nu, \tau) = \h \CF_1^\Phi(\nu, \tau)~.
\ee
As always with perturbative anomalies, the difficulty lies in treating the chiral fermion, $\psi_\alpha$, whose one-loop determinant is not well-defined. On the other hand, we can consider a theory consisting of the chiral multiplet $\Phi$ together with a ``shadow chiral multiplet'' $\b \Phi$ of opposite chirality and gauge charge, whose kinetic operator is the Hermitian conjugate to the one of $\Phi$.%
\footnote{This is not a supersymmetric setup, but this is not an issue. We can think of the shadow chiral as coupling to the ``complex conjugate'' background geometry, so that its partition function is the complex conjugate of the one for $\Phi$.}  Then, the partition function of the pair $\Phi$ and $\b\Phi$ unambiguously defines the {\it absolute value} of $Z^\Phi$. 

In Appendix~\ref{app:chiralfer 4d}, we give an derivation of this absolute value by an explicit heat-kernel computation, generalizing the discussion of the Quillen anomaly in section~\ref{subsec: quillen anomaly 2d}. One finds:
\be\label{abs val res 4d i}
\big| \h \CF_1^\Phi(\nu, \tau)\big| = e^{- {\pi \tau_2\ov 3} \left(a_\psi^3 - {1\ov 2}a_\psi \right)}\, \big| \Gamma_0(\nu, \tau) \big|~, 
\ee
 for the fibering operator, and:
\be\label{abs val res 4d ii}
\big|\h \Pi^\Phi(\nu, \tau)\big|= e^{\pi \tau_2 \left(a_\psi^2 - a_\psi + {1\ov 6}\right)} \, {1\ov|\theta_0(\nu, \tau)|}
\ee
for the flavor flux operator, with $\nu = \tau a_\psi - a_t$. Moreover, from our general discussion in subsection~\ref{subsec: W3d Mgp}, we know the explicit form of the chiral-multiplet supersymmetric effective action at order $\beta_{S^1}$ in the small-circle limit; it is given by \eqref{W3d ax ay} with $\CA_{qqq}=1$, namely:
\be\label{W3d ax ay Phi}
\bW_{\rm 3d}^\Phi(a_\psi, a_t, \tau) = {i \pi \ov 3 \tau^2}\, a_t (\tau a_\psi -a_t) \big( p (a_t - 2 \tau a_\psi) + 3 \m \tau\big)~.
\ee
We then propose that the supersymmetric partition function of the free chiral multiplet is given by:
\be\label{Z Phi full answer}
Z_{\CM_{g,p}\times S^1}^\Phi(a_\psi, a_t, \tau)_\m = e^{-\bW_{\rm 3d}^\Phi(a_\psi, a_t, \tau) }\, \CI^\Phi_{\CM_{g,p}\times S^1}(\nu, \tau)_\m~,
\ee
with $ \CI^\Phi_{\CM_{g,p}\times S^1}$ the holomorphic partition function, including the supersymmetric Casimir energy contributions:
\be
 \CI^\Phi_{\CM_{g,p}\times S^1}(\nu, \tau)_\m= \CF_1(\nu, \tau)^p\, \Pi(\nu, \tau)^\m~.
\ee
One can check that \eqref{Z Phi full answer} reproduces \eqref{abs val res 4d i}-\eqref{abs val res 4d ii}---that is, minus the real part of $\bW_{\rm 3d}$, as given in \eqref{Re W3d}, plus the real part of $2\pi i \CE_{\CM_{g,p}}$ (the supersymmetric Casimir energy) exactly reproduces the factors in the exponents in~\eqref{abs val res 4d i}-\eqref{abs val res 4d ii}. In particular, the absolute value of \eqref{Z Phi full answer}, $|Z_{\CM_{g,p}\times S^1}^\Phi|$, is fully gauge invariant.

\paragraph{The ``gauge-invariant'' supersymmetric partition function.}
Building the partition function of any ``Lagrangian'' gauge theory from the chiral and vector multiplet building blocks, in the usual manner, one then obtains a simple proposal for the complete answer for the supersymmetric partition function in the gauge- and diff-invariant scheme. 

We may state this proposal as a conjecture for any supersymmetric partition function on a product four-manifold $\CM_3 \times S^1$, with geometric parameters $\tau$ and flavor parameters $\nu$. Let $\CI_{\CM_3\times S^1}$ denote the holomorphic partition function (or ``holomorphic $\CM_3$ index''), defined such that there is no order-$\beta_{S^1}$ contribution to $\CI_{\CM_3\times S^1}$ in the 3d limi---as in \eqref{log CIS3 expanded} for $\CM_3= S^3$.

\paragraph{Conjecture:} The ``gauge-invariant'' supersymmetric partition function $Z_{\CM_3 \times S^1}$ is related to the ``holomorphic partition function,'' $\CI_{\CM_3\times S^1}$, by the 3d local functional $\bW_{\rm 3d}$, evaluated on the supersymmetric background  and viewed as a function of the 4d parameters, according to:
\be\label{ZM3S1 conjecture}
Z_{\CM_3 \times S^1}(a_\psi, a_t,Â \tau, \b \tau) = e^{-\bW_{\rm 3d}(a_\psi, a_t, \tau, \b\tau)} \, \CI_{\CM_3\times S^1}(\nu, \tau)~.
\ee
By construction, this supersymmetric partition function is properly gauge-invariant and supersymmetric, in the sense that it is compatible with all the anomalous Ward identities.

\medskip\noindent
In this work, we focussed on the dependence of the partition function on the flavor parameters. The  form \eqref{ZM3S1 conjecture} of the supersymmetric partition function should also account for more subtle anomalous Ward identities in the ``geometric sector,'' which should also contribute to $\bW_{\rm 3d}$, and whose detailed study we leave for future work.

\paragraph{Behavior of $Z_{\CM_{g,p}\times S^1}$ under large gauge transformations.} Finally, we can discuss the behavior of the partition function under large gauge transformations. We again focus on the contributions from the fibering and flux operators in the $\CM_{g,p}\times S^1$ background. Then, under a large gauge transformations across the $S^1$:
\be\label{at shift n}
a_{t, \alpha} \sim a_{t, \alpha}+n_\alpha~,
\ee
we have:
\bea
&Z_{\CM_{g,p} \times S^1}(a_\psi, a_t-n,Â \tau, \b \tau)_\m = e^{\pi i \CA^{\alpha\beta}_{qq} \m_\alpha n_\beta} \, e^{-{\pi i \ov 6} p \CA^\alpha n_\alpha }\,\cr
&\qquad\qquad \times  e^{{\pi i } \CA^{\alpha\beta\gamma}\left({p\ov 3} a_{\psi, \alpha} a_{\psi, \beta} n_\gamma - \m_\alpha a_{\psi, \beta} n_\gamma \right)} \, Z_{\CM_{g,p} \times S^1}(a_\psi, a_t,Â \tau, \b \tau)_\m~.
\eea
We can also consider the large gauge transformation along the $\CM_{g,p}$ Seifert fiber:
\be
a_{\psi, \alpha} \sim a_{\psi, \alpha}+m_\alpha~, \quad \m_\alpha \sim \m_\alpha + m_\alpha p~,
\ee
which gives:
\bea
&Z_{\CM_{g,p} \times S^1}(a_\psi+m, a_t,Â \tau, \b \tau)_{\m+ m p} = e^{\pi i \CA^{\alpha\beta}_{qq}\left( \m_\alpha m_\beta+{p\ov 2} m_\alpha m_\beta\right)} \, e^{-{\pi i \ov 2} p \CA_q^\alpha m_\alpha }\,\cr
&\quad\qquad \times e^{{\pi i} \CA_{qqq}^{\alpha\beta\gamma}\left({p\ov 3} \left(a_{t, \alpha} a_{\psi, \beta} m_\gamma -  a_{t, \alpha} m_\beta m_\gamma \right)- \m_\alpha a_{t,Â \beta} m_\gamma \right)} \, Z_{\CM_{g,p} \times S^1}(a_\psi, a_t,Â \tau, \b \tau)_\m~.
\eea
For $p=0$, we find the behavior of the supersymmetric $\Sigma_g\times T^2$ partition function under large gauge transformations, which nicely agrees with the general expectations discussed in Appendix~\ref{subsec:large gauge 4d}.

In the case $p=1$, $\m=0$, we also find a very simple behavior for the $S^3 \times S^1$ supersymmetric partition function under large gauge transformations across the circle, namely:
\be\label{S3S1 lgt}
Z_{S^3 \times S^1}(\nu+n, \tau) =  e^{-{\pi i \ov 6} \CA^\alpha n_\alpha }  e^{{\pi i  p \ov 3} \CA^{\alpha\beta\gamma}a_{\psi, \alpha} a_{\psi, \beta} n_\gamma} \, Z_{S^3 \times S^1}(\nu, \tau)~,
\ee
for any $n\in \Z$. This is the result up to terms that depend on the $R$-charge; the full result can be obtained from \eqref{S3S1 lgt} by simply allowing the flavor indices $\alpha, \beta, \cdots$ in the exponent to run over the value ``$R$'' as well, at least formally. A more complete account of the $R$-symmetry dependence of partition functions, which is tied to the ``geometric sector,'' will be given elsewhere.

%%%%%%%%%%%%%%%%%%%%%%%%%%%%%%%%%%%%%%%%%%%%%%%%
\section*{Acknowledgements} We would like to thank Benjamin Assel,  Christopher Beem,  Greg Moore, Sameer Murthy, Boris Pioline, James Sparks and Antony Speranza for interesting discussions and correspondence. We especially thank Zohar Komargodski, Ioannis Papadimitriou and Brian Willett for their insightful comments and feedback.
CC is a Royal Society University Research Fellow and a Research Fellow at St John's College, Oxford. 
This research was supported in part by Perimeter Institute for
Theoretical Physics. Research at Perimeter Institute is supported by the Government of Canada through Industry Canada and by the Province of Ontario through the Ministry of Economic Development \& Innovation. The work of H.K. is supported by ERC Consolidator Grant 682608 Higgs bundles: Supersymmetric Gauge Theories and Geometry (HIGGSBNDL).

%%%%%%%%%%%%%%%%%%%%%%%%%%%%%%%%%%%%%%%%%%%%%%%%

%%%%%%%%%%%%%%%%%%%%%%%%%%%%%%%%%%%%%%%%%%%%%%%%%%%%%%%%%

\appendix

\section{Conventions in various dimensions}
In this appendix, we briefly summarize our conventions in various dimensions. In 4d and 3d, our conventions are the same as in \cite{Closset:2013vra}.

\subsection{Four-dimensional conventions}\label{Appendix: 4d}
We denote by $X^\mu$ the real coordinates, with $\mu=1, \cdots, 4$. We denote the real coframe one-forms by $e^{(a)}$, with the frame indices $a=1, \cdots 4$, and:
\be
ds^2(\CM_4) = \sum_{a=1}^4 (e^{(a)})^2~, \qquad   e_\mu^{(a)} e^{(b)}_\nu g^{\mu\nu}= \delta^{ab}~.
\ee
Then, the complex coframe one-forms is given by:
\be
e^1\equiv e^{(1)} + i e^{(2)}~, \quad e^{\b 1}\equiv e^{(1)} - i e^{(2)}~,\quad
e^2\equiv e^{(3)} + i e^{(4)}~, \quad e^{\b 2}\equiv e^{(3)} - i e^{(4)}~.
\ee
Four-dimensional spinors are written as Weyl spinors, denoted by $\psi= \psi_\alpha$ and 
$\t\psi =\t\psi^\alphadot$  for positive- and negative-chirality spinors, respectively, with the indices in their natural positions (in the standard Wess-and-Bagger conventions \cite{Wess:1992cp}). 
We use the Euclidean $\sigma^\mu$-matrices (in the real frame basis):
\be
(\sigma^{a}_{\alpha \betadot}) = (\sigma^1, \sigma^2, \sigma^3, -i {\bf 1})~,\qquad 
(\t\sigma^{a\, \alphadot \beta}) = (-\sigma^1, -\sigma^2, -\sigma^3, -i {\bf 1})~, 
\ee
for $a=1, \cdots, 4$, respectively, with $\sigma^i$ the Pauli matrices and ${\bf 1}$ the $2\times 2$ identity matrix (and, of course, $\sigma^\mu \equiv e^\mu_a \sigma^a$ in terms of the inverse vierbein).
We also define the antisymmetric combinations:
\be
\sigma^{\mu\nu} = {1\ov 4} (\sigma^{\mu}\t\sigma^\nu -\sigma^{\nu}\t\sigma^\mu)~, \qquad
\t\sigma^{\mu\nu} = {1\ov 4} (\t\sigma^{\mu}\sigma^\nu -\t\sigma^{\nu}\sigma^\mu)~,
\ee
which are proportional to the Lorentz symmetry generators in the Weyl spinor representations; they are self-dual and anti-self dual, respectively:
\be
\half\epsilon^{\mu\nu \rho\sigma} \sigma_{\rho\sigma} = \sigma^{\mu\nu}~, \qquad
\half\epsilon^{\mu\nu \rho\sigma} \t\sigma_{\rho\sigma} = -\t\sigma^{\mu\nu}~.
\ee 
Note the following useful identities:
\bea
&\sigma^\mu \t\sigma^\nu+\sigma^\mu \t\sigma^\nu= - 2 g^{\mu\nu}~, \qquad 
&&\t\sigma^\mu \sigma^\nu+\t\sigma^\mu \sigma^\nu= - 2 g^{\mu\nu}~,\cr
&\sigma^\mu\t\sigma^{\nu\rho} + \sigma^{\nu\rho} \sigma^\mu = \epsilon^{\mu\nu\rho\lambda} \sigma_\lambda~, 
&&\t\sigma^\mu \sigma^{\nu\rho} + \t\sigma^{\nu\rho} \t\sigma^\mu = -  \epsilon^{\mu\nu\rho\lambda} \t\sigma_\lambda~.
\eea
Spinor indices of either chirality are raised and lowered with $\ep^{\alpha\beta}$, $\ep_{\alpha\beta}$ or $\ep^{\alphadot\betadot}$, $\ep_{\alphadot\betadot}$, with $\ep^{12}= \ep_{21}=1$. Implicit spinor indices are contracted in the usual conventions, $\eta \psi \equiv \eta^\alpha \psi_\alpha$ and $\t\eta \t\psi \equiv \t\eta_\alphadot \t\psi^\alphadot$.
For any two spinors $\eta_\alpha$ and $\t\eta_\alphadot$ (which are independent from each other, in Euclidean signature), we have:
\be
\eta^\alpha \eta^\beta = -\half \ep^{\alpha\beta} \eta\eta~, \qquad\quad
\t\eta^\alphadot \t\eta^\betadot = \half \ep^{\alphadot\betadot} \t\eta\t\eta~, \qquad\quad
\t\eta^\alphadot  \eta^\alpha = \half \t\sigma^{\mu \alphadot \alpha} \, \eta\sigma_\mu\t\eta~.
\ee
Various other Fierz indentities can be derived straightforwardly.~\footnote{One can directly adapt various identities from \protect\cite{Wess:1992cp} by replacing $\eta^{\mu\nu}$ and $\epsilon^{\mu\nu\rho\sigma}$ there with $g^{\mu\nu}$ and $- i \epsilon^{\mu\nu\rho\sigma}$, respectively, to account for the change in space-time signature.}

\subsection{Three-dimensional conventions and dimensional reduction from 4d}\label{Appendix: 3d}
We denote by $x^\mu$ ($\mu=1,2,3$) the three-dimensional real coordinates. The three-dimensional real frame is denoted by:%
\footnote{Note that the we use coordinate and frame indices $\mu, \nu$ and $a, b, \cdots$, respectively, both for the four-dimensional and three-dimensional quantities. This should not lead to any confusion.}
\be
\h e^a =(\h e^1, \h e^2, \h e^3) = (\eta~,\, e^{(3)}~, \, e^{(4)})~, \qquad  a=1, 2, 3~.
\ee
This is obtained by dimensional reduction from 4d along the ``$e^{(2)}$ direction,'' in our conventions.
We  denote by:
\be
\h e^{(0)}\equiv e^{(2)}
\ee
the frame covector in that direction. Note that the 4d and 3d $\epsilon$-tensor are the related by $\epsilon^{abc}= - \epsilon^{0abc}$, in the frame basis. (The dimensional reduction is discussed in detail in \cite{Closset:2012ru}, which we follow.) In terms of local coordinates, we will have:
\be
x^\mu= (x^1, x^2, x^3) = (X^1, X^3, X^4)~.
\ee
The three-dimensional spinors are two-component Dirac spinors, denoted by either $\psi_\alpha$ or $\t \psi_\alpha$. The reduction of the 4d Weyl spinors give 3d spinors, with the index position given as:
\be
\psi_\alpha  |_{\rm 3d}= \psi_\alpha~, \qquad \t\psi^\alphadot |_{\rm 3d}= \t\psi^\alpha~. 
\ee
The three-dimensional $\gamma$-matrices are (note the default index positions):
\be\label{def gamma 3d}
({\gamma^a})_\alpha^{\phantom{\alpha}\beta}=(\sigma^3,-\sigma^1, - \sigma^2)~.
\ee
They satisfy $\gamma^\mu \gamma^\nu = g^{\mu\nu} + i \epsilon^{\mu\nu\rho}\gamma_\rho$. The 3d $\gamma^\mu$-matrices are obtained from the 4d $\sigma^\mu$ matrices according to:
\be
 \sigma^a_{\alpha \dot \beta} = \gamma^a_{\alpha\beta}~, \qquad  \sigma^0_{\alpha \dot \beta} = i \ep_{\alpha\beta}~,\qquad (\t \sigma^a)^{\dot \beta \alpha} =(\gamma^a)^{\beta \alpha}~, \qquad   (\t \sigma^0)^{\dot \beta \alpha }  = i \ep^{\beta\alpha}~.
\ee
One has to be careful with the signs when reducing from 4d to 3d, due in particular to the fact that the 4d ``dotted indices'' become ``undotted'' in 4d; for instance:
\be
\t \zeta \t \lambda |_{\rm 3d} = \t \zeta_\alphadot \t \lambda^\alphadot |_{\rm 3d} = - \t\zeta^\alpha \t\lambda_\alpha=- \t \zeta \t \lambda~.
\ee

\subsection{Two-dimensional and one-dimensional conventions}\label{Appendix: 2d}
In two dimensions, we denote the real coordinates by $X^\mu = (x, y)$ and we pick a complex coordinate $w= x+ \tau y$. The $\gamma$-matrices can be chosen as:
\be
(\gamma^\mu)= (-\sigma^1, -\sigma^2)~, \qquad \gamma^3= \sigma^3= \mat{1 & 0\\ 0 & -1}~.
\ee
 The 2d Dirac spinors decomposes into Weyl spinors of positive and negative chiralities, which we denote by $\psi_-$ and $\psi_+$, respectively---namely:
 \be
 \psi_\alpha= \mat{\psi_-\\ \psi_+}~.
 \ee
 They have 2d spin $\half$ and $-\half$, respectively. We then reduce to one dimension along the coordinate $x$. The 1d fermions do not carry any space-time index, and are simply denoted by:
 \be
\psi_-  |_{\rm 1d}= \psi~, \qquad \eta_+ |_{\rm 1d}= \eta~.
 \ee

\section{$\theta$-function and elliptic $\Gamma$-function}\label{app:theta}
In this appendix, we collect some useful definitions and properties for some $\theta$- and $\Gamma$-functions used in the main text. 
We will use the standard notation:
\be
q= e^{2 \pi i \tau}~, \qquad y= e^{2 \pi i \nu}~.
\ee

\paragraph{The function $\theta(\nu, \tau)$.} Let us define the ``reduced'' theta-function:
\be\label{def theta0}
\theta_0(\nu, \tau) \equiv \prod_{k=0}^\infty (1-q^k y)(1- q^{k+1} y^{-1})~,
\ee
with $\nu\in \C$ and $\tau$ valued on the upper-half-plane. 
The function \eqref{def theta0} has  simple zeros at $\nu = n + m \tau$, with $n, m\in \Z$. Then, the ``ordinary'' $\theta$-function is defined as:
\be
\theta(\nu, \tau) \equiv e^{\pi i \tau\ov 6} e^{-\pi i \nu}   \prod_{k=0}^\infty (1-q^k y)(1- q^{k+1} y^{-1})~.
\ee
It  satisfies:
\be
\theta(\nu+n + m\tau, \tau) = (-1)^{n+m} e^{-2 \pi i m \nu - \pi i m^2 \tau}\, \theta(\nu, \tau)~,
\ee
for any $n, m\in \Z$, and $\theta(-\nu, \tau) = -\theta(\nu,\tau)$. Under modular transformations, we have:
\be
 \theta(\nu; \tau+1)= e^{\pi i \ov 6}\theta(\nu; \tau)~, \qquad\quad
  \theta\Big({\nu\ov \tau}; -{1\ov \tau}\Big) = e^{2\pi i \left({\nu^2 \ov 2\tau}-{1\ov 4}\right)}\, \theta(\nu, \tau)~,
\ee
for the generator $T$  and $S$ of $SL(2, \Z)$,  respectively. Let us also define the $\eta$-function:
We have:
\be
\eta(\tau) = e^{\pi i \tau\ov 12} \prod_{k=1}^\infty (1-q^k)~.
\ee
It transforms as:
\be
\eta(\tau+1)= e^{\pi i \ov 12} \eta(\tau)~, \qquad\quad \eta\Big(-{1\ov \tau}\Big)=  \sqrt{-i \tau} \eta(\tau)~,
\ee
under $SL(2, \Z)$.

\paragraph{Restricted elliptic gamma function, $\Gamma_0(\nu, \tau)$.}
The elliptic gamma function $\Gamma_e(\nu, \tau,\sigma)$ can be defined as:
\be
\Gamma_e(\nu, \tau,\sigma) = \prod_{j,k=0}^\infty \frac{1-y^{-1}{\bf p}^{j+1}{\bf q}^{k+1}}{1-y {\bf p}^j {\bf q}^k}\ ,
\ee
where ${\bf p}=e^{2\pi i \sigma}$ and  ${\bf q}=  e^{2\pi i \tau}$. See {\it e.g.} \cite{1999math......7061F}  for a discussion of some of its properties. In this paper, we focus on the following ``restricted value'' of the elliptic gamma function, for ${\bf p}={\bf q}=q$:
\be\label{def Gamma0 app}
\Gamma_0(\nu, \tau) \equiv \Gamma_e(\nu+\tau, \tau,\tau) = \prod_{k=0}^\infty \left(\frac{1-y^{-1}q^{k+1}}{1-y q^{k+1}}\right)^{k+1}\ .
\ee
It satisfies:
\be
\Gamma_0(\nu+n+m\tau, \tau) = (-y)^{-\frac{m(m+1)}{2}}q^{-\frac16 m(m^2-1)}\theta_0(\nu,\tau)^m\Gamma_0(\nu,\tau)~,
\ee
for any $n, m\in \Z$.

\section{One-loop determinants and $\zeta$-function regularization}\label{app:chiralfer}
In this Appendix, we compute explicitly the (gauge-invariant) absolute value of one-loop determinants for chiral fermions and their superpartners.  We first review the computation of the determinant of the Dirac operator $D_\bw$ on $T^2$, following the computation of Ray and Singer in \cite{10.2307/1970909}. We then generalize that procedure in order to compute the absolute value of the one-loop determinant of a 4d $\CN=1$ chiral multiplet $\Phi$ on $\CM_{g,p} \times S^1$.

\subsection{Free fermions on $T^2$}\label{app:chiralfer 2d}
Consider a free chiral fermion $\lambda_-$ (of positive chirality)  on $T^2$, with complex coordinate $w=x + \tau y$, coupled to the background $U(1)$ flat connection parametrised by a constant:
\be
\nu = a_x \tau -a_y~.
\ee
We consider the determinant of the Dirac operator $D_{\bw}$, twisted by the background connection:
\be
D_{\bw} = \partial_{\bw} -i a_{\bw} = \partial_{\bw} -\frac{\nu}{2\tau_2}~,
\ee
with the periodic boundary condition along two circles.
We may then {\it define} the absolute value squared of the ``determinant'' of $D_\bw$ as:
\be
|\det(D_\bw)|^2 \equiv \det(D_w D_\bw)~.
\ee
Physically, this corresponds to the partition function for a pair of fermions of opposite chiralities, which admits an obviously gauge-invariant regularization.
The determinant of the Hermitian  operator $D_w D_\bw$ can be written as an infinite product:
\be
\det(D_w D_{\bw}) = \prod_{m,n\in \mathbb{Z}} \lambda_{m,n}~, \qquad  \qquad \lambda_{m,n} = \frac{1}{\tau_2^2}|\nu+m-\tau n|^2~.
\ee
Alternatively, one can consider  the untwisted Laplacian $\Delta \equiv  \partial_w\partial_{\bw}$ with the twisted boundary condition along the two circles:
\be\label{boundary condition}
\psi(w+2\pi)= e^{-2\pi i a_x} \psi(w)~, \qquad \quad
\psi(w+2\pi \tau) = e^{-2 \pi i a_y} \psi(w)~.
\ee
We would like to compute the effective action, defined as
\be
S_{\rm eff}(\nu, \tau) = - \sum_{n, m\in \Z}\log \lambda_{n, m}= \frac{d}{ds}\zeta_{\Delta}(s)\Big|_{s=0}~,
\ee
where  we introduced the zeta function for the Laplacian:
\bea\label{zeta reg}
\zeta_{\Delta}(s) &\equiv \sum_{m,n\in\mathbb{Z}}\lambda_{m,n}^{-s}\\
& = \frac{1}{\Gamma(s)} \int_{0}^\infty dt ~t^{s-1}~\text{tr}(e^{-t\Delta})\ .
\eea
In order to compute this, let us consider the heat kernel for the Laplacian:
\be
G(w, z, t) = \langle w, \bw |~ e^{- t \Delta}~ |z, \bz \rangle\ ,
\ee
which satisfies 
\be
(\d_t -  \Delta_w)~ G(w, z, t) = \begin{cases} \delta^2(w-z) &  {\rm if} \;  t=0 \\ 0 & {\rm if} \; t>0  \end{cases}\ .
\ee
On the complex plane, $\C$, the solution can be written as:
\be
G_0(w, z, t) = {1\ov 4 \pi t} e^{-{|w-z|^2\ov 4 t}}~.
\ee
The heat kernel on $T^2$ can be easily obtained by implementing the twisted boundary condition \eqref{boundary condition}. One can check that the following solution has the desired periodicities:
\be
G(w,z,t) = \sum_{m,n\in \mathbb{Z}}\frac{1}{4\pi t} e^{2\pi i (n a_x + ma_y)} \exp\left(-\frac{1}{4t}\left|w-z- 2\pi(n+m\tau )\right|^2\right)\ .
\ee
We can now write the trace in \eqref{zeta reg} as:
\be\label{trace1}
\text{tr}(e^{-t\Delta})=  \sum_{n, m\in \Z} {\pi \tau_2 \ov t} e^{2\pi i (na_x + m a_y)} e^{-{\pi^2|n+m\tau|^2\ov t}}~.
\ee
Removing the divergence from the term $m=n=0$, we can perform the $t$ integral when $\text{Re}(s)<1$, to obtain:
\bea
\zeta_{\Delta}(s) = 
{\pi\tau_2\Gamma(1-s)\ov \Gamma(s)}  \sum_{n^2+m^2>0}\; e^{2\pi i (n a_x+ma_y)}\left(\frac{1}{\pi^2|m\tau+n|^2}\right)^{1-s}\ .
\eea
One can show the summation uniformly converges when Re$(s)<\frac12$, once we first sum over $n \in \mathbb{Z}$. Taking the derivative with respect to $s$, we have:
\be\label{summation 2}
\zeta_{\Delta}'(0)= \frac{2\tau_2}{\pi}\sum_{n=1}^\infty {\cos(2\pi n a_x)\ov n^2}  +  \sum_{m\neq 0}\frac{1}{|m|}e^{2\pi i m a_y}\sum_{n\in \mathbb{Z}}e^{2\pi i n a_x} \frac{|m|\tau_2}{\pi|m\tau+n|^2}\ .
\ee
The first term gives:%
\footnote{On the second line here, we chose a particular branch for the dilogarithm, so that the final expression, in \protect\eqref{full answer Seff} below, is invariant under the large gauge transformation.}
\bea
&{2\tau_2\ov \pi}\sum_{n=1}^\infty {\cos(2\pi n a_x)\ov n^2}  &=&\;  {\tau_2\ov \pi}\Big(\dilog(e^{2\pi i a_x})+\dilog(e^{-2\pi i a_x})\Big)\\
& &=&  \; 2\pi \tau_2 \left(a_x^2- a_x+ {1\ov 6}\right)~.
\eea
The second term in \eqref{summation 2} can be computed using the Poisson summation formula:
\be\label{Poisson}
\sum_{n\in \mathbb{Z}}f(\nu+n) = \sum_{k\in \mathbb{Z}} e^{-2\pi i k \nu} \hat f(k)\ ,\qquad  \hat f(k) = \int_{-\infty}^\infty f(\xi) e^{2\pi i \xi k} d\xi\ .
\ee
for any function $f(\nu)$ which is periodic under $\nu\rightarrow \nu+1$. This gives:
\bea
& \sum_{m\neq 0}\frac{1}{|m|}e^{2\pi i m a_y}\sum_{n\in \mathbb{Z}}e^{2\pi i n a_x} \frac{|m|\tau_2}{\pi|m\tau+n|^2} \\
&\qquad= \sum_{m\neq 0}\frac{1}{|m|}e^{2\pi i m a_y}\sum_{k\in \mathbb{Z}} e^{-2\pi (|m|\tau_2 |a_x+k|+im\tau_1(a_x+k))}\\
&\qquad= -\sum_{k=0}^\infty \log|1-q^k e^{2\pi i (-a_y+\tau a_x)}|^2 -\sum_{k=0}^\infty \log|1-q^{k+1}e^{2\pi i (a_y - \tau a_x )}|\\
&\qquad= -\log|\theta_0(\nu, \tau)|^2~,
\eea
with $\theta_0(\nu, \tau)$ defined as in \eqref{def theta0}.
We then arrive at the expression:
\be\label{full answer Seff}
e^{-S_{\rm eff}(\nu, \tau)}= e^{-2\pi \tau_2\left( a_x^2- a_x+ {1\ov 6}\right)}\; |\theta_0(\nu, \tau)|^2~.
\ee
As discussed in section \ref{subsec: quillen anomaly 2d}, an holomorphic square root of \eqref{full answer Seff} does not exist. Instead, we find:
\be
\det(D_\bw) = e^{-2\pi \tau_2\left(\frac{a_x^2}{2}- \frac{a_x}{2}+ {1\ov 12}\right)}\;\theta_0(\nu, \tau)~,
\ee
up to a phase factor, which should be compatible with the $U(1)$ anomaly.

\subsection{4d $\CN=1$ chiral multiplet on $\CM_{g,p}\times S^1$}\label{app:chiralfer 4d}
We can run a similar computation for a chiral multiplet (of $R$-charge $r=1$) on $\CM_{g,p} \times S^1$. As reviewed in section \ref{subsec: 4d Z revisit}, the partition function on that space can be factorized into contributions from ``fibering and flux operators.'' The flux operator computation is essentially identical to the $T^2$ computation above. We then focus on the  fibering operator contribution.

The fibering operator for the chiral multiplet, $\h\CF_1^\Phi$, is given formally by the infinite product \cite{Closset:2017bse}:
\be
\h\cF_1^{\Phi}(\nu) \equiv \prod_{n,m\in \mathbb{Z}}\left[\frac{1}{\tau_2^2}(\nu+m-\tau n)\right]^{n}\ .
\ee
We will compute the absolute value of this expression, by a careful zeta-function regularization of the quantity:
\be\label{Fibering squared}
\left|\h\cF_1^{\Phi}(\nu)\right|^2 \equiv \prod_{n,m\in \mathbb{Z}} (\lambda_{m,n})^{n}\ ,~~~\text{where } \lambda_{m,n}= \frac{1}{\tau_2^2}|\nu+m-\tau n|^2\ .  
\ee
We will write $\nu= \tau a_\psi - a_t$, to match the 4d notation in the main text. 
Following the 2d computation in the last section, we introduce the zeta function:
\be\label{zetaF}
\zeta_{F}(s) = \sum_{m,n\in \mathbb{Z}} n\lambda_{m,n}^{-s}= \frac{1}{\Gamma(s)} \int_0^{\infty} dt~t^{s-1} \sum_{m,n\in \mathbb{Z}} n e^{-t\lambda_{m,n}}\ ,
\ee
so that we have:
\be
\log|\h\CF_1^{\Phi}(u)|^2 = -\left.\frac{d}{ds}\zeta_F(s)\right|_{s=0}\ .
\ee
Let us define:
\be
G(t) = \sum_{m,n\in \mathbb{Z}} n e^{-t\lambda_{m,n}}\ .
\ee
In order to use the Poisson summation formula, we decompose $G(t)$ into a linear combination of two functions periodic in $\nu\rightarrow \nu+1$ and $\nu\rightarrow \nu+\tau$:
\be
G(t) = G^{(1)}(t)+a_\psi \, G^{(2)}(t)\ ,
\ee
where:
\be
G^{(1)}(t) = \sum_{m,n\in \mathbb{Z}}(-a_\psi+n) e^{-t\lambda_{m,n}}~, \qquad
G^{(2)}(t) = \sum_{m,n\in\mathbb{Z}} e^{-t\lambda_{m,n}}~.
\ee
Note that $G^{(2)}(t)$ is equivalent to \eqref{trace1}. On the other hand, $G^{(1)}(t)$ can be computed by applying the Poisson summation formula \eqref{Poisson} twice. We find:
\be
G^{(1)}(t) = \frac{ i\pi^2\tau_2}{t^2}\sum_{m,n\in\mathbb{Z}} (n+m\tau_1) e^{2\pi i(na_\psi+ma_t)}e^{-\pi^2|n+m\tau|^2/t}\ .
\ee
We can also decompose the zeta function \eqref{zetaF} accordingly:
\be
\zeta_F(s) = \zeta_F^{(1)}(s) + a_\psi \, \zeta_F^{(2)}(s)\ ,
\ee
with:
\be
\zeta_F^{(i)}(s) = \frac{1}{\Gamma(s)} \int_0^\infty dt~ t^{s-1} G^{(i)}(t)\ ,\qquad i=1,2\ .
\ee
The $t$ integral for $\zeta_F^{(1)}(s)$ converges provided that Re$(s)<2$. We have
\be
\zeta_F^{(1)}(s) = \frac{i\pi^2 \tau_2 \Gamma(2-s)}{\Gamma(s)}\sum_{n^2+m^2>0}(n+m\tau_1) e^{2\pi i (ma_t +na_\psi)} \left(\frac{1}{\pi^2|n+m\tau|^2}\right)^{2-s}\ .
\ee
As in the 2d case, one can show that the infinite sum uniformly converges when Re$(s)<\frac{3}{2}$, if we sum over $n\in\mathbb{Z}$ first with fixed $m$:
\bea\nn
\zeta_F^{(1)}(s) =& \frac{i\pi^2\tau_2\Gamma(2-s)}{\Gamma(s)}\sum_{n\neq 0} n e^{2\pi i n a_\psi} \left(\frac{1}{\pi^2n^2}\right)^{2-s}\\
&+\frac{i\pi^2\tau_2\Gamma(2-s)}{\Gamma(s)}\sum_{m\neq 0} e^{2\pi i ma_t}\sum_{n\in\mathbb{Z}}(n+m\tau_1)e^{2\pi i na_\psi} \left(\frac{1}{\pi^2|n+m\tau|^2}\right)^{2-s}\ .
\eea
We have:
\bea\label{zeta 4d interm 1}
\left.\frac{d}{ds}\zeta_F^{(1)}(s)\right|_{s=0} =~ & \frac{i\tau_2}{\pi^2} \sum_{n\neq 0}\frac{e^{2\pi i n a_\psi}}{n^3} \\
&+\frac{i\tau_2}{\pi^2}\sum_{m\neq 0}e^{2\pi im a_t} \sum_{n\in\mathbb{Z}} e^{2\pi in a_\psi} \frac{(n+m\tau_1)}{|n+m\tau|^4}\ .
\eea
The first line evaluates to:
\bea
\frac{i\tau_2}{\pi^2}\sum_{n\neq 0}\frac{e^{2\pi i n a_\psi}}{n^3} 
&= \frac{i\tau_2}{\pi^2} \left(\text{Li}_3(e^{2\pi i a_\psi})-\text{Li}_3(e^{-2\pi i a_\psi})\right) \\
& = -\frac{i\tau_2}{\pi^2} \left(\frac16 (2\pi i a_\psi -i\pi)^3 + \frac{\pi^2}{6}(2\pi i a_\psi - i\pi)\right) \\
& = -2\pi\tau_2\left(\frac23 a_\psi^3 -a_\psi^2 + \frac13 a_\psi\right)~.
\eea
and the second line in \eqref{zeta 4d interm 1} can be again computed using the Poisson summation formula:
\bea
&\frac{i\tau_2}{\pi^2} \sum_{m\neq 0}e^{2\pi i m a_t}\sum_{n=-\infty}^{\infty} e^{2\pi i n a_\psi}\frac{(n+m\tau_1)}{|n+m\tau|^4}\\
=&~ \sum_{k\in \mathbb{Z}} \sum_{m\neq 0}\frac{e^{2\pi im a_t}}{|m|}(-a_\psi-k)e^{-2\pi\left[|m|\tau_2|a_\psi+k|+im\tau_1(a_\psi+k)\right]}\\
=&-\sum_{k=1}^{\infty}\log |1-y^{-1}q^k|^{2k} + \sum_{k=1}^\infty\log|1-y q^{k}|^{2k} + a_\psi\log|\theta_0(\nu,\tau)|^2\\
=&-\log |\Gamma_0(\nu, \tau)|^2 +a_\psi\log|\theta_0(\nu, \tau)|^2 \ ,
\eea
with $\Gamma_0(\nu, \tau)$ defined as in \eqref{def Gamma0 app}.
Combining all the terms, we have:
\bea
-\left.\frac{d}{ds}\zeta_F(s)\right|_{s=0} &= -\left.\frac{d}{ds}\left(\zeta^{(1)}_F(s)+a_\psi \zeta^{(2)}_F(s)\right)\right|_{s=0} \\
& = -2\pi \tau_2 \left(\frac{a^3_\psi}{3}-\frac{a_\psi}{6}\right) + \log |\Gamma_0(\nu, \tau)|^2\ .
\eea
Therefore, we found:
\be
\left|\h\cF_1^{\Phi}(\nu,\tau)\right|^2 = \exp\left[-2\pi\tau_2\left(\frac{a^3_\psi}{3}-\frac{a_\psi}{6}\right)\right] |\Gamma_0(\nu,\tau)|^2\ .
\ee
This expression does not admit a holomorphic square root. We have:
\be
\h\cF_1^{\Phi}(\nu, \tau) = \exp\left[-\pi\tau_2\left(\frac{a^3_\psi}{3}-\frac{a_\psi}{6}\right)\right] \Gamma_0(\nu, \tau)~,
\ee
up to a phase factor.

%%%%%%%%%%%%%%%%%%%%%%%%%%%%%%%%%

%%%%%%%%%%%%%%%%%%%%%%%%%%%%%%%%%%%%%%%%%%%%%%%%%%%%%%%%%
\section{Large-gauge transformations and the anomaly polynomial}
In the presence of 't Hooft anomalies, the partition function $Z[a]$, viewed as a functional over the space of background gauge fields $a_\mu$, $\fM$, is not a ``function'' over $\fM$ but rather a non-trivial section of a line bundle:
\be
\SL \rightarrow \fM~.
\ee
In the case of free fermions, the partition function is the ``determinant'' of a Dirac operator, and $\SL$ is then know as the determinant line bundle. The anomalies of the theory are encoded invariantly in terms of the non-trivial topology of $\SL$  \cite{Freed:1986hv}.

In this appendix, we compute explicitly the first Chern class of the determinant line bundle is some simple cases, with $Z[a]$ depending only on background flat connections. The first Chern class $c_1(\SL)$ can be obtained by pushing-foward the anomaly polynomial  onto space-time \cite{Freed:1986hv}, and it is thus determined directly by the anomalies. In this approach, it is easy to see that, even as we restrict ourselves to a subspace of flat gauge fields $a_\mu$, $\fM_0 \subset \fM$, the anomaly still implies that the partition function must transform non-trivially under large gauge transformations.

\subsection{Determinant line bundles on $J(T^2)$ and torus partition function}\label{large gauge 2d}

Consider a 2d $\CN=(0,2)$ supersymmetric theory coupled to a $U(1)$ background gauge multiplet $\CV_F$. In the presence of a t'Hooft anomaly, the partition function $Z(\nu,\tau)$, with a fixed complex structure $\tau$, can be thought of as a section of a holomorphic line bundle $\SL$ over the Jacobian torus:
\be
\nu~\in~J(T^2)\cong T^2~,
\ee
with $\nu = \tau a_x - a_y$ as in section~\ref{sec: 2d anomaly}.
Let $\CN$ be the product space $J\times T^2$, with the projection map:%
\footnote{We denote by $T^2$ the physical spacetime, and by $J$ the Jacobian.}
\be
J\times T^2\overset{\pi}{\longrightarrow} J\ .
\ee
Then, $\SL$ can be obtained from the Poincar\'e line bundle $\CL$ over $J\times T^2$ of degree zero.% 
\footnote{The Poincar\'e line bundle $\CL$ for a curve $\Sigma$ is defined by the property that, when we restrict to a point $L\in J$, it reduces to $L$ on $\Sigma\sim \Sigma\times \{L\}$. $\CL$ is defined up to a pull-back of a line bundle $\CR$ on $J$, {\it i.e.}, $\CL\sim \CL'\otimes \pi^* \CR$. See \protect\cite{arbarello1985geometry} for a detailed explanation.}
The first Chern class of the ``determinant line bundle,'' $\SL$, is obtained by pushing-forward the 4-form anomaly polynomial on $T^2$, according to \cite{Freed:1986hv}:
\be\label{first chern}
c_1(\SL) = \int_{T^2} \left.\text{ch}(\CL)\wedge \hat A(T\CN)\right|_{\text{4-form}}\ .
\ee
In particular, under the large gauge transformation:
\be
\nu\rightarrow \nu+1\ ,~~ \nu\rightarrow \nu+\tau\ ,
\ee
the $T^2$ partition function should pick up a phase determined by the curvature of the line bundle \eqref{first chern}. Since the tangent bundle on $T^2\times J$ is trivial, \eqref{first chern} can be written as:
\be
c_1(\SL) = \int_{T^2} \frac12 c_1(\CL)^2~.
\ee
We can decompose $c_1(\CL)$ as:
\be
c_1(\CL) = \delta_{0,2} + \delta_{1,1} + \delta_{2,0}\ ,
\ee
where $\delta_{p,q}$ is a $p$-form along $T^2$ and a $q$-form along $J$. Now, we have:
\be
c_1(\SL) = \frac12\int_{T^2}\Big( \delta^2_{1,1} + 2\delta_{2,0}\delta_{0,2}\Big)~.
\ee
Restricting ourselves to flat gauge fields on $T^2$, $f= da=0$, we are left with:
\be\label{c1 t2}
c_1(\SL) = \frac12\int_{T^2} (\delta_{1,1})^2 \ .
\ee
This can be computed by explicitly constructing the locally holomorphic section of $\CL$~\cite{alvarez2002beyond}. One can show that:
\be
c_1(\SL) =\CA_{qq}\, \omega~,
\ee
where $\CA_{qq}$ is the quadratic 't Hooft anomaly, and $\omega$ is the normalised K\"ahler form on $J$. For a theory coupled to a general abelian background gauge multiplet, the parameter space is a product of the Jacobian tori for each $U(1)_\alpha$, and the formula generalises to:
\be
c_1(\SL) = 2\pi\CA^{\alpha\beta} da_{x,\alpha} da_{y,\beta}\ ,
\ee
where $\nu_\alpha = a_{x,\alpha}\tau -a_{y,\alpha}$.
In particular, the connection on the line bundle $\SL$ can be locally chosen as:
\be\label{connection1}
A = 2\pi\CA^{\alpha\beta}\left(c\cdot a_{\alpha, x} da_{\beta, y} - (1-c)\cdot a_{\beta, y} da_{\alpha,x}\right)~,
\ee
for any $c\in \mathbb{R}$. This implies that a section $Z(\n)$ of the line bundle $\SL$ picks up a phase under the large gauge transformation along a 1-cycle $\gamma$, as:
\be
\phi_\gamma:Z(\nu) \rightarrow e^{i\Lambda_{\gamma}(a_x, a_y)} Z(\nu)\ ,
\ee
when the connection transforms as $\delta_\gamma A = d\Lambda_\gamma$.
More explicitly, we have
\bea
&Z(\nu_\alpha+1,\tau) = e^{2\pi i (1-c)\CA^{\alpha\beta}a_{\beta,x}} Z(\nu, \tau)~,\\
&Z(\nu_\alpha+\tau, \tau) = e^{2\pi i c \CA^{\alpha\beta}a_{\beta,y}} Z(\nu,\tau)~.
\eea
For 2d $\CN=(0,2)$ supersymmetric theories, the consistency of the gauge anomaly with supersymmetry fixes the constant $c=\frac12$, as explained in the main text.

\paragraph{Modular transformation}
The relation between the gravitational anomaly and the 
behaviour of the partition function under the $SL(2,\mathbb{Z})$ modular transformation:
\bea
&T~:(a_x, a_y,\tau)\rightarrow (a_x, a_y+ a_x, \tau+1)~,\\
&S~:(a_x, a_y,\tau)\rightarrow (a_y, -a_x, -1/\tau)~,
\eea
can be understood in a similar way, as explained in detail in \cite{Seiberg:2018ntt}. We view the parameter space $Y$ as a fibration:
\be
J \rightarrow Y\rightarrow \mathbb{H}/SL(2,\mathbb{Z})\ ,
\ee 
where the fiber at each point $\tau \in \mathbb{H}/SL(2,\mathbb{Z})$ is the Jacobian $J$ (with the complex structure induced by $\tau$). Now let $\CN$ be the product $T^2\times Y$.
Again, the first chern class of $\SL$ over $Y$ can be computed from the anomaly polynomial:
\bea\label{chern modular}
c_1(\SL) &=& \int_{T^2}\text{ch}(\CL)\wedge \hat A(T\CN)\Big|_{\text{4-form}}\\
&=& \frac12 c_1(\CL)^2 + \frac{1}{24} \int_{T^2}p_1(T\CN)
\eea
In the absence of the flavour symmetry, the first Chern class is determined by the gravitational term:
\be
c_1(\SL) = \frac{k_g}{2} \lambda\ ,
\ee
where $\lambda$ is the first Chern class of the Hodge bundle over $\mathbb{H}/SL(2,\mathbb{Z})$. This implies that the partition function, $Z(\tau)$, transforms as the one-dimensional representation of $SL(2,\mathbb{Z})$: 
\be
Z(\tau+1) = e^{-i\pi k_g/6}Z(\tau)\ ,~~Z(-1/\tau) = e^{i\pi k_g/2}Z(\tau)\ . 
\ee 
Now, let us couple the theory to the flat background gauge for the flavour symmetry, $\nu\in J$. The first term of \eqref{chern modular} implies that the connection on the line bundle restricted to the fiber direction can be locally written as \eqref{connection1}, which transforms under the $T$ operation as:
\be
\delta_T A = \pi \CA^{\alpha\beta} (2c-1)d(a_{\alpha,x}a_{\beta,x}) \ ,
\ee
and under $S$ as:
\be
\delta_S A = -2\pi \CA^{\alpha\beta} (2c-1) d(a_{\alpha,x}a_{\beta,y})\ .
\ee
Therefore the section $Z(\nu,\tau)$ of $\SL\rightarrow Y$ transforms as:
\bea
&Z(a_x,a_y+a_x,\tau+1) = e^{-i\pi k_g/6} e^{\pi i \CA^{\alpha\beta}(2c-1)a_{\alpha,x}a_{\beta,x}}Z(a_x,a_y,\tau)~,\cr
&Z(a_y, -a_x, -1/\tau) = e^{i\pi k_g/2} e^{-2\pi i \CA^{\alpha\beta}(2c-1)a_{\alpha,x}a_{\beta,y}}Z(a_x,a_y,\tau)~,
\eea
under the $T$ and $S$ transformations, respectively.
In particular, given the ``supersymmetric'' value $c=\frac12$, the flavour dependence drops out and we reproduce the simple formulae \eqref{modularT1} and \eqref{modularS1}.

\subsection{Large gauge transformations for the $4d$ partition function}\label{subsec:large gauge 4d}
Consider now the supersymmetric partition function on $\CM_{g,p}\times S^1$. Now, a supersymmetric  $U(1)$ background gauge field is characterized by an integer $\m\in \Z$ (if $p=0$) or $\m\in \Z_p$ (if $p\neq 0$), and by a complex parameter $\nu$, with the identifications:
\be\label{large gauge 4d}
(\nu,\m)\rightarrow (\nu+1,\m)\ ,\qquad
(\nu,\m)\rightarrow (\nu+\tau, \m+p)~,
\ee
under large gauge transformations.

Here, we will only discuss the case $p=0$, corresponding to $\Sigma_g \times T^2$. Then, $\nu$ is valued in the Jacobian torus $J(T^2)$, and the discussion above generalizes easily. The $\Sigma_g\times T^2$ partition function can be built out of the (non-holomorphic) flavor flux operator and handle-gluing operator. As a special case of the discussion in the main text, the flux operator of a free chiral is given by:
\be\label{Pi nonholo app}
\h \Pi^\Phi(\nu, \tau)={e^{-\pi i \left(\tau a_\psi^2- a_\psi a_t \right)}\ov \theta(\nu, \tau)}~,
\ee
with $\nu = \tau a_\psi - a_t$, and similarly for the handle-gluing operator. A general partition function can be constructed from those building blocks. By explicit computation, one can check that the non-holomorphic flux operators of a general 4d $\CN=1$ gauge theory transform as:
\bea\label{app flux}
&\hat\Pi_\alpha(\nu_\alpha+1, \tau) = (-1)^{\CA^{\alpha\beta}}e^{-\pi i \CA^{\alpha\beta\gamma}a_{\psi,\gamma}}\hat\Pi_\alpha(\nu, \tau)\ ,\\
&\hat\Pi_\alpha(\nu_\beta+\tau, \tau) = (-1)^{\CA^{\alpha\beta}}e^{-\pi i\CA^{\alpha\beta\gamma}a_{t,\gamma}} \hat\Pi_\alpha(\nu, \tau)~.
\eea
These relations can be understood as before, on general grounds (except for the sign, which is more subtle).

 For a fixed background gauge flux $-\m$ on the base Riemann surface $\Sigma_g$, following the discussion in section \ref{large gauge 2d}, we consider the product space $\CN_6  = J\times \Sigma_g \times T^2$ with the projection $\pi$ onto $J$:
\be
J\times \Sigma_g \times T^2 \overset{\pi}{\longrightarrow} J.
\ee
The determinant line bundle $\SL$ on $J$ can be obtained from the line bundle $\cL_{-\m}$ on $\CN_6$:
\be
\cL_{-\m} = L_{-\m}\otimes \CL\ ,
\ee
where $L_{-\m}$ is the pull-back of the line bundle of degree $-\m$ on $\Sigma_g$, and $\CL$ is the pull-back of the Poincar\'e line bundle of degree zero for $T^2$. The first Chern class of $\SL$ can be computed from the 6-form anomaly polynomial, according to the formula:
\be
c_1(\SL) = \int_{\Sigma_g\times T^2}\left.\text{ch}(\CL_\m)\wedge \hat A(T\CN_6)\right|_{\text{6-form}}\ .
\ee
Since $p_1(T\CN_6)=0$, we have:
\be
c_1(\SL) = \frac16\int_{\Sigma_g\times T^2} c_1(\CL_{-\m})^3\ .
\ee 
It is convenient to decompose the integrand into a linear combination of type $\delta_{p,q,r}$, the $p$- ,$q$- and $r$-forms along the $\Sigma_g$, $T^2$ and $J$, respectively: 
\be
c_1(\CL_{-\m}) = \delta_{2,0,0} + \delta_{0,2,0}+ \delta_{0,0,2} + \delta_{1,1,0} + \delta_{0,1,1} +\delta_{1,0,1}\ . 
\ee
Since $\delta_{0,2,0}= \delta_{1,1,0}=0$ for our background, the expression reduces to:
\be
c_1(\SL) = \frac12 \int_{\Sigma_g\times T^2} \delta_{2,0,0}\delta_{0,1,1}^2\ .
\ee
The integration over $\Sigma_g$ gives 
\be
\int_{\Sigma_g} \delta_{2,0,0} =- \CA_{qqq} \m - \CA_{qqR}(g-1)~,
\ee
in terms of the 't Hooft anomaly coefficients. The integration over $T^2$ can be done as in \eqref{c1 t2}. We conclude that, for a theory coupled to a general abelian background gauge field, we obtain
\be\label{c1 SL 4d res}
c_1(\SL) =- \CA^{\alpha\beta\gamma} \m^\alpha da_{x,\beta}da_{y,\gamma} - \CA^{R\alpha\beta} (g-1) da_{x,\alpha} da_{y,\beta}~.
\ee
Note that this is linear in $\m$. This then explains the large gauge transformations \eqref{app flux} for the flux operator, exactly as in section~\ref{large gauge 2d}. The second term in \eqref{c1 SL 4d res} similarly determines the transformations properties of the handle-gluing operator.

%%%%%%%%%%%%%%%%%%%%%%%%%%%%%%%%%

\bibliographystyle{utphys}
\bibliography{bibholo}{}

\end{document}